\documentclass[fleqn,usenatbib]{mnras}
\usepackage{multirow}
\usepackage{newtxtext,newtxmath}
\usepackage{longtable}
\usepackage{indentfirst}
\usepackage{multicol}
\usepackage{graphicx}
\usepackage{subcaption}
\usepackage{caption}
\usepackage{diagbox}
\usepackage[T1]{fontenc}
\usepackage{threeparttable}
\DeclareRobustCommand{\VAN}[3]{#2}
\let\VANthebibliography\thebibliography
\def\thebibliography{\DeclareRobustCommand{\VAN}[3]{##3}\VANthebibliography}
\usepackage{graphicx}
\usepackage{tablefootnote}
\usepackage{amsmath}	

\title[Catalogue of PN abundances with the VLT/FORS2]{A Catalogue of Planetary Nebulae Chemical Abundances in the Galactic Bulge}

\author[S. Tan et al.]{
Shuyu Tan,$^{1}$\thanks{E-mail: shuyut@hku.hk (ST)}
Quentin A. Parker,$^{1}$\thanks{E-mail: quentinp@hku.hk (QAP) } Albert A. Zijlstra$^{2}$ and Bryan Rees$^{2}$
\\
$^{1}$The Laboratory for Space Research, Faculty of Science, The University of Hong Kong, Cyberport 4, Hong Kong\\
$^{2}$Jodrell Bank Centre for Astrophysics, The University of Manchester, Oxford Road, M13 9PL, Manchester, UK\\}

\date{Accepted XXX. Received YYY; in original form ZZZ}

\pubyear{2015}

\begin{document}
\label{firstpage}
\pagerange{\pageref{firstpage}--\pageref{lastpage}}
\maketitle

\begin{abstract}
In this, the third of a series of papers, we present well determined chemical abundances for 124 Planetary nebulae (PNe) in the Galactic bulge from deep, long-slit FORS2 spectra from the 8.2~m ESO Very Large telescope (VLT). Prior to this work there were 
only $\sim240$ bulge PNe with chemical abundances previously determined over a $\sim50$ year period and of highly
variable quality. For 34 of these PNe we are presenting their abundances for the first time which adds $\sim14$\% to the 
available sample of bulge PNe abundances. The interstellar reddening, physical conditions (electron densities, $n_{\mathrm{e}}$, temperatures, $T_{\mathrm{e}}$), and chemical compositions are derived as single values for each PN but also using different line diagnostics. Selected comparisons with the best literature fluxes for 75 PNe in common reveals that these significant new data are robust, reliable and internally self-consistent forming the largest independent, high quality and well understood derivation of PNe abundances currently available for study. 

\end{abstract}



\section{Introduction}
\noindent
Planetary nebulae (PNe) form the most luminous phase in the evolution of low- to intermediate-mass stars with masses ranging 
from $\sim0.8$ to $8M_{\odot}$, and luminosity typically falling between $0.5$ and $1.0\times10^{4} L_{\odot}$. 
The majority of their luminosity is emitted in a few, very bright emission lines, including several forbidden collisionally-excited lines (CELs) and the hydrogen Balmer series lines. {The use of CELs and optical recombination lines (ORLs) allows the accurate measurement of elemental abundances of both light elements like helium and nitrogen, as well as heavier elements such as oxygen, sulphur, and neon in PNe.} While helium ORLs are usually the most prominent in nebulae, deeper spectra can also detect ORLs of other heavier elements, making ORLs {an alternative} to CELs in deriving elemental abundances in PNe. {Still, interpreting the results warrants extra caution due to remaining issues with PNe abundance determination, such as the discrepancy between chemical abundances derived from CELs and ORLs \citep[see][and references therein]{peimbert2017nebular}, the sulfur anomaly in PNe \citep[a deficit in sulfur abundance,][]{henry2004sulfur} and the self-enrichment of oxygen in carbon-rich dust PNe \citep{delgado2015oxygen}.}

As the luminosity of PNe emission lines is typically on the order of $10^{4} L_{\odot}$, they are easily visible across the Galaxy and beyond where individual main sequence stars are too faint. 
Additionally, PNe provide a detectable population for study close to the crowded Galactic Centre \citep{durand1998kinematics}. PNe are also
key representatives of late-stage stellar evolution for low to intermediate-mass stars, where their residual 
cores end their lives as white dwarfs on the cooling track. The measured abundances of PNe reflect both the results of 
interior nucleosynthesis \citep[e.g. He, N and C,][]{chiappini2003oxygen, henry2004sulfur} in their 
progenitors and also the interstellar abundances at the time the PN progenitor stars were born from their 
molecular clouds and enriched with elements only produced in more massive stars (such as sulphur and 
argon). In these ways, PNe are the best abundance tracers of relatively old stellar populations in our own and 
external galaxies in the local group, e.g. \citet{arnaboldi2014planetary}. 

The Galactic bulge is one key such old stellar population. It contains a large number of PNe, e.g. 
\citep{2006MNRAS.373...79P} that are of sufficient surface brightness for detailed abundance study. Previously 
available  Bulge PNe abundance determinations have suggested a solar to slightly sub-solar 
metallicity \citep{ratag1992abundances, chiappini2009abundances, 
pottasch2015abundances}. On the other hand, spectroscopy of red giants and microlensed main sequence stars in the bulge show 
a rather wide abundance range. \citet{johnson2014light} report that bulge stars show a range of [Fe/H] between 
$-0.8$ and 0.4, and $\alpha$ enhancements between 0.0 and 0.3. Uncertainties increase at higher metallicity 
where molecular blending becomes an issue. 
\citet{bensby2013chemical} also report a wide range of bulge metallicities. 
\citet{chiappini2009abundances} conclude that oxygen abundances of PNe 
are 0.3 dex lower than expected for their progenitor bulge stellar
population. Due to nucleosynthesis limitations in lower mass stars PNe can only measure [O/H] rather than 
[Fe/H], but \citet{smith2014zinc} and \citet{smith2017abundances} determined
zinc abundances (expected to mirror Fe) for a small sample and confirm the tendency towards solar abundances.
The observed differences between observed stellar and PNe abundances in the bulge remains problematic 
and impacts the question of the origin of the bulge, as either a pseudobulge, arising from scattered 
disk stars, or an old classical bulge. Hence, getting a better handle on a wider sample of Bulge PNe 
abundances could have significant implications and was a major motivation for this work. 

In this study, we assume that chemical abundances determined for any individual PN are representative of the entire nebula. We recognise that internal abundance variations may exist, albeit likely modest, e.g., \citet{1988A&A...191..128M, 2013MNRAS.434.1513D, mari2023low}. For most extant literature, and for all comparisons made in this study, each PNe produces a single, overall abundance estimate determined from an integrated 1-D spectrum across the spectrograph slit. We know PNe are complex, structured and resolved 3-D sources with ionisation stratification, internal condensations, and other features. Factors such as spectrograph slit widths and positions, in terms of orientation and offsets from any 
central star, can vary between different works and finally that observing conditions 
(seeing, transparency, airmass and etc.) also differ. Hence, different literature compilations may not sample exactly the same part of a given PN. Integral Field Unit (IFU) work, such as \citet{2013MNRAS.434.1513D} and \citet{garcia2022muse}, can cover entire PNe, or at least representative fractions of the full projected form, to create 3-D spectroscopic data cubes. Existing IFU PNe studies have already demonstrated 
that physical conditions and chemical abundances can vary across a given PN, and these variations do not usually correspond directly to those observed in line fluxes. Instead, the final physical conditions derived from observations with different slit positions predominantly reflect the results in regions 
covered by the slit where the line 
fluxes are strong, potentially leading to discrepancies in results. Until direct point-to-point mapping of 
physical parameters and chemical abundances based on IFU observations are available for large samples of PNe from 
different studies, this type of research remains essential. {\bf Furthermore all PNe in this study are compact ($\leq$10~arcseconds across) so a decent, central fraction of each PN is always sampled.}

\section{The formation of the Galactic Bulge}
To help understand the context of abundances determined for a significant sample of bulge PNe, a brief 
introduction to current understanding of the formation of the Galactic bulge is provided. 
There are currently two main scenarios \citep[e.g.][]{raha1991dynamical, 
debattista2004bulges, brooks2016bulge, fisher2016observational}. The classical scenario involves 
the gravitational collapse of primordial gas and/or the hierarchical merger of sub-clumps, 
leading to rapid star formation and disk accretion. 
This type of bulge is dominated by an old, metal-rich, spheroidal population characterised by an 
enhancement of $\alpha$-elements \citep[e.g.][]{wyse1997, zoccali2008metal, bensby2011b, johnson2011}. 
Alternatively, the secular evolution of the disk, driven by non-axisymmetric structures such as bars, can slowly 
bring gas to the centre, turning it into a spheroid via the buckling instability or resonant thickening event, 
forming a so-called "pseudobulge" \citep[][for a review]{combes1981model, combes1993formation, 
athanassoula2005nature, sellwood2014secular}. 

Pseudobulges form at a slower rate, with longer star formation timescales and have younger stellar populations 
than classical bulges \citep[e.g.][and references therein]{feltzing1999age, loon2003infrared, kormendy2004secular}. 
The bulge's vertical metallicity gradient was initially interpreted as due to dissipative collapse 
during the formation of a classical bulge. However, recent kinematic constraints suggest the bulge could have a 
composite nature, with a substantial fraction of pseudobulge, or it might purely be a 
pseudobulge \citep{shen2010, di2014mapping}. Further evidence for a pseudobulge comes from the 
kinematic and chemical properties of the bulge stellar populations which show a 
concordance with the stellar populations identified in the inner disk of the Galaxy 
\citep{babusiaux2010insights, gonzalez2011inner, bensby2011first, uttenthaler2012constraining}. 

The metallicity distribution functions (MDF) obtained from red giant branch (RGB), red clump (RC) and M giant stars 
in different bulge regions exhibit similar peaky structures (see Fig.~4 in \citet{barbuy2018chemodynamical}).  
\citet{ness2013argos} observed three predominant MDF peaks derived from 28,000 ARGOS bulge stars 
\citep{2013MNRAS.428.3660F}. These are associated respectively with the pseudobulge, the vertically thicker 
pseudobulge, and the inner thick disc, indicating a predominant pseudobulge fraction. 
Numerical simulations in \citet{martinez2013metallicity} suggest that bar and buckling instabilities could produce 
vertical and longitudinal metallicity gradients. This is supported by observational evidence from the photometric 
metallicity map of RGB stars in \citet{gonzalez2013reddening}.  

\citet{bensby2017chemical} performed an age analysis of 90 microlensed dwarf stars to investigate the evolutionary 
history of the bulge. The results indicate a considerable fraction of young stars (26\% of the sample are younger 
than 5~Gyr), leaving limited room for a classical bulge. However, the sample size is modest, and observations of 
stars could be biased towards metal-rich components or mixed with foreground objects. 
Therefore, an independent line of detailed abundance measurement evidence from carefully selected bulge PNe 
could be crucial to inform this debate. The accurate determination of bulge PNe membership can be achieved 
through a range of powerful selection criteria, as 
outlined by \citet{rees2013alignment} and utilized in this study (detailed in Paper~I). These criteria include:
\begin{enumerate}
    \item the PN's location within the inner 10$^{\circ}$ of the Galactic Centre,
    \item a measured angular size greater than 2 arcseconds but less than 35 arcseconds; see 
    \citet{acker2006400},
    \item availability of the PN's radio flux at $5\ \mathrm{GHz}$ in the range of 4.2 mJy to 59.1 mJy 
    \citep{acker1992strasbourg, siodmiak2001analysis}.
\end{enumerate}

These criteria effectively exclude foreground contamination, as demonstrated by \citet{stasinska1994extensive} and 
\citet{rees2011study}. Indeed, our independent evaluation of potential foreground contamination determined by examining HASH PNe in two zones either side of the Bulge in Galactic longitude that satisfy our Bulge selection criteria, indicate we might expect up to 15\% contamination. In a previous study using a limited sample of bulge PNe, \citet{escudero2001abundances} found 
a possible vertical abundance gradient that suggested a few PNe with low N/O 
ratios could have originated from old, lower mass progenitors. Later, an attempt at PNe age determination was made 
for 31 objects in \citet{gesicki2014accelerated} based on central-star masses derived from photo-ionization 
modelling. Even though this is a small sample, a similar fraction of young stars to dwarf stars was found. Chemical 
abundances used in both studies are sub-samples of data provided by \citet{chiappini2009abundances}. 

However, to make real progress, determination of accurate chemical abundances for a larger sample 
of bulge PNe is needed and as derived from deep and high signal-to-noise ratio (s/n) spectra. 
Such an enlarged sample can better characterize the nature of underlying stellar populations 
of both PNe and stars, and so help us understand the chemical evolution in the bulge. 

In this, the third in a series of papers that present different sets of results for 136 compact, 
confirmed PNe within a $10\times10$ degree region of the Galactic bulge, we report 
our detailed abundance determinations. Our in depth analysis of these results are provided 
in Paper IV, Tan et al., in preparation.
The abundances reported here derive from deep, medium-resolution spectroscopy from our VLT/FOSR2 
observations from a very careful and homogeneous data reduction. 
The forensic selection of this well-defined sample and the evaluation of the associated PN imaging 
data and underlying morphology were discussed in Paper~I \citep{tan2023morphologies} while in Paper~II we present 
results of a remarkable 5$\sigma$ PN major axis alignment 
signal but only for a special subset of the bulge PNe sample that host short period binaries \citep{tan2023stars}. 

Sec.~\ref{sec:observation} describes the observations and data reduction. The excellent
consistency and quality of our data shown in Sec.~\ref{sec:short_long} and \ref{sec:comp_lit} 
is demonstrated through reliable previous literature studies of 75 objects in common with our sample. 
The derived plasma diagnostics and elemental abundances for the sample are given in Sec.~\ref{sec:res}. 
We present an in-depth discussion in Sec.~\ref{sec:discussion} and our 
summary and conclusions in Sec.~\ref{sec:conclusion}.

\section{Observations} 
\label{sec:observation}
\subsection{Long-slit spectroscopy with the VLT/FORS2}

\noindent The long-slit spectroscopic observations used in this study
were conducted with the FORS2 instrument on the 8~m ESO VLT/UT1 \citep{appenzeller1998successful} 
between April 2015 and July 2018 under program IDs 095.D-0270(A), 097.D-
0024(A), 099.D-0163(A), and 0101.D-0192(A) with PI Rees and co-Is Zijlstra and Parker. 
Although the observations were conducted as `filler' programs, where observations 
could be performed under any conditions (e.g. poor seeing, low transparency due to clouds and/or 
high humidity) a decent fraction are of very high quality. The observed sample of 138 PNe are 
between 2 to 10 arcsec in angular size and fall within the inner 10 degrees of the Galactic 
Centre. All are considered bulge members based on criteria in \citet{zijlstra1990radio} and 
\citet{rees2013alignment}. Among the nearly 4000  currently confirmed Galactic PNe  contained 
in the ``Gold Standard'' HASH database\footnote{HASH: online at \url{http://www.hashpn.space}. 
HASH federates available multi-wavelength imaging, spectroscopic and other data for all known 
Galactic and Magellanic Cloud PNe.},\citep[e.g.][]{2016JPhCS.728c2008P,2022FrASS...9.5287P}, 
the objects PNG~005.9-02.6 and PNG~007.5+04.3 from the original program
target selection are now classified as symbiotic stars due to their 
mid-infrared images and spectroscopic features. These are now excluded from this study, thus leaving 136 PNe. 

Table~\ref{table:nights} provides a summary of the observing dates, exposure times, and the number of objects 
covered in each VLT program. Unfortunately, for 12 high surface brightness objects observed, several bright but 
important emission lines for standard abundance determinations were saturated even in the shortest exposures, and no 
line flux measurements were available in the literature. Consequently, we excluded these nebulae from further 
analysis in our study. 

\begin{table}
\centering
\caption{Summary of the VLT observing programs.}
\label{table:nights}
\begin{tabular}{llccc} 
\hline
\multirow{2}{*}{Program ID}   & \multirow{2}{*}{Obs. Date}  & \# of PNe & \multicolumn{2}{c}{Amount of Time}  \\ \cline{4-5}
&&Observed&[nights]&[hours]
\\
\hline
095.D-0270(A)  & Apr - Aug 2015  & 62                 & 25  & 74   \\
097.D-0024(A)  & May - Sep 2016  & 42                 & 23 & 54   \\
099.D-0163(A)  & Jun - Sep 2017 & 16                 & 9 & 19   \\
101.D-0192(A) & May - Jul 2018 & 23                 & 11 & 31  \\ 
\hline
Total:      &        & 138$^{*}$ & 68  & 178  \\
\hline
\end{tabular}
\begin{tablenotes}
      \small 
      \item * The observations of faint PNe PNG~005.8-06.1, PNG~007.5+04.3, PNG~355.9-04.2, 
      PNG~357.3+04.0 and PNG~358.9+03.4 were repeated due to either observing conditions or issues 
      with the slit position.
    \end{tablenotes}
\end{table}
To date, approximately 240 bulge PNe\footnote{This is compiled over all major works on Galactic and Galactic bulge PNe and included \citet{wang2007elemental}, \citet{chiappini2009abundances} and \citet{stanghellini2018galactic} which provides a comprehensive combined data set of earlier works dedicated to PN abundances and also \citet{pottasch2015abundances}. Note that the criteria for bulge membership may vary between different studies.} have had their chemical abundances determined. Our study provides new abundance data for 34 PNe that were previously unreported, effectively increasing the available sample of bulge PNe for study by 14\%. Additionally, 25\% of the existing literature results are considered highly unreliable due to an uncertainty greater than 0.3~dex. Our work has achieved an abundance precision typically better than 0.3~dex including for 6 PNe in common with the poorly determined examples.

\subsection{Spectroscopic instrumental configuration}
The VLT spectroscopic detector was a mosaic of two $2\text{k}\times4\text{k}$ MIT/LL CCDs  
with $2\times2$ on-chip binning (resulting in 0.25 arcsec/pixel). The instrument was used in 
Long Slit Spectroscopy (LSS) mode with a slit measuring $0.5^{\mathrm{"}}\times6.8^{\mathrm{'}}$ on the sky. 
Two grisms, GRIS1200B+97 (G1200B) and GRIS600RI+19 (G600RI) were employed to provide low–medium resolution optical spectroscopy.
The G1200B grism covers the blue spectral range from $~3360$-$5110$~\AA~ with a medium spectral resolution up to 1420, while the G600RI grism, used with GG435 blocking filter, covers the red spectral range from $\sim$5120-8450~\AA~ with a lower resolution up to 1000. As limited by the colour range of standard stars used for flux calibration, the combined wavelength coverage of the spectra was $3700$-$8450$~\AA~ for 46 PNe and $3750$-$8450$~\AA~ for 90 PNe due to small differences in instrumental set-up over the 3 year period of the VLT observing program. Exposure times varied from 2 to to 1500 seconds. Each PNe target had two exposures with the blue and red grisms respectively typically of 30s and 1000s to be sensitive to very bright 
and faint emission lines and in order to avoid both saturation or no line detection depending on the surface 
brightness of the PNe. 

\subsection{Data reduction}
\label{data_red}
\noindent The two-dimensional long-slit spectra were reduced with a typical multi-step method. First, 
cosmic rays were removed using a \texttt{Python} implementation of the \textsc{L.A.Cosmic} algorithm 
\citep{van2001cosmic}. A standard reduction procedure including bias subtraction and wavelength calibration was then 
carried out using the ESO pipeline with a \texttt{Reflex} workflow \citep{freudling2013automated}. As the bright 
skylines, e.g. [O~{\sc i}] $\lambda\lambda$5577, 6300 and 6363, become strongly 
inhomogeneous in the spatial direction in longer wavelength exposures, a careful sky subtraction was performed. This 
was via spline fitting to lines in carefully-chosen sky windows below and above the object spectra using a 
\texttt{figaro} routine \citep{shortridge2014starlink} in \textsc{starlink} 
\citep{warren2014ccdpack}. In addition, as the Galactic bulge has a dense stellar population, 
the spectra are almost always contaminated by other stars that fall within the width and along the length of the 
slit. Such contaminating sources and any PN central star continuum emissions were carefully removed where practical 
with the \texttt{iraf.continuum} routine. The sky-subtracted, continuum-removed spectra were then corrected for  
atmospheric extinction and flux-calibrated using the ESO pipeline.

{The flat field lamp used for the GRIS1200B grism has a documented instability in its spectral energy distribution (SED), which could introduce wavelength-dependent variations across the blue spectra. We decided not to perform an SED normalization on the response curve for the blue spectra (\texttt{-use\_flat\_sed} = false) to avoid substantial systematic distortions in calibrated fluxes even though this is the default pipeline option. This approach is in line with the FORS2 manual's recommendation. By comparing our reductions with and without SED normalization, we found that this issue can lead to overestimated blue emission line fluxes, particularly for those with wavelengths in the central region of the blue spectral coverage important for abundance determination, such as H$\gamma$ and [O~{\sc iii}]~$\lambda$4363.} {This overestimation could amount to up to 20\% with a moderate wavelength dependence that could results in a median decrease of 160~K in $T_{\mathrm{e}}$([O~{\sc iii}]).}

To maximise the signal-to-noise, 
especially for weaker lines, a 2-D frame pixel was attributed to an emission line only when it and at least half of 
its neighbouring pixels are brighter than 2$\sigma$ of the background noise. Remaining pixels were considered part 
of the background and set to zero. This is following a similar methodology used by \citet{gorny2009planetary}. The 
final, reduced, extracted 1-D spectra were obtained after summing the frame perpendicular to the dispersion 
direction. The mean systematic errors associated with the wavelength calibration and flux calibration propagated 
through the pipeline are $\sim$2\%.

For more than 85\% of the PNe in our sample we detected over 60 different emission lines from their long exposure 
spectra. All emission lines fluxes were then measured from the extracted 1-D spectra using the automated line 
fitting algorithm \citep[\textsc{alfa};][]{wesson2016alfa}. The reliability
of \textsc{alfa} for measuring PNe line fluxes has been demonstrated in previous studies \citep[e.g.][]
{sowicka2017planetary, boffin2018nature}. \textsc{alfa} performs Gaussian fits to input spectrum and drives the 
final line fluxes by constructing and optimising a 
synthetic spectrum with a generic algorithm. The genetic parameters were tested with grids 
of values and the configuration that gives the smallest combined residuals 
was employed. Errors in emission line fluxes were estimated from the RMS values of residuals after 
subtracting the fitted spectrum. Flux measurement for some PNe emission lines with \textsc{alfa} can be unreliable 
due to ineffective deblending of the [N~{\sc ii}] $\lambda$6548 and H$\alpha$ lines in low-excitation nebulae at 
these modest spectral resolutions. Here the [N~{\sc ii}] emission is much weaker than H$\alpha$. 
To address this we first measured the [N~{\sc ii}] $\lambda$6583/$\lambda$6548 flux ratios and compared them with 
their theoretical {value of 3.05 \citep{storey2000theoretical}} to identify spectra that were affected. In these cases, careful manual measurements of [N~{\sc ii}] $\lambda$6548, H$\alpha$ and [N~{\sc ii}]~$\lambda$6583 line 
fluxes were carried out using the \texttt{iraf.splot} deblending tool. 

The typical uncertainty in our line flux measurement is $\sim$7\%. For some fainter emission lines and the [N~{\sc 
ii}] $\lambda$6458 line, which can suffer from strong blending with H$\alpha$, the uncertainties in line flux 
measurements could be up to 30\%. Fortunately, multiple exposures of the same PNe are often available, albeit with 
different exposure times. This facilitates cross-checking of line fluxes measured from different spectra and allows 
identification of cases where cosmic-rays impinge on an emission line during a particular exposure. Fluxes of all 
main hydrogen Balmer lines, including H$\alpha$, H$\beta$, H$\gamma$, and H$\delta$, 
and the ratios of common forbidden line doublets, e.g. [O~{\sc ii}]~$\lambda\lambda$3726,29, 
[S~{\sc ii}]~$\lambda\lambda$6716,31 and [O~{\sc ii}]~$\lambda\lambda$7319,30 were assessed 
using \texttt{iraf.splot}. Manual verification was employed when necessary to eliminate any 
inaccurately identified lines or faint central star attributes in some PNe spectra.

In cases where multiple spectra  were taken, the longest exposures, typically of 1000 seconds, 
were used because of their higher s/n and detection of weaker lines. Here some strong emission lines 
can be saturated such as [O~{\sc iii}]~$\lambda$5007 and H$\beta$ in the blue arm and 
[N~{\sc ii}]~$\lambda$6583 and H$\alpha$ in the red arm. The  [O~{\sc iii}]~$\lambda$5007 
and [N~{\sc ii}]~$\lambda$6583 saturated lines can usually be scaled using the weaker component if unsaturated
in combination with the theoretically predicted line ratios or by using 
the shorter exposure measurements. A scaling factor was estimated using other bright, unsaturated lines in the same 
arm. This correction method is highly effective with associated uncertainties usually less than 2\%. For a few 
objects where the brightest emission lines are saturated in both the long and shorter exposures, literature line 
ratios for the same PNe were used for the correction, provided that the emission line flux ratios (e.g. [N~{\sc ii}] 
$\lambda\lambda$6548,83, [He~{\sc i}] $\lambda\lambda$5875,6678, [S~{\sc ii}] $\lambda\lambda$6716,31) agree within 
the uncertainty estimates. 

\begin{figure}
    \centering
    \begin{subfigure}[b]{0.44\textwidth}
         \centering
\includegraphics[width=\textwidth]{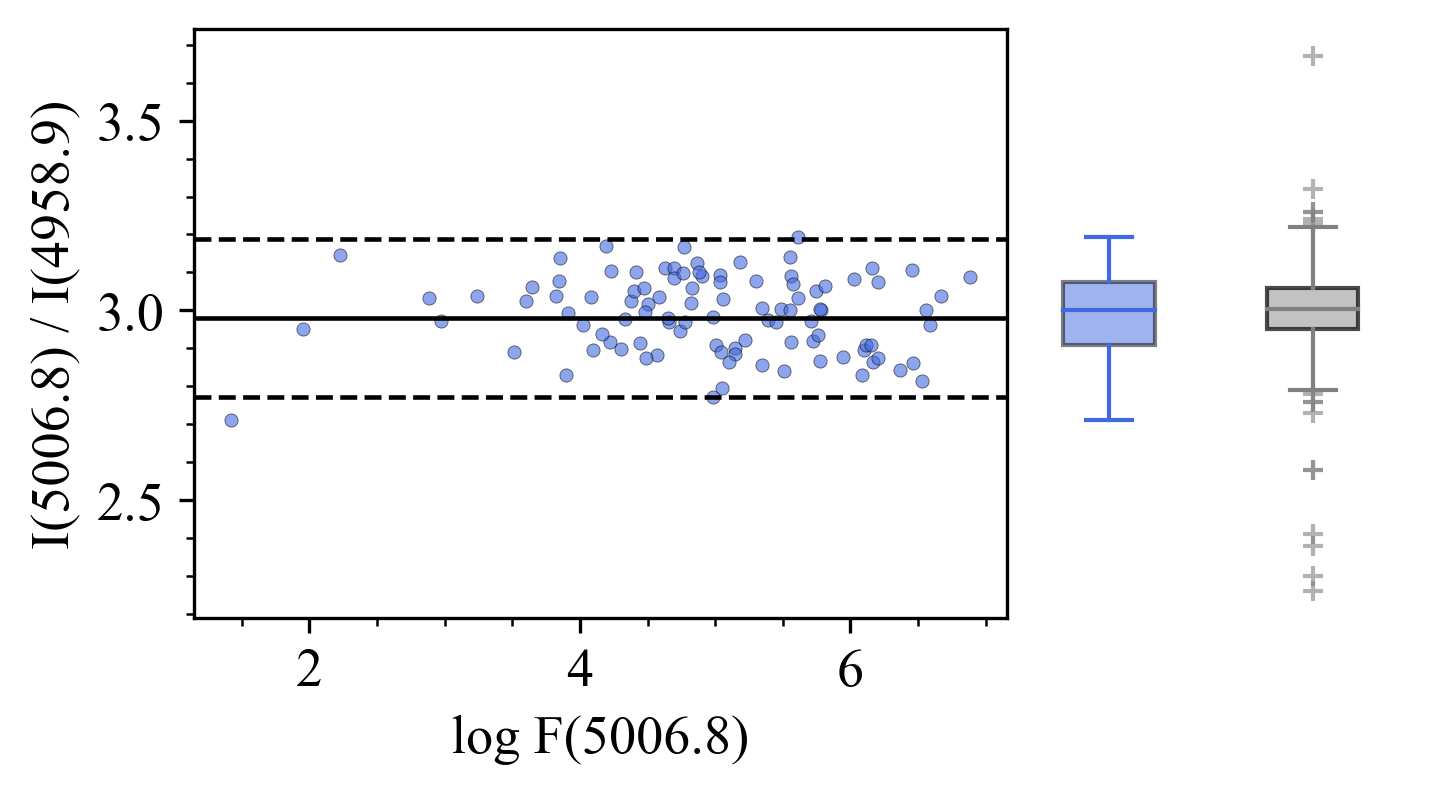}
         \caption{[O~{\sc iii}] line ratios for 100 PNe from our sample}
         \label{fig:oiii}
     \end{subfigure}
    \begin{subfigure}[b]{0.44\textwidth}
         \centering
         \includegraphics[width=\textwidth]{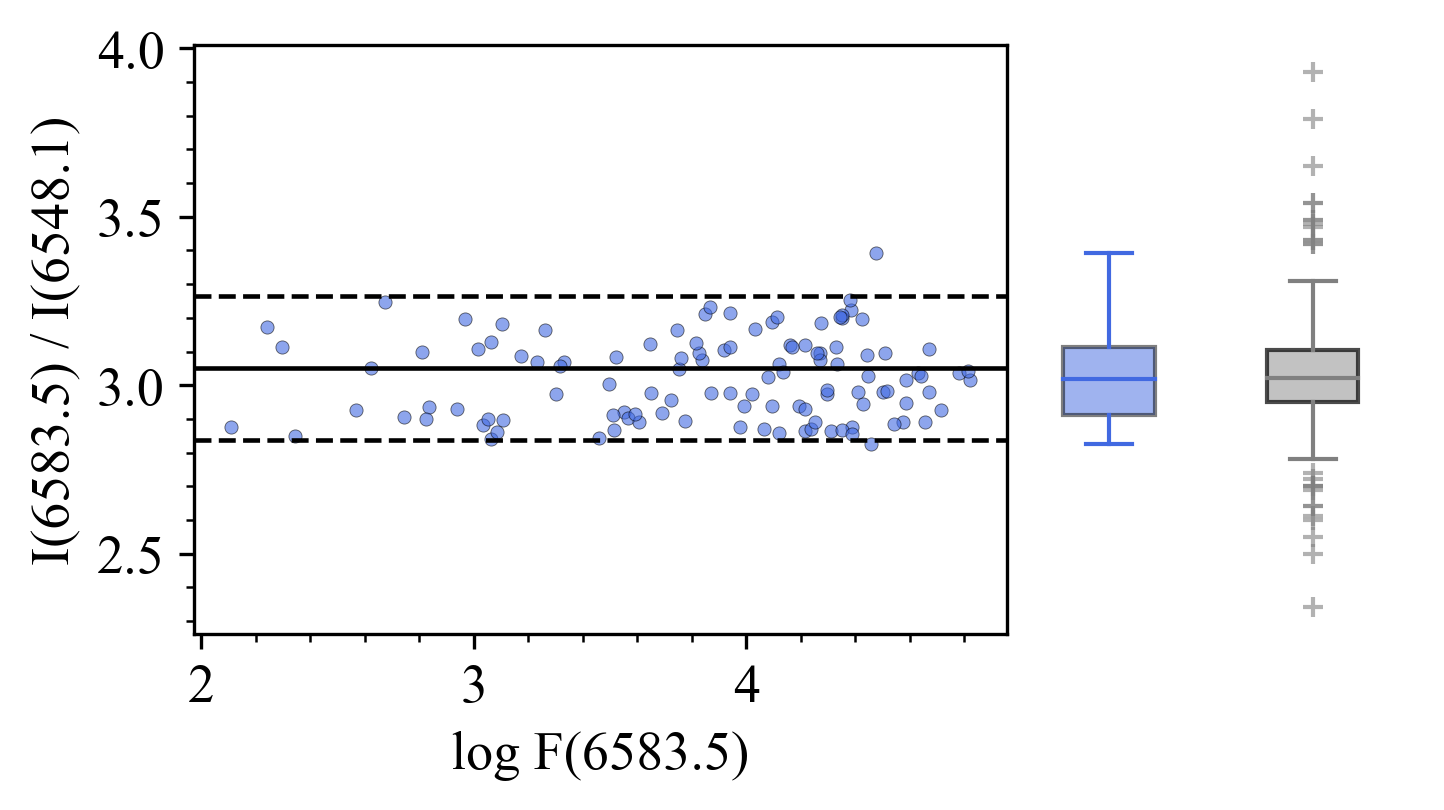}
         \caption{[N~{\sc ii}] line ratios for 114 PNe from our sample}
         \label{fig:nii}
     \end{subfigure}
\caption{Comparison of measured flux ratios of [O~{\sc iii}] and [N~{\sc ii}] forbidden emission lines from our 
observations with their theoretical values. The left panels shows de-reddened line ratios plotted 
against the [O~{\sc iii}]~$\lambda$5007 or [N~{\sc ii} $\lambda$6584 flux in $10^{-16} \mathrm{erg~cm^{-2}s^{-1}}$ 
on a log scale, with saturated spectra excluded. The black solid line represents the theoretical 
line ratio value of 2.98 and 3.05, 
respectively, while the dashed lines indicate a 7\% deviation corresponding to the combined typical uncertainty in 
the data reduction and spectrum fitting. The right panels shows box plots that compare our results (blue) with 
those for 124 Galactic PNe from \citet{rodriguez2020impact} (grey). The boxes depict the 25th to 75th 
percentiles, with the median indicated by a horizontal line. The whiskers extend to the 10th and 90th percentile, 
while outliers are represented by individual crosses. The results show excellent agreement to 
within the anticipated 7\% errors and no bias as a function of flux, providing a solid base 
level confidence in the reliability of the data reduction.}
\end{figure}

Fig.~\ref{fig:oiii} and \ref{fig:nii} display the [O~{\sc iii}] $\lambda$5007/$\lambda$4959 and [N~{\sc ii}] 
$\lambda$6583/$\lambda$6548 flux ratios obtained from the unsaturated spectra, which were used to assess 
uncertainties associated with the data reduction. We corrected for the effects of interstellar extinction 
by using the hydrogen Balmer ratio (see Sec.\ref{sec:phy_paras} for details), since observed line ratios 
would be slightly higher than theoretical values. The combined 7\% uncertainty interval estimated from 
the individual 2\% data reduction uncertainties of the two sets of emission lines is also presented. Our results 
showed that the determined [O~{\sc iii}] line ratios agree with the theoretical value within 
the uncertainties, without any systematic bias as a function of line flux. This provides a base level confidence in 
both the quality of our data reduction process and spectra. Similarly, the [N~{\sc ii}] ratios did not exhibit any 
significant bias, except for one outlier where a larger deviation of the line intensity ratio was observed. This 
deviation was due to a very weak [N~{\sc ii}] emission in comparison to H$\alpha$ (F$($H$\alpha)/$F([N~{\sc 
ii}]~$\lambda6583)>~25$) in this particular PN, and the associated uncertainties.

Relative line intensities were used to determine chemical abundances in this work without the need 
for absolute fluxes. Measured line fluxes were scaled to I$\mathrm{\left(H\beta\right)} = 100$, 
which is a common practice. Complete line intensity lists of emission used for this abundance analysis for each PNe in our sample are available in online supplementary materials.

\subsection{Examples of reduced ESO 8~m VLT 1-D PNe spectra}
\begin{figure*}
\centering
    \includegraphics[width = \textwidth]{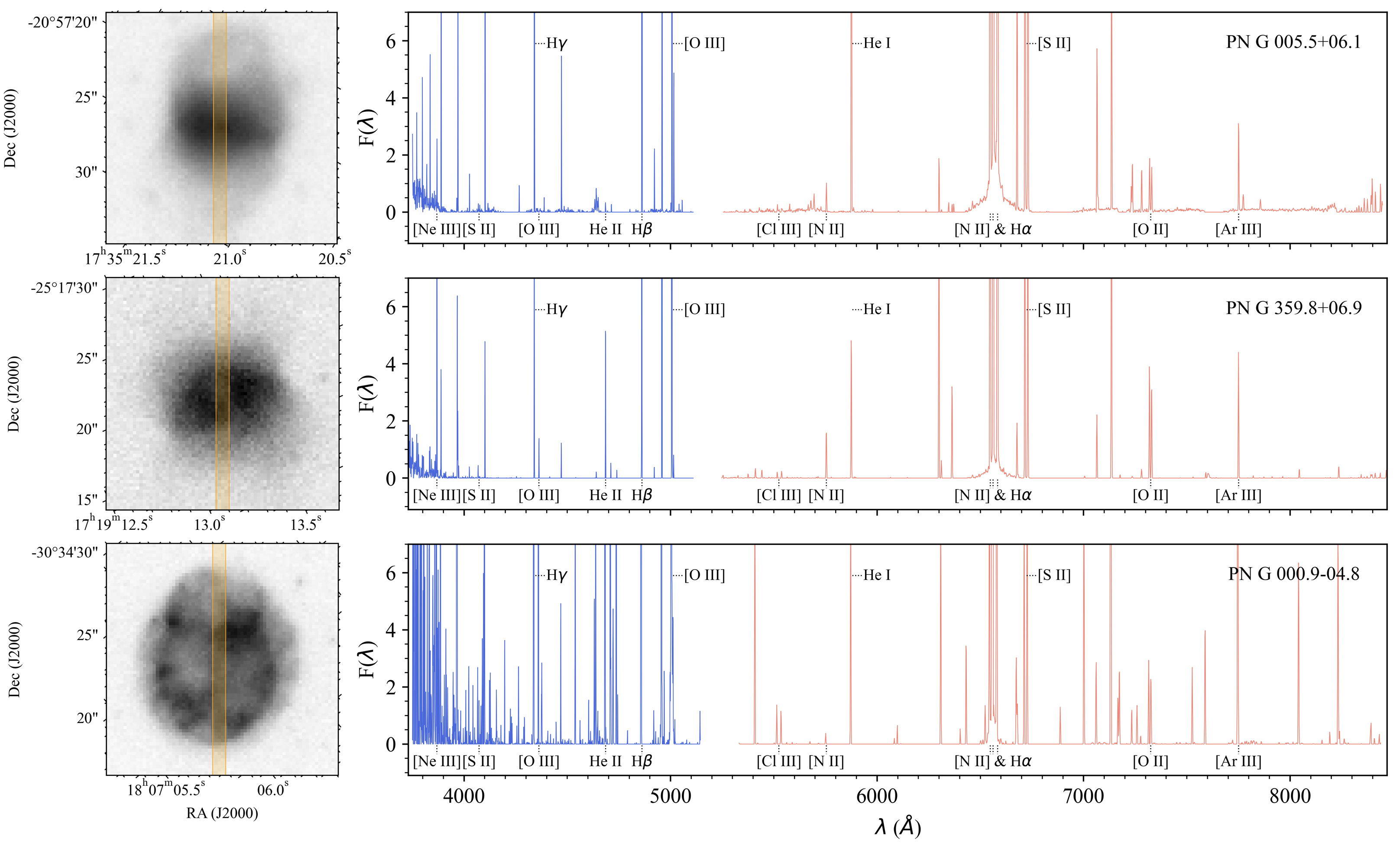}
    \caption{Illustrative examples of optical VLT/FORS2 spectra of three PNe with 
    different degrees of excitation. The upper panel displays the observations of a low 
    excitation PN, PNG~005.5+06.1, with an acquisition image on the left and a reduced 1-D spectrum 
    on the right. The PN image presented is 18~arcsec on a side and the slit position and width is 
    indicated by the orange shaded region. The spectroscopic observations from the blue and red arms 
    are distinguished using blue and red colours. The fluxes are given in units of $10^{-16} \mathrm{erg~cm^{-2} 
    s^{-1}}$. Some selected nebulae emission lines are also identified. The middle and lower panels display the 
    observations of medium and high excitation objects, PNG~359.8+06.9 and PNG~000.9-04.8, respectively 
    with the same display scheme.}
    \label{1-D-spectra}
\end{figure*}

Fig.~\ref{1-D-spectra} shows fully reduced blue and red arm ESO 8~m VLT 1-D spectra for three example 
PNe from our sample; PNG~005.5+06.1, PNG~359.8+06.9, and PNG~000.9-04.8. These are 
selected to represent PNe with low, medium, 
and high excitation characteristics and feature a typical range of emission lines under those conditions. 
The figure shows at left the narrow-band acquisition image for each PNe together with the relative placement 
of the spectrograph slit and then the resultant 1-D spectrum to the right. Key emission lines are identified.

For PNG~005.5+06.1, \citet{acker1991estimation} reported the only spectroscopic observations previously available and as
obtained with the ESO 1.52~m telescope. They detected 16 emission lines but no lines from ions of Ne and Cl. 
Our much higher s/n VLT spectra detected 66 emission lines, including [Ne~{\sc iii}] 
and [Cl~{\sc iii}] lines, thus enabling the determination of Ne and Cl abundances for the first time.
The best available spectra of PNG~359.8+06.9 was reported in \citet{escudero2004new} and was also observed 
with 1.52~m ESO telescope, detecting 29 emission lines. Our VLT spectra revealed 56 emission 
lines, notably detecting [S~{\sc ii}]~$\lambda$4069 and [Cl~{\sc iii}]~$\lambda\lambda$5517,37 lines, 
which also allowed proper determination of [S~{\sc ii}] temperature and [Cl~{\sc iii}] density estimates. 
The best previously available spectra of PNG~000.9-04.8 are those presented in \citet{gorny2009planetary} 
from the CTIO 4~m telescope that provided 25 CELs and ORLs of hydrogen and helium. Our VLT spectra detected 87 
lines, providing better spectral coverage and importantly, 
detecting recombination lines of heavier elements for this object for the first time.

\subsection{Chemical abundance determination using \textsc{neat}}
\label{sec:phy_paras}
\noindent
The physical conditions and chemical abundances for the PNe sample (refer Sec.~\ref{sec:res}) were determined 
iteratively using the Nebular Empirical Analysis Tool \citep[\textsc{neat}, version 2.3.46;][]
{wesson2012understanding}\footnote{A manual for \textsc{nest} is available at 
\url{https://nebulousresearch.org/codes/neat/manual}}. The robustness of \textsc{neat} in deriving abundances with 
the VLT/FORS2 spectroscopy has been demonstrated in previous works, e.g. \citet{jones2016ngc} and 
\citet{wesson2018confirmation}. The chemical abundance determination process adopted in \textsc{neat} is described 
below.

\subsubsection{Extinction correction}
\noindent
Prior to further work, a correction for the extinction due to absorption and scattering of light 
by interstellar dust grains was applied using the 
\citet{fitzpatrick1999correcting} curve and H$\beta$, H$\gamma$, H$\delta$ lines. The logarithmic extinction at 
H$\beta$ (extinction coefficient), c(H$\beta$), in \textsc{neat}, is computed by comparing the observed 
H$\beta$/H$\gamma$ and H$\gamma$/H$\delta$ line ratios with their intrinsic values, assuming an electron 
temperature, $T_{\mathrm{e}}$, of 10,000~K and an electron density, $n_{\mathrm{e}}$, of 1000~$\mathrm{cm^{-3}}$, as 
commonly employed. Afterwards, the line intensities were de-reddened, 
and the electron temperatures and densities were actually determined as outlined later. 
The updated $T_{\mathrm{e}}$ and $n_{\mathrm{e}}$ values were then used to recalculate the 
intrinsic Balmer line ratios, and the line intensities were de-reddened using the 
revised c(H$\beta$) value. These steps are repeated until the values converge.

As the blue and red spectra from the two spectrograph arms are non-overlapping, the H$\alpha$ line was used to scale 
the red spectrum such that the H$\alpha$/H$\beta$ ratio yield an extinction coefficient consistent with the estimate 
from the blue spectrum. This scaling factor comes from possible slight changes in either slit position or seeing 
conditions when taking multiple exposures of the same object 
and reacquiring blue or red spectra. Typically, our derived scale factor values are within $\pm$20\% 
of unity. Rarely, slightly larger or smaller factors arise from larger shifts in slit coverage or 
larger variations in observing conditions. This is usually for cases where there was a few months 
between repeat observations of the same PN. Our [N~{\sc ii}] line ratios agree very well with 
theoretical values (see in Fig.~\ref{fig:nii}), demonstrating the accuracy of both the [N~{\sc ii}] 
and H$\alpha$ line flux measurements. The s/n ratio of the hydrogen Balmer lines used for this 
purpose are generally $> 100$, so the uncertainty caused by such scaling is small.

\subsubsection{Physical conditions}
In \textsc{neat}, a three-zone ionization scheme was implemented to categorise emission 
lines based on their ionization potential (IP): those with IP $< 20$~eV are designated 
as the low ionization zone, those with 20~eV $<$ IP $<$ 45~eV 
are classified  as the medium ionization zone, and lines with IP $> 45$~eV are assigned to the high ionization zone. 
Table~\ref{tab:three_zone} presents the representative values of $T_{\mathrm{e}}$ and $n_{\mathrm{e}}$ 
for each zone  \citep[e.g.][]{osterbrock2006astrophysics}. 
The electron density measurements from the two main sets of different diagnostic lines used here are given equal 
weight. {In determining $T_{\mathrm{e}}$, the [N~{\sc ii}] diagnostic is given a dominant weight for the low-ionization zone due to its lower dependence on electron densities compared to other diagnostic lines \citep{mendez2023density}, while [O~{\sc iii}] is prioritised for the medium-ionisation zone because of the higher brightness of the lines and the broader sensitivity range they offer \citep{proxauf2014upgrading}.}
\textsc{neat} also uses other temperature and density diagnostics from other ionic species in the different 
ionisation zones used as shown in Table~\ref{tab:three_zone} and these are all given equal weight.
\begin{table}
    \centering
        \caption{Plasma diagnostics used for the low, medium and high ionisation zone as applied by \textsc{neat}.}
        \label{tab:three_zone}
{\def\arraystretch{1.5}
    \begin{tabular}{lll} 
    \hline
Zone & Temperature diagnostics & Density diagnostics\\
\hline Low & [N~{\sc ii}] $\frac{\lambda\lambda6548+84}{\lambda5755}$  &[O~{\sc ii}] $\lambda$3726/$\lambda$3729  \\
&[O~{\sc ii}] $\frac{\lambda\lambda7319+30}{\lambda\lambda3726+29}$& [S~{\sc ii}] $\lambda$6716/$\lambda$6731
\\ & [S~{\sc ii}] $\frac{\lambda\lambda6716+31}{\lambda\lambda4068+76}$ &   \\
\hline Medium & [O~{\sc iii}]$\frac{\lambda4959+\lambda5007}{\lambda4363}$ & {[Cl~{\sc iii}$] 
\lambda5517/\lambda5537$} \\
&[Ar~{\sc iii}] $\frac{\lambda\lambda7135+7751}{\lambda5192}$& {[Ar {\sc iv}$] \lambda4711/\lambda4740$}
\\
\hline High & [Ar~{\sc v}] $\frac{\lambda6435+\lambda7005}{\lambda4625}$ & - \\
\hline
\end{tabular}}
\end{table}

\textsc{neat} primarily uses atomic data obtained from the \textsc{chianti 9.0} database \citep{dere1997chianti, 
dere2019chianti} for CELs, except for O$^{+}$ and S$^{2+}$, which have documented errors in \textsc{chianti} data, 
\citet{kisielius2009electron, wesson2012understanding}. The transition probabilities and collision strengths of 
O$^{+}$ were adopted respectively from \citet{zeippen1982transition} and \citet{pradhan1976collision}. For S$^{2+}$, 
the transition probabilities from \citet{mendoza1982transition} and collision strengths from 
\citet{mendoza1983transition} were used. {The atomic data for recombination lines (ORLs) in \textsc{neat} are from multiple sources, with details available in Table~1 of \citet{wesson2012understanding}.}
\textsc{neat} derived the uncertainties in electron temperatures and densities through a Monte 
Carlo approach with 10,000 realisations.

Recombination excitation contributes to the total observed intensities of the [N~{\sc ii}]~$\lambda$5755, 
the [O~{\sc ii}]~$\lambda\lambda$7319,30 and the [O~{\sc iii}]~$\lambda$4363 auroral lines. 
These were corrected in \textsc{neat} according to equations (1-3) in \citet{liu2000ngc}. The recombination 
contribution to [N~{\sc ii}] and [O~{\sc ii}] lines could be significant as most nitrogen and oxygen are in the form of N$^{2+}$ and O$^{2+}$. {As a result, such corrections may lead to a decrease in $T_{\mathrm{e}}$([N~{\sc ii}]) by up to 20\% as well as a change in O$^{+}$ ionic abundance by up to 0.3 dex according to \citep{rodriguez2020impact}. Ignoring these corrections, as in earlier studies with low s/r spectra, could impact the accuracy of abundances results as the O$^{+}$ ionic abundance is key in the ICF formula for most elements.} {We employed the physical parameters obtained from CELs to calculate the ionic abundances from recombination lines used in equations (1-3) in \citet{liu2000ngc}. However, according to studies such as \citet{liu2006optical} and \citet{yuan2011three}, ORLs might originate from cooler regions within the nebulae. In some cases of extremely high abundance discrepancies, for example, in \cite{liu2006chemical} albeit with large uncertainties, the electron temperature derived from the hydrogen Balmer jump ($\lambda<3645$\AA) for Hf~2-2 is as low as $\sim 1000$~K. Similarly, \citet{garcia2022muse} estimates electron temperatures in ORL emission regions for three objects to be around 4000K, based on a comparison of 2-D temperature maps derived from [N~{\sc ii}] and [S~{\sc iii}] lines. Unfortunately, such methods for electron temperature determination could not be performed with our observations.}

For the [N~{\sc ii}]~$\lambda$5755 emission line, we detected at least one of the N~{\sc ii} lines from multiplet V3 and 3d-4f transitions in more than half of PNe with the VLT/FORS2, along with other multiplets such as V12 and V20, as well as singlets V5 and V28 in our deeper spectra. {Since stronger V3 and 3d-4f lines were better detected, we used the flux-weighted N$^{2+}$ ionic abundances derived from them for the [N~{\sc ii}]~$\lambda$5755 recombination correction. When either the ORL contribution to [N~{\sc ii}]~$\lambda$5755 estimated from V3 or 3d-4f transitions exceeds the observed intensity, the other is used. Through a detailed analysis, \citet{rodriguez2020impact} demonstrated that ORLs have a minimal contribution to [N~{\sc ii}]~$\lambda$5755 intensities for objects with low degrees of ionization ($\log(\text{O}^{2+}/\text{O}^{+})<1$). Therefore, for objects with an with a estimated contribution of ORLs leading to a decrease greater than 10\% in $T_{\mathrm{e}}$, we do not apply these corrections if the He$^{2+}$ emission lines are reliably observed in our spectra and the resulting $\log(\text{O}^{2+}/\text{O}^{+})<1.4$. This is because the observed ORLs may be significantly contaminated by continuum fluorescence excitation \citep{escalante2005n, escalante2012excitation}.} We derived the recombination O$^{+}$ abundance using the V1, V2 and V10 multiplets.

\subsubsection{Chemical abundances}
Ionic abundances were first determined from the de-reddened collisionally excited line intensities listed in 
Table.\ref{tab:line_list}. The total elemental abundances (Sec.~\ref{sec:res}) were obtained by multiplying the sum 
of ionic abundances with the corresponding ionization correction factors (ICFs) to account for the contribution of 
unobserved ions. In our case, we adopt the ICF scheme derived in \citet{delgado2014ionization}, hereafter DMS14, for 
all elements, except for N, for which we used the classical N/O~$=$~N$^{+}$/O$^{+}$ relation in 
\citet{kingsburgh1994elemental} (KB94), following the recommendations of \citet{delgado2015oxygen}. 

\begin{table}
\centering
\caption{The list of collisional excitation lines used for the ionic abundance determinations in this work.}
\label{tab:line_list}
\begin{tabular}{ll}
\hline $\mathrm{X}^{\mathrm{i}+}$ & Line \\
\hline $\mathrm{He}^{+}$ & {${\mathrm{He}}$~{\sc i}~$\lambda \lambda 4471,5876,6678$} \\
$\mathrm{He}^{2+}$ & {${\mathrm{He}}$~{\sc ii}$~\lambda \lambda 4686,5412$} \\
$\mathrm{N}^{+}$ & {$[\mathrm{N}$~{\sc ii}$]~\lambda \lambda 5755, 6548, 6584$} \\
$\mathrm{O}^{+}$ & {$[\mathrm{O}$~{\sc ii}$]~\lambda \lambda 3726,3729,7319,7330$} \\
$\mathrm{O}^{2+}$ & {$[\mathrm{O}$~{\sc iii}$]~\lambda 4959,5007$} \\
$\mathrm{Ne}^{2+}$ & {$[\mathrm{Ne}$~{\sc iii}$]~\lambda 3868$} \\
$\mathrm{S}^{+}$ & {$[\mathrm{S}$~{\sc ii}$]~\lambda \lambda 6716,6731$} \\
$\mathrm{S}^{2+}$ & {$[\mathrm{S}$~{\sc iii}$]~\lambda 6312$} \\
$\mathrm{Cl}^{2+}$ & {$[\mathrm{Cl}$~{\sc iii}$]~\lambda \lambda 5517,5537$} \\
$\mathrm{Cl}^{3+}$ & {$[\mathrm{Cl}$~{\sc iv}]$~\lambda 7531,8046$} \\
$\mathrm{Ar}^{2+}$ & {$[\mathrm{Ar}$~{\sc iii}$]~\lambda \lambda 7135,7751$} \\
$\mathrm{Ar}^{3+}$ & {$[\mathrm{Ar}$~{\sc iv}$]~\lambda\lambda 4711, 4740$} \\
$\mathrm{Ar}^{4+}$ & {$[\mathrm{Ar}$~{\sc v}$]~\lambda\lambda 6435,7005$} \\
\hline
\end{tabular}
\end{table}

\section{Evaluation of Data Quality}
\label{sec:short_long}
To ensure the internal consistency and quality of our data, we compared chemical abundances and physical 
parameters derived from both the long and shorter exposure spectra of the same PNe, as well as 
estimates of ionic abundances obtained 
using different emission lines of the same ionic species. This internal comparison is crucial in assessing the 
quality of our spectra and the reliability of measurements and estimations derived from them. This is followed in 
Sec.~5 with an external comparison with literature values. Chemical abundances are expressed on a 
logarithmic scale, where $\log\mathrm{(H)}=12$.

\subsection{Comparison of results from the short and long exposures}
\label{sec:short}

{Our observations include short-exposure spectra in both blue and red arms for 80 PNe. As the observation configurations show minimal variation, the chemical abundances derived from these spectra are expected to be consistent with their long-exposure counterparts due to the same amount of interstellar extinction and the emission from the same object region passing through the slit. However, background noise, stellar continuum variations, or low-level instrumental variabilities could result in small discrepancies. Here, we compare the chemical abundances independently derived from short and long exposures of the same PNe to assess the consistency of the results under different s/n levels of spectroscopic observations.}

\subsubsection{Extinction coefficients}
\label{sec:extinc_LS}
The short exposure spectra of three PNe in our sample, PNG~000.1+02.6, PNG~000.3+06.9 and PNG~359.8+05.2, did not yield a detection of H$\gamma$. Additionally, the short exposure spectra of 4 objects PNG~356.5-03.6, PNG~357.1+04.4, PNG~357.9-03.8 and PNG~358.5+02.9 had low s/n for H$\gamma$ which prevented accurate measurements. This caused a severe overestimation of their extinction coefficients and so we were unable to derive chemical abundances for these combined seven PNe using the short exposures to compare with their well-determined longer exposure counterparts. These PNe were excluded from our comparison. 

The extinction coefficient, c(H$\beta$), estimated from the short and long exposures are in good agreement, with a 
median difference of $-0.07$. Minor discrepancies can also arise due to slight changes of 
theoretical hydrogen Balmer line ratios with different physical parameters. For example, 
in the low-density limit, increasing the electron temperature from 5,000~K to 10,000~K could 
result in a shift of 3\% in H$\beta$/H$\gamma$ ratio \citep[see Table~4.2 in][]{osterbrock2006astrophysics}. 

\subsubsection{Nebular plasma diagnostics}
\label{sec:sl_phy}
The nebula physical parameters for our PNe, including electron densities and temperatures, determined from 
the shorter exposure spectra generally exhibit good agreement with their longer exposure counterparts, 
with no systematic differences and with discrepancies typically well below 0.1~dex. However, due 
to the lower detection levels of weaker plasma diagnostic lines in the shorter exposure spectra, 
the discrepancy for individual PNe can exceed 0.5 dex.

A comparison of the electron temperature estimates from the two commonly used diagnostics, $T_{\mathrm{e}}$([N~{\sc 
ii}]) and $T_{\mathrm{e}}$([O~{\sc iii}]), obtained from the short and long exposures, shows a generally good 
agreement. The median differences are $0.004\pm0.03$ and $-0.01_{-0.10}^{+0.01}$~dex, respectively, with errors 
originating from the 16th and 84th percentile values. For the other two temperature diagnostics with larger spectral 
separations, [O~{\sc ii}] and [S~{\sc ii}], larger deviations were observed, with median differences of 
$-0.06_{-0.7}^{+0.08}$ and $-0.06_{-0.12}^{+0.08}$~dex, respectively. This can be attributed to varying estimates of 
extinction corrections as pointed in Sec.\ref{sec:extinc_LS}. The median differences in resulting electron 
temperatures adopted for the low- and medium-ionization zones, when comparing short exposures to long exposures, are 
$0.002_{-0.06}^{+0.04}$ and $-0.01_{-0.08}^{+0.03}$~dex, respectively.

\begin{figure}
    \centering
    \includegraphics[width=0.42\textwidth]{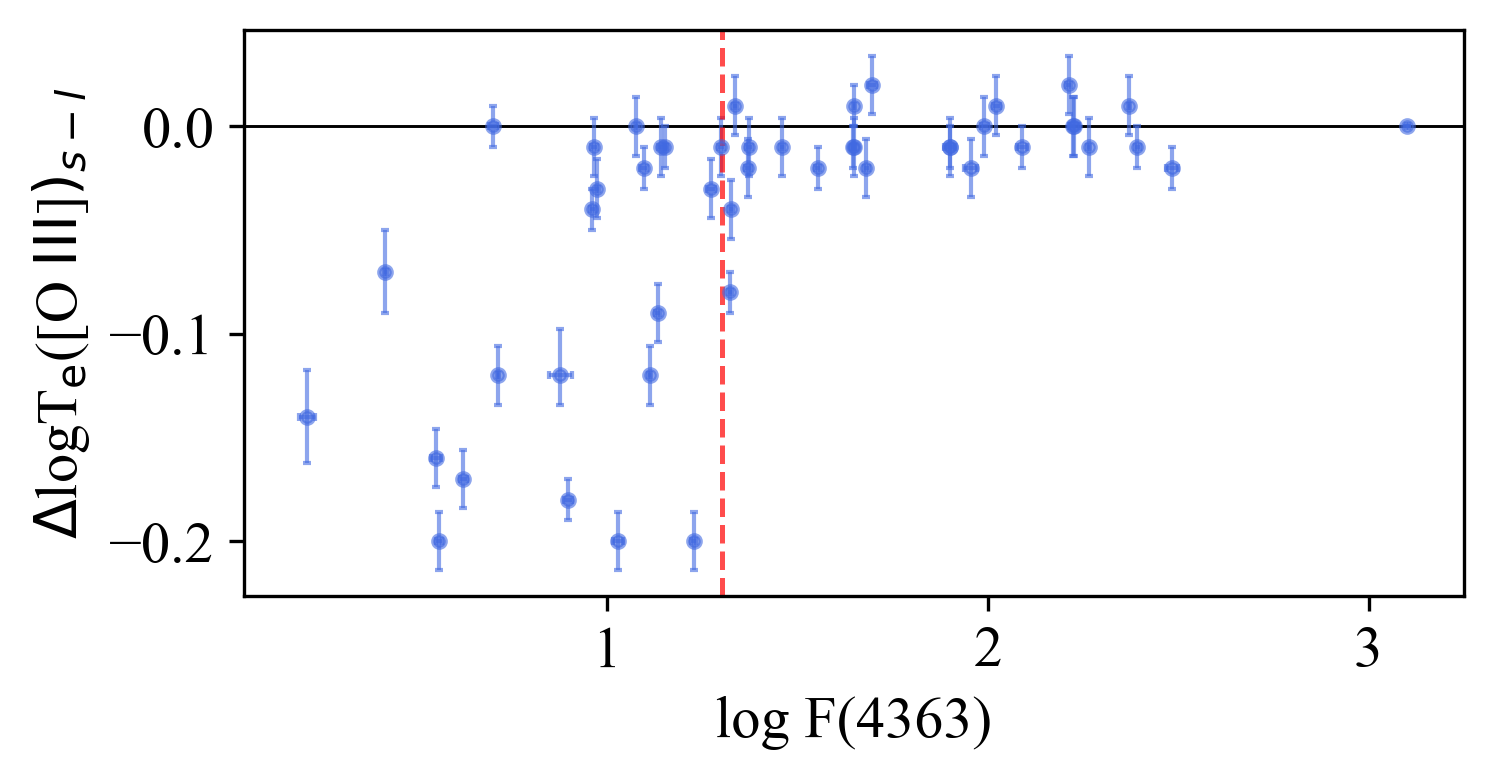}
    \caption{The relation between [O~{\sc iii}]~$\lambda$4363 raw fluxes measured from the long exposure and the 
    differences in electron temperatures $\Delta$log~T$_e$ derived from diagnostic [O~{\sc iii}] lines using short 
    and long exposures. The $\lambda$4363 raw fluxes, F(4363) are in the units of $10^{-16} 
    \mathrm{erg~cm^{-2}s^{-1}}$. The error bars on the data points represent the measurement uncertainty. Both axes 
    are presented on a log scale. It is clear the $\Delta$log~T$_e$ values are close to zero until the [O~{\sc 
    iii}]~$\lambda$4363 line flux drops below 20 $\times 10^{-16} \mathrm{erg~cm^{-2}s^{-1}}$, which is indicated by the red dashed line.}
    \label{fig:oiii_temp}
\end{figure}

We noticed a systematically lower $T_{\mathrm{e}}$([O~{\sc iii}]) estimates from the short-exposures spectra. 
Fig.~\ref{fig:oiii_temp} illustrates the difference in $T_{\mathrm{e}}$([O~{\sc iii}]) obtained from the short and 
long exposures as a function of F($\lambda$4363) in the unit of $10^{-16} \mathrm{erg~cm^{-2}s^{-1}}$ measured from 
the long exposures, which we use as a reference for the intrinsic value. The plot reveals that the underestimation 
of $T_{\mathrm{e}}$([O~{\sc iii}]) could be significant when the F($\lambda$4363) line flux measured from long 
exposures is less than $\sim20\times 10^{-16}\mathrm{erg~cm^{-2}s^{-1}}$. Upon inspecting the raw spectroscopic 
images, the pixel values corresponding to the [O~{\sc iii}]~$\lambda$4363 emission line were close to the background 
level. Therefore, when the F($\lambda$4363) line is measured from the short exposures, this leads to a lower 
$T_{\mathrm{e}}$([O~{\sc iii}]) estimate in these cases.

The electron densities, $n_{\mathrm{e}}$, derived from short and long exposures exhibit slightly larger 
discrepancies compared to those in $T_{\mathrm{e}}$ while the measurement uncertainties in electron densities are 
usually larger. For the low-ionization zone, the $n_{\mathrm{e}}$ values obtained from [O~{\sc ii}] and [S~{\sc ii}] diagnostic lines in short and long exposure spectra generally agree within 2$\sigma$. The $n_{\mathrm{e}}$([S~{\sc ii}]) values derived from the short-exposure spectra tend to be higher, which might be due to the different 
extinction corrections applied. Regarding the medium-ionization zone, the $n_{\mathrm{e}}$ values derived from [Cl~
{\sc iii}] and [Ar~{\sc iv}] density diagnostic lines agree within 2$\sigma$ for 72\% and 68\% of the objects, 
respectively, as the emission lines for these elements are usually weak. In several cases, line intensities measured 
from the short-exposure spectra yielded abnormally low densities due to lower s/n. The median differences in 
resulting $n_{\mathrm{e}}$ values are $0.01_{-0.11}^{+0.23}$ for the low-ionization zone and $0.03_{-0.20}^{+0.59}$ 
dex for the medium-ionization zones.

\subsubsection{Chemical abundance from available lines in short and long exposure spectra}
\label{sec:long_short_res}
We now compare the elemental abundances derived separately from the short- and long-exposure spectra, wherever 
possible. Summary results are given in Table~\ref{tab:abun_ls_diff}. The abundance values of a given element are represented on a log scale where $\log$(H)$=12$. Column~2 of the table presents the median differences in derived elemental abundances, along with 
the uncertainties based on the 16th percentile and 84th percentile values. The number of PNe used in the calculation 
is indicated in brackets.
Overall, the results from short- and long-exposure spectra show a good agreement. Short-exposure spectra tend to 
yield slightly higher abundance estimates for these elements, with a median difference typically below 0.1 dex. The 
discrepancies in He/H are minimal, whereas heavier elements display differences greater than 0.5 dex for some 
individual PNe. 

\renewcommand\arraystretch{1.35}
\begin{table}
 \centering
 \caption{The median difference between elemental abundances derived from the short-exposure and long-exposure 
 spectra. Column~1 displays the element concerned. Column~2 provides the median difference, accompanied by the 16th 
 percentile and 84th percentile values, as well as the number of PNe used in the calculation (in brackets). 
 Columns~3 and 4 list the number of PNe with results that agree within 2$\sigma$ and within 0.2 dex, respectively, 
 along with the corresponding fractions in parentheses.}
    \label{tab:abun_ls_diff}
    \begin{tabular}{cccc}
    \hline
   12$+\log$((X/H)  & $\Delta_{s-l}$  &  $\Delta_{s-l}$ $<$ 2$\sigma$ & |$\Delta_{s-l}$| $<$ 0.2 dex \\\hline
    He & $-0.00_{-0.04}^{+0.04}$ (70) &  46 [0.66] & 69 [0.99] \\ N & $\phantom{-}0.02_{-0.18}^{+0.48}$ (64) &  34 [0.53] & 39 [0.61] \\ 
O & $\phantom{-}0.05_{-0.17}^{+0.28}$ (70) &  42 [0.60] & 43 [0.61] \\ 
Ne & $\phantom{-}0.14_{-0.14}^{+0.53}$ (56) &  26 [0.46] & 29 [0.52] \\ 
S & $\phantom{-}0.05_{-0.41}^{+0.42}$ (69) &  44 [0.64] & 39 [0.57] \\ 
Ar & $\phantom{-}0.04_{-0.16}^{+0.24}$ (70) &  35 [0.50] & 46 [0.66] \\ 
Cl & $\phantom{-}0.11_{-0.20}^{+0.50}$ (62) &  36 [0.58] & 35 [0.56]  \vspace{0.07cm} \\
   \hline
    \end{tabular}
\end{table}

The discrepancies in the elemental abundances are attributed to s/n and the number of detectable emission lines for 
a given ionic species. The elemental abundance is calculated by multiplying the sum of ionic abundances with the 
corresponding ICF. To better understand these discrepancies, we examined the correlation between them and 
differences in individual ionic abundances and ICFs. We regard either the ionic abundance or ICF as 
responsible for the elemental abundance discrepancy when a strong correlation coefficient ($r>0.5$) is observed. 

The observed discrepancies in He/H, O/H, S/H, Ne/H, Ar/H, and Cl/H abundances mainly stem from differences in the 
abundances of the dominant ions: He$^{+}$/H$^{+}$, O$^{2+}$/H$^{+}$, Ne$^{2+}$/H$^{+}$, S$^{2}$/H$^{+}$, 
Ar$^{2+}$/H$^{+}$, and Cl$^{2+}$/H$^{+}$. The calculation of ionic abundances relies solely on the emission line 
intensity of the ion measured from the spectra and the corresponding line emissivities determined by the 
physical parameters \citep{osterbrock2006astrophysics}. Overall, the agreement in He/H is generally good, with a 
median difference in He$^{+}$/H$^{+}$ of $0_{-0.06}^{+0.02}$~dex, although some results deviate by more than 0.2 dex 
due to measurements of weak He~{\sc i} lines or large deviations in electron 
temperature ($\Delta T_{\mathrm{e}} >0.5$~dex). Similarly, short exposures tend to measure more abundant Cl$^{2+}$, 
primarily due to [Cl~{\sc iii}]~$\lambda\lambda$5517,37 lines being at the red end with poor-s/n, resulting in 
ineffective removal of background noise pixels. The determination of ionic abundance ratios for O$^{2+}$/H$^{+}$, 
Ne$^{2+}$/H$^{+}$, S$^{2+}$/H$^{+}$, Ar$^{2+}$/H$^{+}$ and Cl$^{2+}$/H$^{+}$ depends on lines emitted from medium-
ionization regions. The strong emission lines from these ions make the differences in measured line intensities 
between short and long-exposure spectra insignificant in terms of resulting ionic abundances. Our investigation 
discovered that changes in these abundances exhibit an extremely strong correlation with variations in the derived 
electron temperatures in the medium-ionization zone, which is consistent with the formula used in \textsc{neat} to 
calculate ionic abundances. Notably, since Ne$^{2+}$ is the only observable ionisation stage of Ne in our optical 
spectra, a systematic reduction in electron temperature computed from short exposures, as discussed previously in 
Sec.\ref{sec:sl_phy}, results in higher Ne/H values obtained from the short exposures.

We found that the discrepancies in the N/H estimates are mainly associated with the differences in ICFs. Since the 
dominant ionization stages of nitrogen, N$^{2+}$, manifest in the UV and far-infrared \citep[e.g.][]
{wang2007elemental} and are not detectable in our optical spectra, the ICF accounting for such unobserved ionization 
stages of nitrogen, icf(N), is typically much larger than unity, compared to other elements. The determination of 
icf(N) relies on the abundance ratios of O$^{+}$/N$^{+}$. Similar to the previously discussed findings, the 
difference in N$^{+}$/H$^{+}$ and O$^{+}$/H$^{+}$ are strongly correlated with the difference in electron temperatures in the 
low-ionization zone, typically used $T_{\mathrm{e}}$([N~{\sc ii}]). However, accounting for the recombination 
intensities of [N~{\sc ii}] $\lambda$5755 and [O~{\sc ii}]~$\lambda$7319,30 introduces further complexities. Thus, 
precise determination of the ionic abundances of N$^{2+}$ and O$^{2+}$ through ORLs in extensive spectra is 
essential in accurately calculating both electron temperature and N/H values.

In summary, chemical abundances derived from the short- and long-exposure spectra agree within 2$\sigma$ for over 
half of the PNe in this study, while a similar fraction of objects show a consistency better 
than 0.2 dex. This suggests the agreement between high-quality measurements and our error estimation from Monte-
Carlo simulations is reasonable. We found that the discrepancy in ionic abundances derived 
for He, O, Ne, S, Ar or the ICFs applied (for N) primarily result from differences in the measured physical 
parameters that arise when having to deal with lower s/n spectra. We found that short-exposure spectra tend to give 
an underestimation of $T_{\mathrm{e}}$ in the medium-ionization 
zone resulting in slight over-estimations of the abundances of O, Ne and Ar. The differences can be large. 
Nevertheless, for many PNe our short-exposure spectra can still effectively detect the key emission lines at 
sufficient s/n that are used to determine the abundance of the dominated ions as the equivalent long-exposures. As 
the wide $\Delta T_{\mathrm{e}}$ estimates show and their critical correlation with the derived abundances between 
the short and long exposures, accurate measurement of plasma diagnostics line ratios from the high s/n spectra is 
essential to the reliability of chemical abundance determinations. The determination of N/H could suffer from large 
systematic uncertainties as correction for the recombination contribution 
is needed. In addition, as discussed in \citet{wesson2018confirmation}, different ICF schemes could result in a 
difference up to 20\% in N/H. Such errors in ICFs were included in the DMS14 scheme we used in this study. As a 
result, uncertainties associated with N/H could be larger than that derived using other schemes. This could 
explain why fewer determinations of N/H agree within 0.2 dex. 

\subsection{Comparison of the results obtained with the He~{\sc{i}} lines and [O~{\sc{ii}}] lines}
\noindent Multiple He~{\sc{i}} lines are available across the optical spectra for many PNe 
sampled in both the blue and red spectrograph arms. A comparison between the He$^{+}$/H$^{+}$ derived from 
these lines can be used to establish the consistency of the results derived from the two arms. 
Additionally, O$^{+}$ abundances can be derived from either the blue [O~{\sc ii}] 
$\lambda\lambda$3726,29 or the red [O~{\sc ii}] $\lambda\lambda$7319,30 doublet. Standard star flux 
calibration is not available for most of the PNe in our sample in the wavelength range of the 
blue [O~{\sc ii}] lines, but a comparison between of the results from different lines of the same 
element can help elucidate factors that cause discrepancy and help to assess the reliability of 
the O$^{+}$ abundances derived from the red doublet.

\subsubsection{The He~{\sc{i}} lines}
\label{sec:he_lines}
The three He~{\sc{i}} lines at $\lambda\lambda$4471, 5876, 6678 were used in our abundances derivation 
with weights of 1, 3 and 1 respectively. Two other usually well-represented lines, He~{\sc{i}} 
$\lambda\lambda$3889, 7065 result from the decays to 2s$^{3}$S and 2p$^{3}$P that are sensitive to 
opacity effects and so are not used as their observed intensities significantly deviate from the case 
B approximation predictions \citep{porter2005theoretical, blagrave2007deviations, del2022helium}. 
The photoionization models in \citet{rodriguez2020impact} show that He~{\sc{i}} lines arising from 
the transitions 2$^{1}$P$^{\mathrm{o}}$-n$^{1}$D and 2$^{3}$P$^{\mathrm{o}}$-n$^{3}$D, i.e. 
He~{\sc{i}} $\lambda$4026, $\lambda$4388, $\lambda$4471, $\lambda$4922, $\lambda$5876 and 
$\lambda$6678, are {mostly insensitive} to optical depth effects so the abundances derived from these lines can be realistically compared with the adopted He$^{+}$/H$^{+}$ results. 

\begin{figure}
    \centering
    \includegraphics[width=0.42\textwidth]{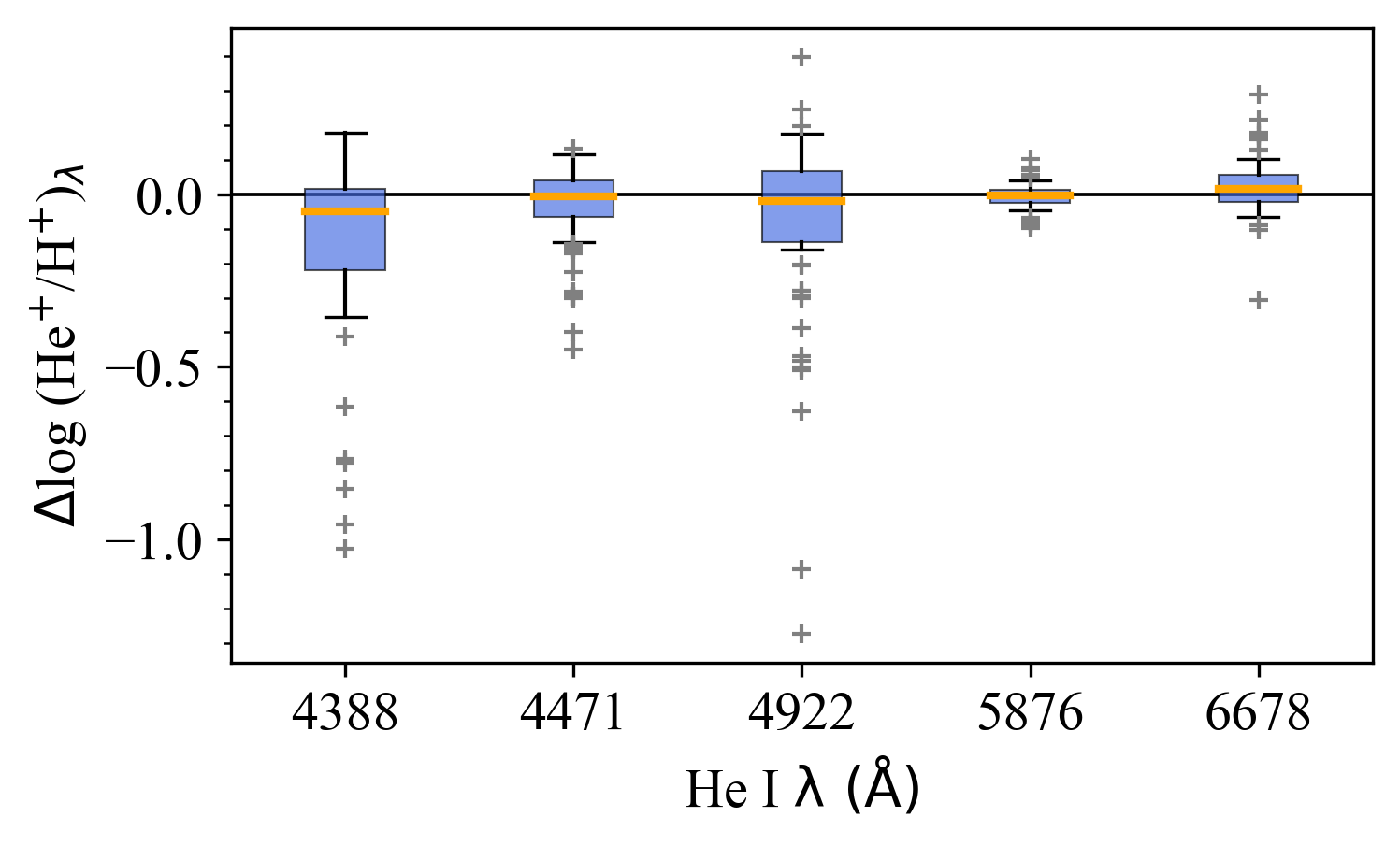}   
    \caption{{Box plots showing the comparison of He$^{+}$ abundances derived from individual He~{\sc i} lines and the final adopted values for the PNe sample. The final adopted He$^{+}$/H$^{+}$ value for any PNe is the average of the results derived from He~{\sc i}~$\lambda\lambda$4471, 5876 and 6678, having corresponding weights of 1, 3 and 1, respectively. Each box spans from the 16th to 84th percentiles, with the median value denoted by orange lines. Whiskers extend to represent the 1st and 99th percentiles, while grey crosses indicate outliers.}}
    \label{fig:he_comp}
\end{figure}

The box plots of the differences between He$^{+}$ abundances derived from each He~{\sc{i}} line and 
the weighted average abundance implied by the three brightest He~{\sc{i}} lines, $\lambda$4471,
$\lambda$5876 and $\lambda$6678 {for the PNe sample} are shown in 
Fig.~\ref{fig:he_comp}. The median values of the differences between the results from each line and the final adopted value are generally within 0.04 dex. Therefore, each of 
these He~{\sc i} lines can provide an unbiased estimation of He$^{+}$/H$^{+}$, as demonstrated by the photoionization models in \citet{rodriguez2020impact}.
The spread of discrepancies in the two He~{\sc i} lines, He~{\sc i} $\lambda\lambda$4388, 4922, not used in the derivation, is wider with many more outliers and with some cases showing 
$\Delta\log$~(He$^{+}$/H$^{+}$)~$> 0.5$. This is owing to the larger errors from measuring 
these weaker He~{\sc{i}} emission lines. 

\begin{figure}
    \centering
    \includegraphics[width=0.42\textwidth]{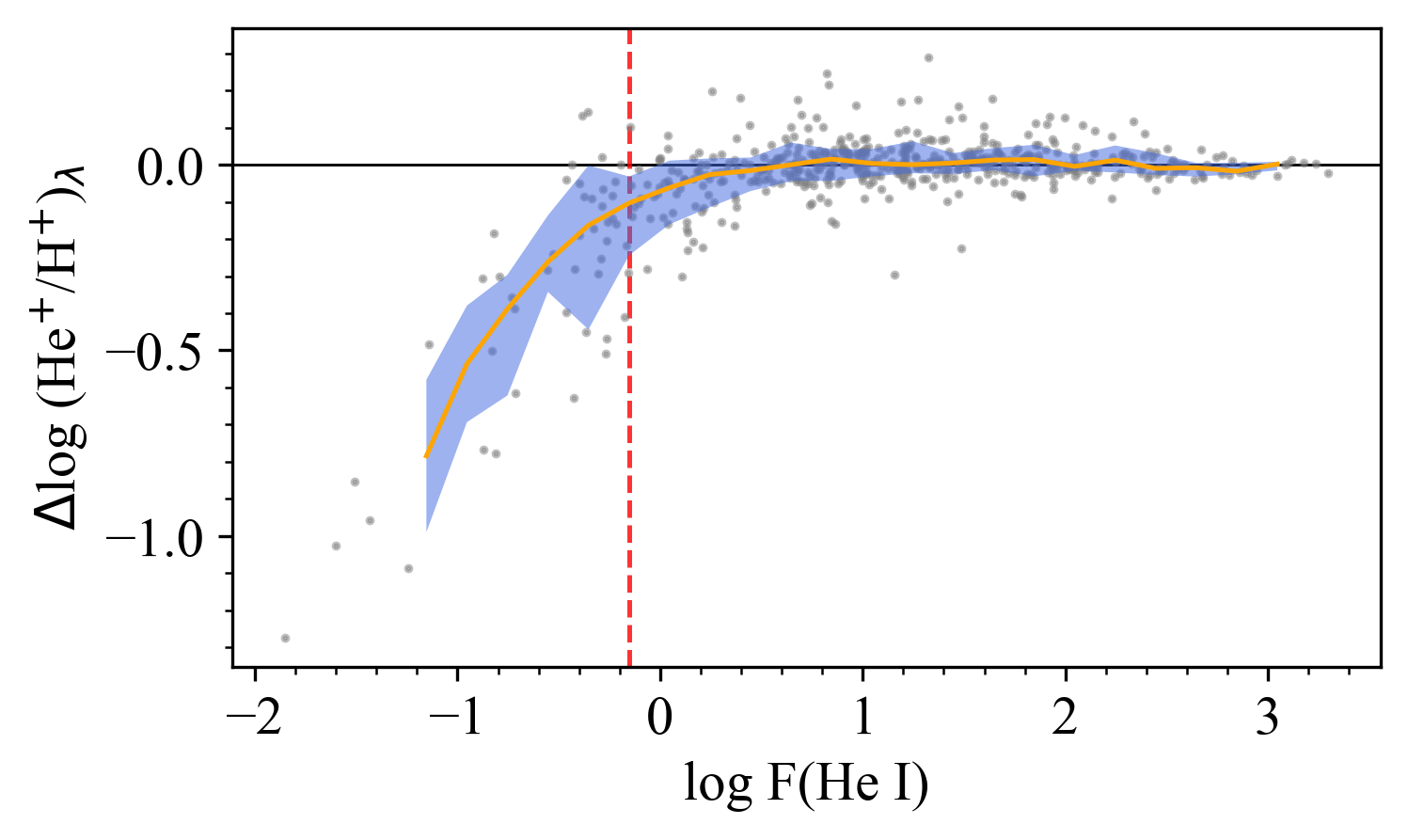}
    \caption{{The differences between $\log$~(He$^{+}$/H$^{+}$) values derived from each He~{\sc i} line and the final adopted values are plotted against their raw fluxes ($\log$~F(He~{\sc i}) in the units of $10^{-16} \mathrm{erg~cm^{-2}s^{-1}}$) for all measured He~{\sc{i}} lines across the PNe sample. The light blue shaded region encloses the 16th and 84th percentile of $\Delta\log$~(He$^{+}$/H$^{+}$) in 0.2 dex bins in log~F(He~{\sc i}), while the orange curve indicates the median value. The vertical red dashed line at $\log$~F(He~{\sc i})~$=0.15$ shows the demarcation where the 84th percentile value of  each bin becomes positive.}}
    \label{fig:he_flux}
\end{figure}

In Fig.~\ref{fig:he_flux}, we present the same discrepancies in He$^{+}$/H$^{+}$ derived 
from the multiple He~{\sc{i}} lines in each PNe spectrum as a function of the raw line fluxes. 
The orange line shows the median difference in $\log$~(He$^{+}$/H$^{+}$) as a function of He~{\sc i} 
line fluxes in bins of 0.2~dex, while the shaded region indicates the 16-84th percentile range 
of the measures. Unsurprisingly, the discrepancy increases as the flux decreases. 
For lines with log~fluxes $<-1$, the median differences is worse than $-0.5$ dex. The black 
dotted line at $\log$~F(He~{\sc i}) $=-0.15$ marks where the 84th percentile value of each bin 
becomes positive. All emission lines measured with a flux $\log$~F(He~{\sc{i}}) $\leq-0.15$ leads to a result lower than the adopted He$^{+}$/H$^{+}$ value. 

\begin{figure}
    \centering
    \includegraphics[width=0.42\textwidth]{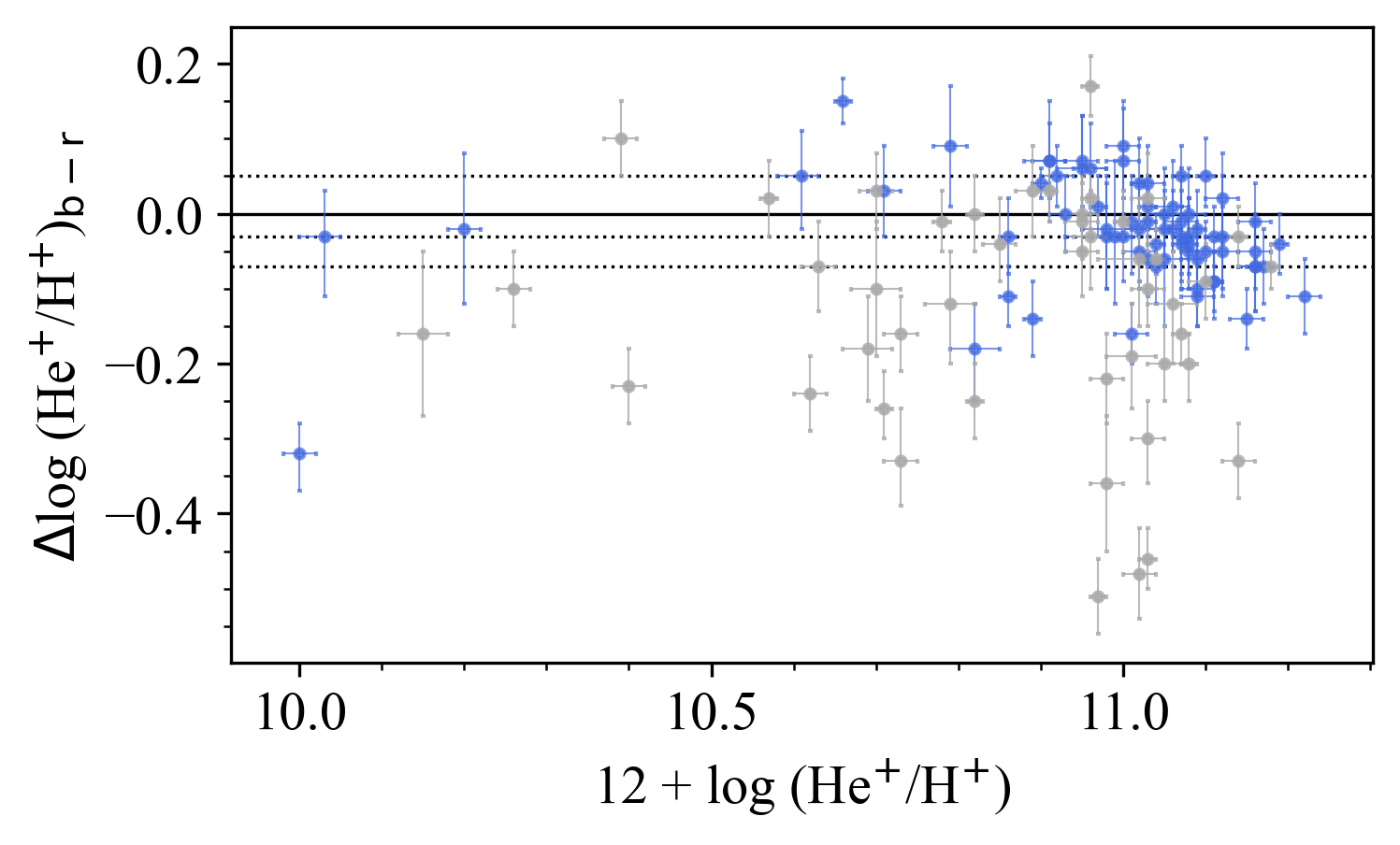}
    \caption{The differences in He$^{+}$ abundances derived from the blue and red spectra. Blue dots correspond to objects with raw fluxes for all He~{\sc i} lines with $\log$~F(He~{\sc i}) $>-0.15$, while grey dots represent those with weaker lines. 
    These unsurprisingly 
    have a greater spread in $\Delta\log$~(He$^{+}$/H$^{+}$). Error bars account for the measurement uncertainties. 
    Grey dashed horizontal lines represent the median difference, as well as the 16th and 84th percentile values.}
    \label{fig:He_BR}
\end{figure}

To further examine whether the He$^{+}$/H$^{+}$ derived from the blue and the red spectral arms are consistent, we averaged the results of He$^{+}$ abundances derived from lines in the blue spectra, namely He~{\sc{i}} $\lambda\lambda$4388, 4471, 4922, and those in the red spectra, namely He~{\sc{i}} $\lambda\lambda$5876, 6678. The differences between the results from the He$^{+}$ abundances from lines in the blue and red spectra are plotted versus the adopted He$^{+}$/H$^{+}$ values in 
Fig.~\ref{fig:He_BR}. Error bars come from the observed measurement uncertainties. In the plot, the blue dots are for PNe with all He~{\sc{i}} line fluxes $>0.7\times 
10^{-16} \mathrm{erg~cm^{-2}s^{-1}}$, 
therefore, with no significant underestimation in He$^{+}$/H$^{+}$. In this sample 86\% of objects have results whose red and blue lines spectra agree within 2$\sigma$. The difference between these results is 
$-0.01^{+0.07}_{-0.06}$~dex, indicating good consistency. The grey points are for PNe where one or more of the He~{\sc{i}} line fluxes are 
less than $0.7\times 10^{-16} \mathrm{erg~cm^{-2}s^{-1}}$. Larger discrepancies are seen, as might be expected with the blue spectra results being generally lower. The median difference is $-0.08^{+0.10}_{-0.18}$ dex with this value 
and associated errors given by the horizontal dotted lines.

The agreement between the He$^{+}$ abundances obtained from the blue and red spectra demonstrates that the scaling factor applied to the red spectra is effective and does not introduce any systematic errors in our abundance calculations, the flux calibration or extinction corrections. It is clear that He~{\sc{i}} emission lines with a measured flux below $0.7\times 10^{-16} 
\mathrm{erg~cm^{-2}s^{-1}}$ result in 
an underestimation of ionic abundances. This finding is also likely to be applicable to other elements. 

\subsubsection{[O~{\sc{ii}}] lines}
\label{o+_comp}
In the analysis of PNe oxygen abundances, previous studies \citep[e.g.][]{viegas1994density, escudero2004new, exter2004abundance} have found that using the [O~{\sc ii}] $\lambda\lambda$7319,30{\footnote{The [O~{\sc ii}]~$\lambda\lambda$7319,30 line pair in fact consists of two separate doublets: [O~{\sc ii}]~$\lambda\lambda$7319,20 and [O~{\sc ii}]~$\lambda\lambda$7330,31.}} red emission line pair typically results in a systematically higher O$^{+}$ abundance of up to 1~dex compared to the [O~{\sc ii}] $\lambda\lambda$3726,29 blue line pair. The discrepancy persists even after subtracting the contribution of recombination [O~{\sc ii}] $\lambda\lambda$7319,30 lines, with differences in the results reaching a factor of 2 or more. Most previous studies on PNe oxygen abundances from optical spectra have predominantly used the $\lambda\lambda$3726,29 pair, e.g. \citet{kingsburgh1994elemental} and \citet{magrini2009planetary}, since they are usually bright and can be detected in the spectra with high s/n.

\begin{figure}
    \centering
    \includegraphics[width=0.42\textwidth]{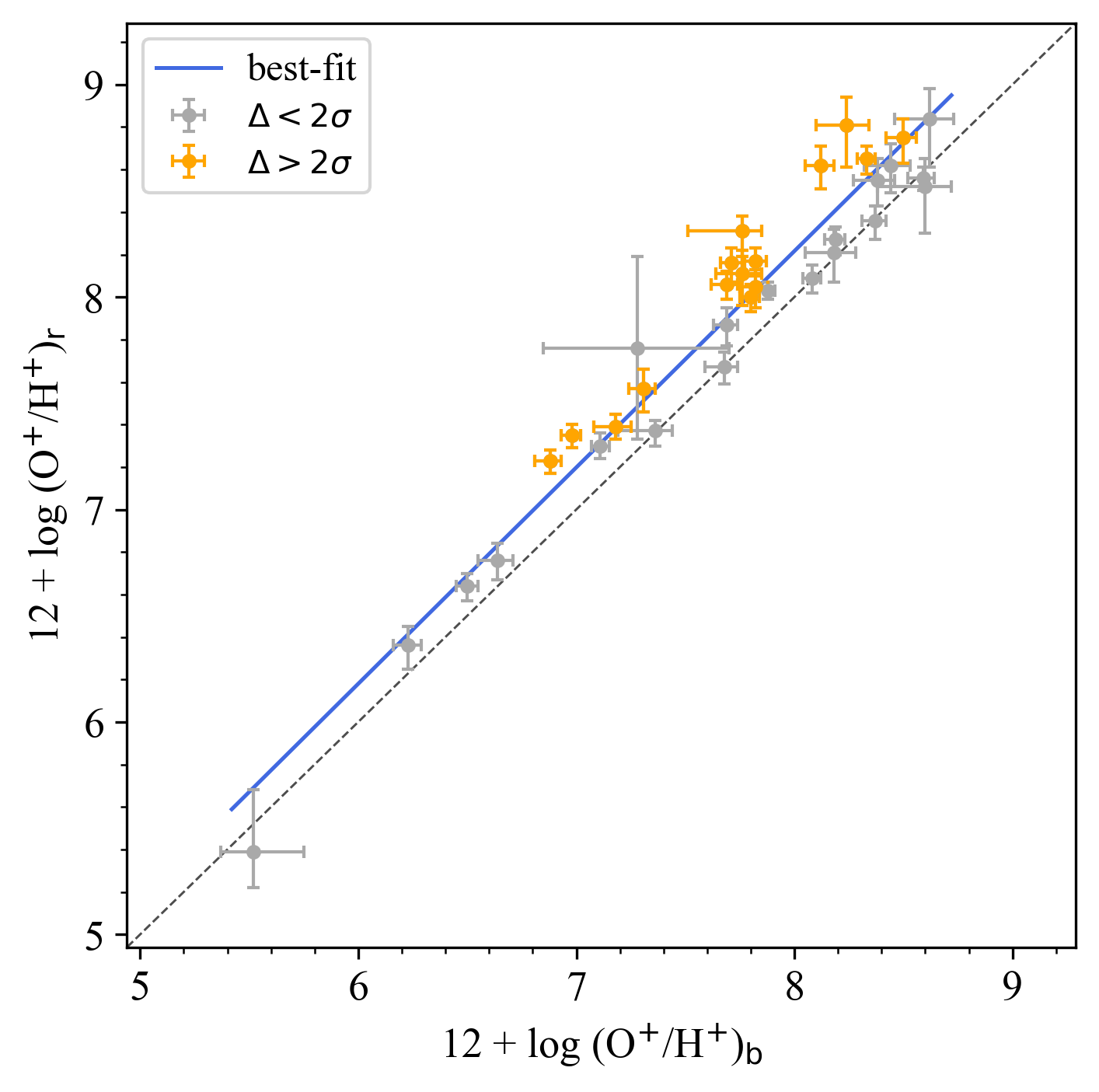}
    \caption{The relationship between ionic abundances of O$^{+}$ derived from the blue [O~{\sc ii}] $\lambda\lambda$3726,29 (subscript 'b') doublet and the red [O~{\sc ii}] $\lambda\lambda$7319,30 
    (subscript 'r') doublet of 34 PNe in this work. The solid blue line shows the best-fit, $y=1.02x+0.07$, 
    while the dotted line represent $y=x$. Orange dots represent the results that do not agree within 2$\sigma$ of $y=x$, taking into account measurement errors.}
    \label{fig:oii_rb}
\end{figure}

In our total sample, 34 PNe have both the blue and red [O~{\sc{ii}}] lines present in their spectra. A comparison between the O$^{+}$/H$^{+}$ derived from the red and the blue doublets is shown in Fig.~\ref{fig:oii_rb}. The red doublet [O~{\sc ii}] lines yield O$^{+}$ abundances consistently exceed those from the blue [O~{\sc ii}] doublets by a median value of 0.19~dex. The orange dots represent data points that deviate from the dotted 1-to-1 correspondence line by more than 2$\sigma$ (given the errors on each point). A generally tight relationship is observed, with the O$^{+}_{\mathrm{b}}$ and O$^{+}_{\mathrm{r}}$ are correlated with a coefficient $r=0.97$. The solid blue line represents the least-squares best-fit between O$^{+}_{\mathrm{r}}$ and O$^{+}_{\mathrm{b}}$, which is derived as $y=1.02x+0.07$ with a $R^{2} = 0.97$.

{We assessed whether uncertainties in the blue [O~{\sc ii}] line flux measurements, which are situated at the edge of our spectroscopic images and could be affected by instrument responsivity issues, could result in this discrepancy. No correlation was found between the raw fluxes of [O~{\sc ii}]~$\lambda\lambda$3726,29 lines and the O$^{+}_{\mathrm{r}}$/O$^{+}_{\mathrm{b}}$ ratios. We noted that higher extinction coefficients estimates obtained using the H$\delta$/H$\gamma$ ratio suggest a potential flux deficit (H$\delta$/H$\gamma$ gives higher c(H$\beta$) than H$\beta$/H$\gamma$ for 85\% of PNe) at the blue end. However, the difference between the measured H$\delta$ intensity and the expected value based on c(H$\beta$) obtained from H$\beta$/H$\gamma$ line ratios also show no correlation with O$^{+}_{\mathrm{r}}$/O$^{+}_{\mathrm{b}}$. Therefore, these factors may not contribute to the discrepancy.}

\begin{figure}
    \centering
    \includegraphics[width=0.42\textwidth]{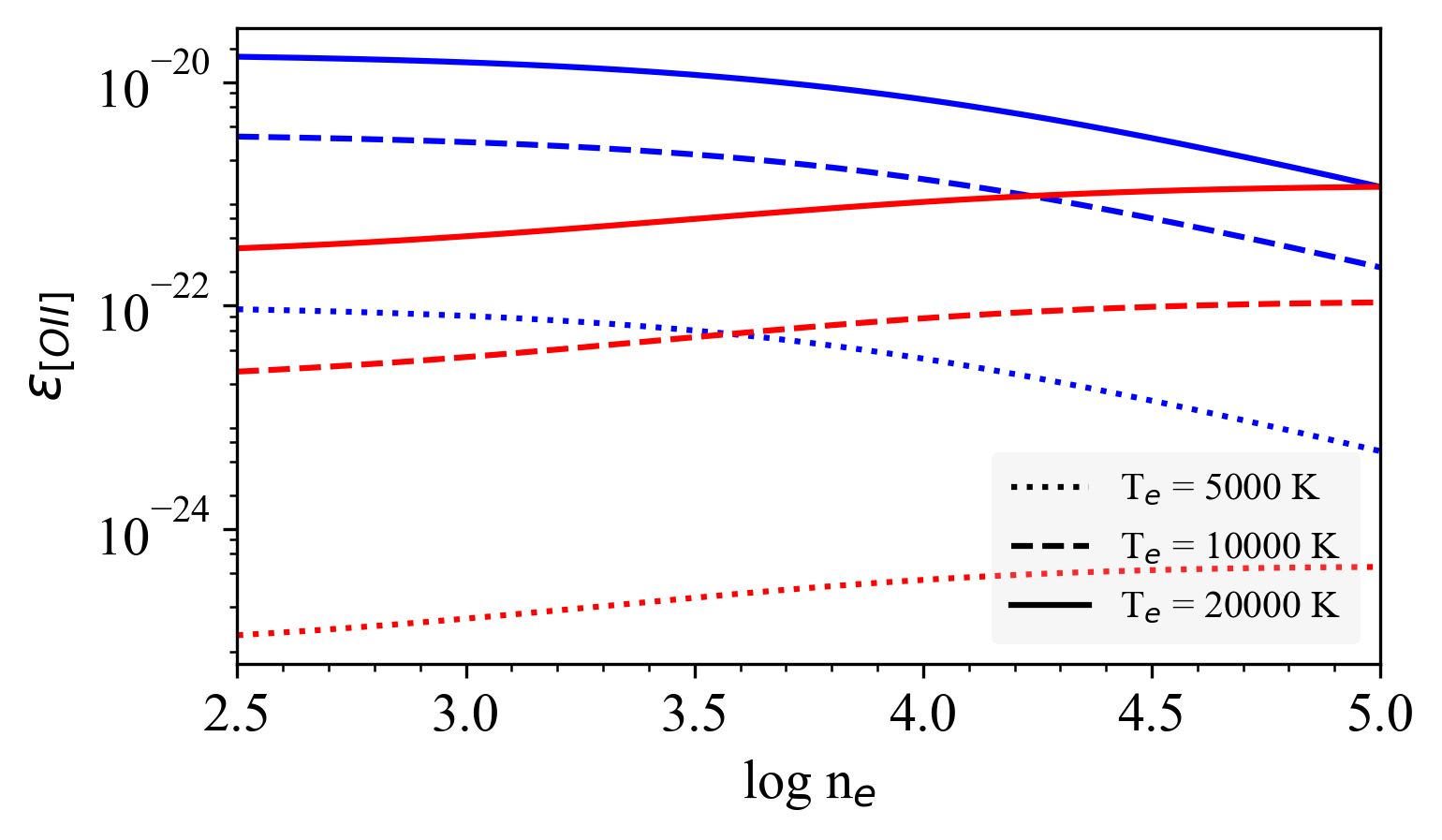}
    \caption{The summed line emissivities, $\epsilon$, of [O~{\sc ii}]~$\lambda\lambda$3726,29 (in blue) and [O~{\sc ii}]~$\lambda\lambda$7319,30 (in red) emission lines, measured in units of erg$\cdot$s$^{-1}\cdot$cm$^{3}$, as a function of electron density in the units of cm$^{-3}$, on a log scale. The data was obtained using the \texttt{getEmissivity} commands in \textsc{PyNeb} using the data of \citet{zeippen1982transition} and \citet{wiese1996atomic} for different transitions. The solid, dashed and dotted curves correspond to electron temperatures of 5000, 10,000, and 20,000~K, respectively.}
    \label{fig:oii_emi}
\end{figure}

{Through a study of H~{\sc ii} regions using deep spectra, \citet{mendez2023density} suggested that the presence of high-density clumps in nebulae account for a discrepancy of $\sim0.1$~dex for H~{\sc ii} regions. While the discrepancy observed in our sample of PNe is larger, it is likely the same underlying cause. Similar to \citet{mendez2023density}, we found that [O~{\sc ii}]~($\lambda\lambda$7319+30)/($\lambda\lambda$3726+29) line ratio consistently gives higher estimates of $T_{\mathrm{e}}$ than [N~{\sc ii}] lines, with a median difference of $0.29^{+0.21}_{-0.30}$~dex. Among the 34 PNe studied, 12 (35\%) have $T_{\mathrm{e}}$([O~{\sc ii}]) at the high-temperature limit of 35,000~K. Fig.~\ref{fig:oii_emi} presents the change in summed blue and red [O~{\sc ii}] line emissivities as a function of $n_{\mathrm{e}}$. High-density clumps could account for the observed discrepancy in two ways: one is that the measured blue [O~{\sc ii}] lines are strongly biased towards low density regions, similar to [S~{\sc ii}], resulting in an underestimation of the overall electron density and a significant overestimation of [O~{\sc ii}] temperatures. Meanwhile, the emissivity for the blue [O~{\sc ii}] lines in the high density regions is overestimated which lead to an underestimation of the O$^{+}_{\mathrm{b}}$ abundances. The second is that, as the red [O~{\sc ii}] lines are biased towards high densities, using the lower $n_{\mathrm{e}}$ values estimated from [O~{\sc ii}] and [S~{\sc ii}] could underestimate emissivity and overestimate abundance.}

{Thus, deriving O$^{+}$ abundances using the blue [O~{\sc ii}] lines may provide fair measurements for the low-density regions while the high-density regions might not be well-represented. In contrast, using the red [O~{\sc ii}] lines suffers from underestimated $n_{\mathrm{e}}$ values, leading to inflated abundances. However, for high-density regions where the red [O~{\sc ii}] lines are enhanced, the actual electron temperature could be lower than the overall $T_{\mathrm{e}}$ estimates using [N~{\sc ii}] lines, their actual emissivity might be lower. This could be a compensate for the accuracy of our final estimations.}

{To summarise, in our results, the red [O~{\sc ii}]~$\lambda\lambda$7319,30 doublet lines consistently yields higher O$^{+}$/H$^{+}$ than the blue [O~{\sc ii}]~$\lambda\lambda$3726,28 doublet, possibly due to the presence of condensations in the nebulae. As such, the O$^{+}_{\mathrm{r}}$ values used for the majority of objects in our study reflect measurements from the dense clumps in the nebulae, if present.}

\section{Abundance Comparisons with literature results}
\label{sec:comp_lit}
In this section, we compare our abundance results with some of the most recent published 
studies, namely WL07, GKA07, GCS09, 
CCM10, PBS15, and SZG17, which refer to \citet{wang2007elemental}, 
\citet{girard2007chemical}, \citet{gorny2009planetary}, \citet{cavichia2010planetary}, \citet{pottasch2015abundances}, and \citet{smith2017abundances}, 
respectively. WL07, GKA07, and CCM10 used 2-m class telescopes, while GCS09 utilised 4-m class 
telescopes and SZG17 used the VLT. Additionally, PBS15 employed both optical spectra from the literature and mid-
infrared observations from \emph{Spitzer}. Most of these studies made use of low- to medium-resolution spectra while 
WL07 utilized high-resolution spectra spanning from 3900~\AA~ to 4980~\AA~ and SZG17 employed high-resolution 
VLT/UVES spectra. We excluded earlier work by \citet{escudero2004new} for comparison, as its sample largely overlaps 
with the later works. Furthermore, we did not include \citet{exter2004abundance}, which used the multi-object 
spectrograph on 1.2-m telescope, as some literature values for physical conditions were adopted, making the data less homogeneous. To place this comparison on the same footing we derived the elemental abundances using 
the same ICF scheme introduced in KB94 as used in most of this literature.

Table~\ref{tab:lit_diff} displays the median differences in elemental abundances, together with uncertainties 
estimated from the 16th and 84th percentile values, of helium, nitrogen, oxygen, neon, sulphur, argon and chlorine for PNe that 
are common between this work and seven other literature samples. The median difference in extinction 
coefficient is also presented in Col.~1. The numbers in brackets following the abundance differences give 
the number of PNe available for comparison for that particular element. The He/H values show excellent agreement, as the emission lines are usually bright and all the ionization stages 
can be observed in the optical spectra. The agreement in N/H, O/H, S/H, and Ar/H among the different 
authors is generally good, with median differences typically below 0.2 dex, except for PBS15 and SZG17, which will be 
discussed later.

\renewcommand\arraystretch{1.34}
\setlength{\tabcolsep}{3.3pt}
\begin{table*}
\caption{Comparison between PNe literature abundances and this work. The differences shown correspond to our results 
minus the literature values, with median values and uncertainties based on the 16th and 84th percentile values. The 
number of objects in common with literature studies is indicated in brackets. An asterisk next to WL07 denotes that the recombination contribution has been subtracted from the [N~{\sc ii}]~$\lambda$5755, [O~{\sc 
ii}]~$\lambda\lambda$7319,30, and [O~{\sc iii}]~$\lambda$4363 intensities. Electron temperatures in PBS15 adopted 
additional diagnostics from mid-infrared spectra.}
\label{tab:lit_diff}
    \centering
    \begin{tabular}{lcccccccc}
    \hline
       $\phantom{^{\star}}$Sample  & $\Delta$c(H$\beta$) & $\Delta\log$(He/H) & $\Delta\log$(N/H) & $\Delta\log$(O/H) & $\Delta\log$(Ne/H) & $\Delta\log$(S/H) & $\Delta\log$(Ar/H) & $\Delta\log$(Cl/H)  \\    \hline
       ${^{\star}}$WL07 (10)  &  
       $\phantom{-}0.11_{-0.21}^{+0.56}$ &  
       $\phantom{-}0.00_{-0.03}^{+0.03}$(10) & 
       $-0.03_{-0.16}^{+0.17}$(10) & 
       $-0.04_{-0.07}^{+0.06}$(10) & 
       $\phantom{-}0.01_{-0.02}^{+0.10}$(10) 
       &  $-0.05_{-0.12}^{+0.12}$(10) & 
       $\phantom{-}0.09_{-0.08}^{+0.10}$(10) & 
       $\phantom{-}0.00_{-0.16}^{+0.01}$(10) 
       \\ 
         $\phantom{^{\star}}$GKA07 (6) $\phantom{0}$ 
         & $-0.07_{-0.17}^{+0.22}$ 
         &  $\phantom{-}0.03_{-0.02}^{+0.02}$(6)$\phantom{0}$ 
         &    $\phantom{-}0.18_{-0.16}^{+0.06}$(6)$\phantom{0}$ 
         &    $\phantom{-}0.08_{-0.10}^{+0.27}$(6)$\phantom{0}$ 
         &  $\phantom{-}0.21_{-0.16}^{+0.10}$(5)$\phantom{0}$ 
         &   $\phantom{-}0.11_{-0.11}^{+0.16}$(6)$\phantom{0}$ 
         &  $\phantom{-}0.15_{-0.22}^{+0.16}$(6)$\phantom{0}$ 
         &   $\phantom{-}0.21_{-0.24}^{+0.17}$(6)$\phantom{0}$
         \\     
         $\phantom{^{\star}}$GCS09 (40)  
         & $\phantom{-}0.01_{-0.18}^{+0.32}$ 
         & $-0.08_{-0.07}^{+0.07}$(40)
         & $-0.14_{-0.15}^{+0.18}$(37)
         &  $\phantom{-}0.08_{-0.14}^{+0.23}$(40)
         &  $\phantom{-}0.22_{-0.23}^{+0.2}$(30) 
         & $-0.01_{-0.28}^{+0.17}$(36)
         & $\phantom{-}0.01_{-0.16}^{+0.19}$(39)
         &  $-1.04_{-0.38}^{+0.38}$(27)
         \\
        $\phantom{^{\star}}$CCM10 (15) 
        & $\phantom{-}0.38_{-0.43}^{+0.32}$
        & $-0.02_{-0.05}^{+0.13}$(15)
        &  $\phantom{-}0.21_{-0.17}^{+0.24}$(14)
        &  $\phantom{-}0.25_{-0.26}^{+0.26}$(15)
        & $\phantom{-}0.32_{-0.19}^{+0.13}$(12)
        &   $\phantom{-}0.15_{-0.46}^{+0.39}$(15)
        & $\phantom{-}0.13_{-0.4}^{+0.15}$(14) &                              -
         \\
       $^{\square}$PBS15 (8)$\phantom{0}$ 
       & $-0.15_{-0.12}^{+0.14}$
       & $\phantom{-}0.05_{-0.06}^{+0.06}$(8)$\phantom{0}$ 
       & $\phantom{-}0.05_{-0.32}^{+0.30}$(8)$\phantom{0}$ 
       & $\phantom{-}0.01_{-1.13}^{+0.16}$(8)$\phantom{0}$ 
       & $-0.21_{-0.25}^{+0.31}$(4)$\phantom{0}$ 
       & $-0.47_{-0.24}^{+0.41}$(8)$\phantom{0}$ 
       & $-0.26_{-1.04}^{+0.18}$(7)$\phantom{0}$ 
       & $-0.58_{-0.09}^{+0.09}$(2)$\phantom{0}$
         \\
        $\phantom{^{\star}}$SZG17 (6)$\phantom{0}$  
        & $-1.07_{-0.27}^{+0.36}$ 
        &   $\phantom{-}0.04_{-0.12}^{+0.17}$(6)$\phantom{0}$ 
        &  $-0.47_{-0.18}^{+0.09}$(6)$\phantom{0}$ 
        &    $\phantom{-}0.01_{-0.31}^{+0.11}$(6)$\phantom{0}$ 
        &      -  
        & $-0.33_{-0.45}^{+0.09}$(5)$\phantom{0}$ 
        &  $\phantom{-}0.48_{-0.45}^{+0.29}$(4)$\phantom{0}$ 
        &  $-0.64_{-0.43}^{+0.49}$(4)$ $
         \vspace{0.05cm}\\ \hline
    \end{tabular}
\end{table*}

First we analyse the elemental abundances. It is noteworthy that the data presented in SGZ17 exhibit significant discordance ($>\pm0.3$~dex) for N/H, S/H, Ar/H and Cl/H in comparison to our work and other published results. The observations were conducted using VLT/UVES, so 
better agreement with our findings might have been anticipated. We also observed systematical higher extinction 
coefficients (by 1.07) in SGZ17 compared to our data. As such inconsistencies are not seen when comparing our 
results with other studies in the literature, it is likely that the extinction coefficients in SGZ17 are 
substantially overestimated. In comparing raw line fluxes relative to H$\beta$ in SZG17 and our observations, we found their study underestimated blue and overestimated red line fluxes, with magnitude of these deviations exhibiting a positive wavelength dependence. In fact, the wavelength coverage in SZG17 is up to 6680~\AA~, and the dominant ions of Ar, S, N and Cl are 
derived from [Ar~{\sc iv}]~$\lambda$4711,40, [S~{\sc iii}]~$\lambda$6312, [N~{\sc ii}]~$\lambda$6548,83 lines and [Cl~{\sc iii}]~$\lambda$5517,37 lines. 
An underestimation of Ar/H is, therefore, likely due to the underestimations of blue [Ar~{\sc iv}] line fluxes. 
Similarly, over-estimations of S/H, N/H and Cl/H could result from the overestimated red [S~{\sc iii}], [N~
{\sc ii}] and [Cl~{\sc iii}] line fluxes. 

{We note that the abundances reported in GKA07 and CCM10 (excluding He/H in CCM10) are consistently lower than our findings. 
In GKA07, the extinction curve by \citet{fitzpatrick1999correcting} and the KB94 ICF scheme were not employed; instead, the curve by \citet{brocklehurst1972interpretation} and ICFs from \citet{aller1984spectra} and 
\citet{samland1992spectrophotometric} were adopted. Beyond that, large differences in extinction coefficients and electron temperatures ($>800$~K) are observed when compared to our results. Our data also shows consistently higher abundances ($\geq$0.13~dex) relative to CCM10 for all elements except helium. Similar discrepancies were noticed in CCM10 when compared with earlier works. This is primarily due to the higher electron temperatures used in their study, particularly for the higher ionization zones. The $T_{\mathrm{e}}$ values derived in CCM10 using [N~{\sc ii}] and [O~{\sc iii}] lines were systematically higher than our results, with median values of 1800 K and 3300 K, respectively. We recalculated the physical parameters and chemical abundances using the de-reddened line intensities provided in GKA07 and CCM10, following the methodology employed in this work, to determine whether the procedure deriving the physical parameters, atomic data and assumptions used could explain the discrepancy. For CCM10, the median discrepancies in $T_{\mathrm{e}}$([N~{\sc ii}]) and $T_{\mathrm{e}}$([O~{\sc iii}]) became to 2300 and 2700~K. The discrepancies in elemental abundances for GKA07 and CCM10 persisted, indicating that errors in their extinction correction and line flux measurements, which might be due to the s/n of their spectra, are likely the main cause of this discrepancy.}

{We observed similar higher electron temperature estimates in GCS09, with median differences of 750~K and 200~K for $T_{\mathrm{e}}$([N~{\sc ii}]) and $T_{\mathrm{e}}$([O~{\sc iii}]), respectively. The use of low-resolution spectra from 4-m class telescopes resulted in smaller discrepancies in derived $T_{\mathrm{e}}$ values compared to CCM10, which used 2-m class telescopes. This highlights the importance of the telescope size in spectral analysis. The values of $T_{\mathrm{e}}$([N~{\sc ii}]) and $T_{\mathrm{e}}$([O~{\sc iii}])  in GCS09 are higher by median values of 550 and $-150$~K when comparing to our results neglecting the recombination corrections for [O~{\sc iii}] and [N~{\sc ii}] auroral lines. When comparing their chemical abundances with our results derived without those recombination corrections, the median differences in O and Ne reduced to less than 0.06~dex. The Cl abundances in GCS09 exhibit the largest discordance, over 1~dex, among all the elements and literature sources examined. We have recalculated the chemical abundances using the de-reddened line intensities in GCS09 and the methods in this study. The median difference in Cl/H significantly reduced to $0.05_{-0.18}^{+0.32}$ and $0.05_{-0.19}^{+0.22}$~dex, respectively compared to our results with and without the recombination corrections. This suggests that the discrepancies in Cl/H are likely due to the atomic data used for Cl. However, the discrepancy in N/H remained after this recalculation even when comparing with our results neglecting the recombination contribution, which is likely attributable to the blending of H$\alpha$ with the [N~{\sc ii}] line, as mentioned in their work.}

\label{sec:pbs15}
PBS15 employed alternative diagnostic lines from mid-infrared spectra, making the corrections for recombination in 
[N~{\sc ii}]~$\lambda$5755, [O~{\sc ii}]~$\lambda\lambda$7319,30, and [O~{III}]~$\lambda$4363 lines unnecessary. The 
agreement in O/H is good, with a median difference of 0.09~dex. One major advantage of using infrared data is the ability to observe all ionization stages of Ne, S, and Ar, thus eliminating the need for ICFs. Our results for Ne/H, S/H and Ar/H are systematically lower than their results by $>0.20$~dex. These discrepancies are within the uncertainties associated with the KB94 ICF scheme as discussed in DMS14 (at maximum, $\sim0.2$, 0.4 and 0.7~dex for S and Ar, respectively). However, apart from potential errors in the ICF estimates, this could reflect an 
underestimation of unobserved ionization stages using ICFs of these elements.

In WL07, the recombination corrections for [N~{\sc ii}], [O~{\sc ii}] and [O~{\sc iii}] auroral lines were carefully addressed. This study also applied similar data reduction approaches and calculation assumptions as in this work. Our results show a good agreement with the abundances from WL07, with discrepancies typically less than 0.05~dex. We compare the ionic abundance measurements for 10 PNe in common to WL07 in the following section to assess the overall quality of our abundance data and explore plausible reasons (such as those mentioned above) for the observed discrepancies.

\subsection{Comparison of ionic abundances with the literature results in WL07}
\label{sec:in_abun_comp}

\begin{figure*}
\caption{The individual differences in elemental and ionic abundances for 10 PNe derived in this 
work and common to WL07 for the elements He, N, O, Ne, S, Ar and Cl respectively. PNG numbers of the 
objects in common with our work are shown in the horizontal axis of the bottom plot for Cl. 
The error bars are only from uncertainties in this work. Blue circles represent elemental abundances while orange triangles, green diamonds, purple crosses and pink stars indicate the ionic abundances 
of X$^{+}$/H$^{+}$, X$^{2+}$/H$^{+}$, X$^{3+}$/H$^{+}$, and X$^{4+}$/H$^{+}$ respectively. Note that the y-axes have different ranges.}
    \centering
    \begin{subfigure}{\textwidth}
    \centering
        \includegraphics[width=13cm]{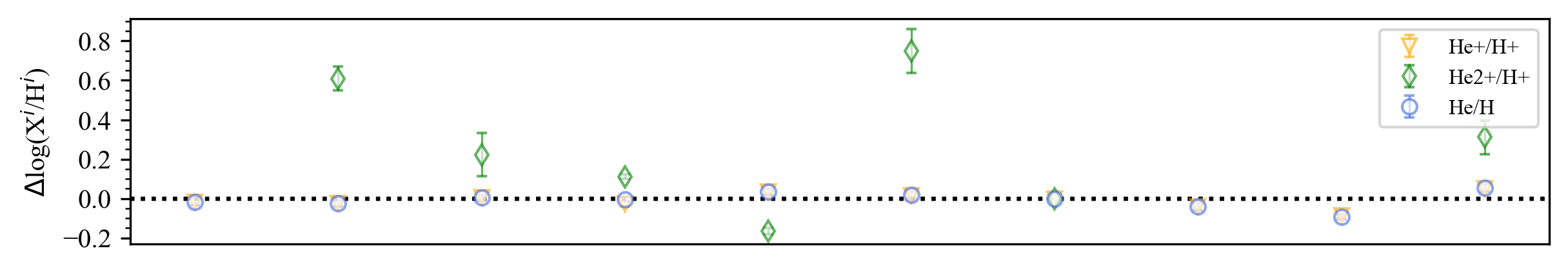}
        \caption{He}
        \label{sub_fig:abun_he}
    \end{subfigure}
    \begin{subfigure}{\textwidth}
    \centering
        \includegraphics[width=13cm]{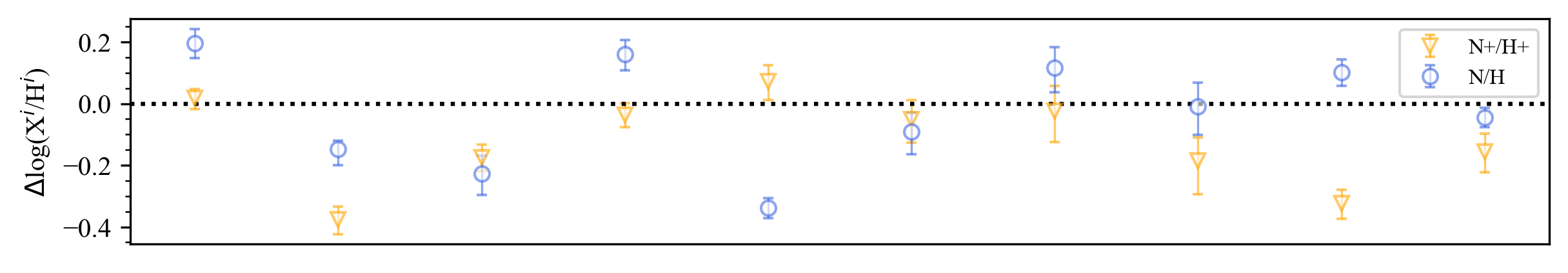}
        \caption{N}
        \label{sub_fig:abun_n}
    \end{subfigure}
    \begin{subfigure}{\textwidth}
    \centering
        \includegraphics[width=13cm]{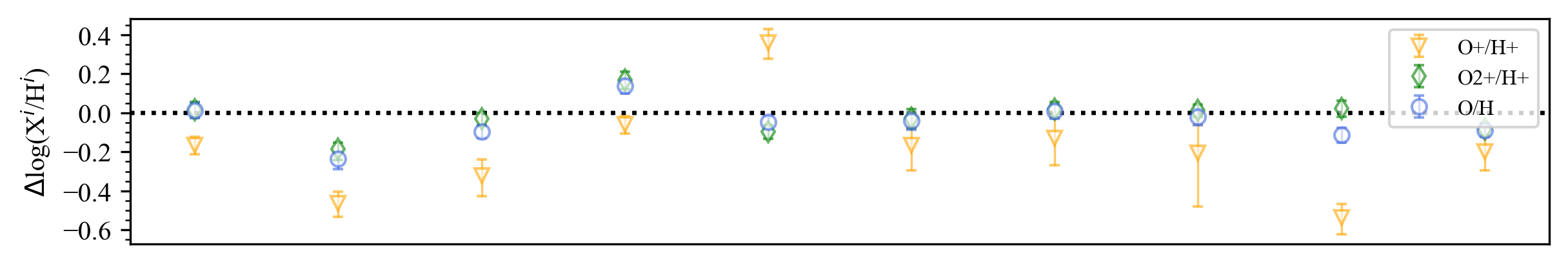}
        \caption{O}
        \label{sub_fig:abun_o}
    \end{subfigure}
    \begin{subfigure}{\textwidth}
    \centering
        \includegraphics[width=13cm]{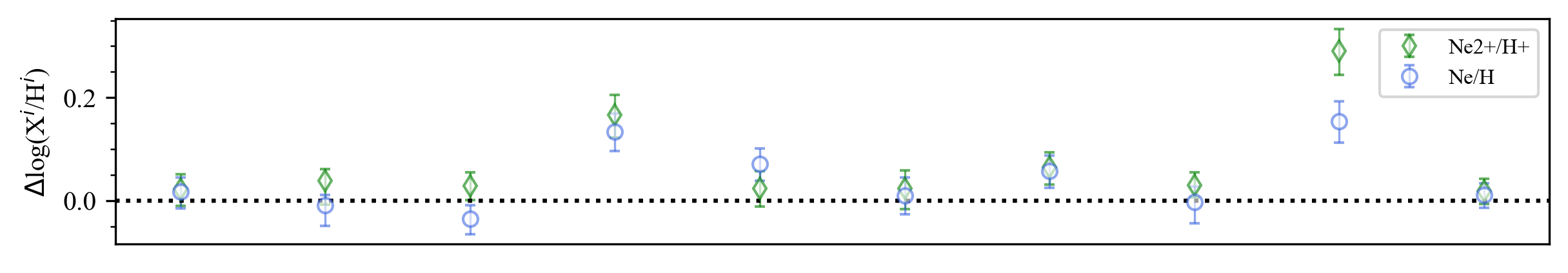}
        \caption{Ne}
        \label{sub_fig:abun_ne}
    \end{subfigure}
    \begin{subfigure}{\textwidth}
    \centering
        \includegraphics[width=13cm]{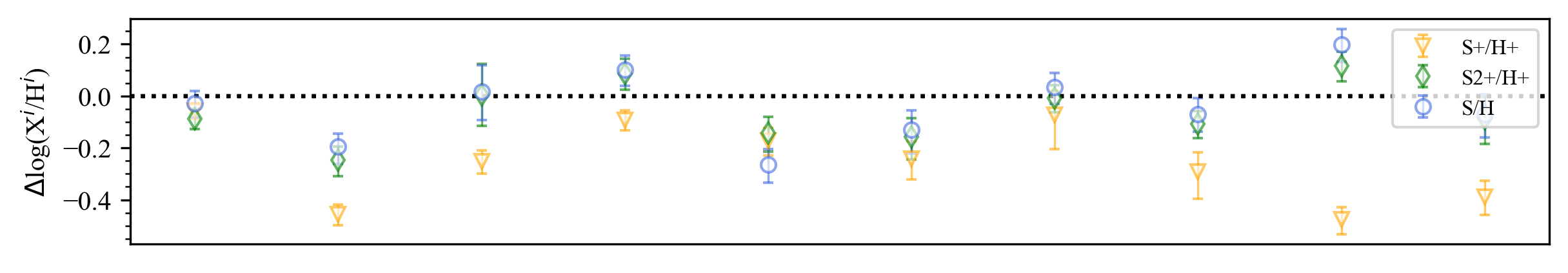}
        \caption{S}
        \label{sub_fig:abun_s}
    \end{subfigure}
    \begin{subfigure}{\textwidth}
    \centering
        \includegraphics[width=13cm]{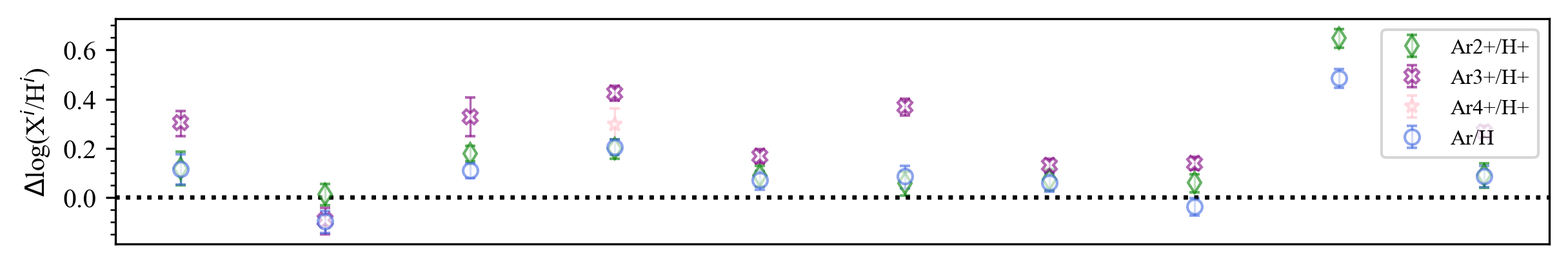}
        \caption{Ar}
        \label{sub_fig:abun_ar}
    \end{subfigure}
    \begin{subfigure}{\textwidth}
    \centering
        \includegraphics[width=13cm]{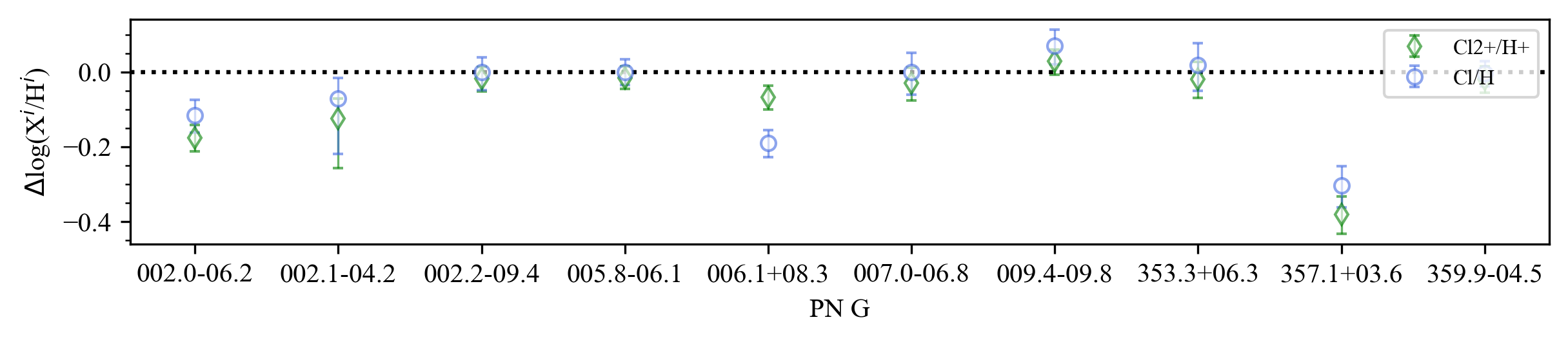}
        \caption{Cl}
        \label{sub_fig:abun_cl}
    \end{subfigure}
    \label{fig:ionic_comp}
\end{figure*}

There are 10 PNe available for direct comparison between this work and that reported in WL07, 
i.e. PNG~002.0-06.2, PNG~002.1-04.2, PNG~002.2-09.4, PNG~005.8-06.1, PNG~006.1+08.3, PNG~007.0-06.8, 
PNG~009.4-09.8, PNG~353.3+06.3, PNG~357.1+03.6, and PNG~359.9-04.5. 
For both the low- and medium-ionization regions, electron temperatures derived in WL07 and in this 
work agree well with most differences below 0.05 dex. Electron densities estimates also usually 
agree within 0.15 dex. 
The differences in ionic and final elemental abundances between our data and WL07 
for the 10 PNe in common are presented in Fig.~\ref{fig:ionic_comp} (a-f), together with our 
measurement uncertainties as error bars. This includes the elements He, N, O, Ne, S, Ar and Cl. 
Brief comments on the observed differences for the plotted ionic species are given below. 
\\\\
\textit{Helium} - the abundance value determined for He is simply the sum of He$^{+}$/H$^{+}$ and He$^{2+}$/H$^{+}$ 
(which is usually 2-3 dex weaker than He$^{+}$/H$^{+}$). The minor differences of $\leq0.01$~dex observed between 
He/H values for 8 PNe are mainly from He$^{+}$/H$^{+}$ measurements, as seen in Fig.~\ref{sub_fig:abun_he}. We weighted the three He~{\sc i} 
lines, He~{\sc i} $\lambda$4471, $\lambda$5876 and $\lambda$6678 by 1:3:1, as done in WL07. The differences in He$^{+}$/H$^{+}$  arise due to variations in the physical parameters within each PNe. The more uncertain measurements for He~{\sc ii}~$\lambda$4686 lines, could account for the larger discrepancies of $>0.5$~dex in He$^{2+}$/H$^{+}$ seen for two PNe.  
\\\\
\textit{Nitrogen} - {the discrepancies in N$^{+}$/H$^{+}$ show a moderately positive correlation with the differences in de-reddened line fluxes and a negative correlation with those in electron temperatures. After comparing our [N~{\sc ii}] line flux measurements with those in WL07, we found that, for most of the objects, the [N~{\sc ii}]~$\lambda$6548,83 line intensities in WL07 are higher than our results by up to 45\%. This could be due to lower overall extinction coefficients estimated in WL07 and a result of an imperfect deblending of the closely spaced H$\alpha$ and [N~{\sc ii}] emission lines in their low-resolution spectra when [N~{\sc ii}] lines are relatively weaker. For $T_{\mathrm{e}}$([N~{\sc ii}]) values, uncertainties from the detecting weak [N~{\sc ii}]~$\lambda$5755 lines and recombination corrections contribute to the discrepancies in individual objects. The difference in N$^{+}$ abundance can partly account for the deviation in N/H. The ICF applied to N also depends on O$^{+}$/H$^{+}$ in the KB94 scheme. As discussed in Sec.~\ref{sec:short}, this also contributes as a large difference in O$^{+}$/H$^{+}$ between our data and the present literature.}
\\\\
\textit{Oxygen} - the ionic abundance largely depends on the abundance of the dominant ion O$^{2+}$. The determination of O$^{2+}$/H$^{+}$ abundance ratios strongly depends on the measurement of the [O~{\sc iii}] $\lambda$4959, 5007 lines which are not only used for the calculation but also for the derivation of physical parameters. As the two [O~{\sc iii}] lines are usually strong and indeed the strongest nebular lines in un-dereddened PNe, their flux measurement are not often badly affected by spectral quality. The differences in line intensities are generally below 20\% when compared with WL07. The O$^{2+}$/H$^{+}$ results usually show good consistency below 0.08~dex. The variations in O$^{2+}$ abundances primarily also stem from differences in $T_{\mathrm{e}}$([O~{\sc iii}]). Upon investigation, these differences in $T_{\mathrm{e}}$([O~{\sc iii}]) show a moderate negative correlation with the difference in extinction coefficients. Most of the O$^{+}$/H$^{+}$ results in WL07 are higher than those in our data, with the discrepancies being highly correlated with the differences in $T_{\mathrm{e}}$([N~{\sc ii}]).
\\\\
\noindent
\textit{Neon} - the observed deviation in neon abundances are small though our results for Ne$^{2+}$/H$^{+}$ 
are generally higher for most PNe. The [Ne~{\sc iii}]~
$\lambda$3868 line intensities in this work are typically higher than those in WL07. The small differences in Ne$^{2+}$/H$^{+}$ seen could be a result of minor line flux measurement errors and different PNe physical parameters derived.
\\\\
\textit{Sulphur} - we observed that the differences in S/H between our study and WL07 are greater than 0.2~dex for three objects, which can be attributed to both the S$^{2+}$/H$^{+}$ and icf(S) determined. In both studies, S$^{2+}$/H$^{+}$ was calculated using the [S~{\sc iii}]~$\lambda$6312 line. We noted that the difference in S$^{2+}$/H$^{+}$ arise from differences in [S~{\sc iii}]~$\lambda$6312 line intensities and electron temperatures. In the comparison, we observed that the line intensities measured in WL07 were consistently higher than our data, typically by $\sim$20\%. The ICF of sulphur, similar to that of N, depends on the O/O$^{+}$ ratio, and the discrepancy in S/H can be attributed to the inconsistent results of O$^{+}$/H$^{+}$. 
\\\\
\textit{Argon} - we observed a systematic increase in both the individual ionic and total elemental abundances of argon compared to the values obtained in WL07, as depicted in Fig.~\ref{sub_fig:abun_ar}. {The median difference in Ar/H of 0.09~dex is the largest among all the elements. This trend was also observed by \citet{gutenkunst2008chemical} when comparing their results with those from other literature sources. The elevated Ar/H values were be attributed to several factors, including the observation of Ar$^{3+}$ ions, and uncertainties in $T_{\mathrm{e}}$. Indeed, [Ar~{\sc iv}] lines were observed in both studies for nine PNe while the object, PNG~357.1+03.6, with no [Ar~{\sc iv}] lines observed in WL07 exhibits the largest discrepancy in Ar/H. The differences in Ar$^{2+}$ abundances could primarily result from higher [Ar~{\sc iii}]~$\lambda$7135 line intensities measured in this study. This is likely due to the poor instrument detection at the red end of the spectral coverage of WL07 ($\sim$7300-7400\AA). The discrepancy diminishes as the electron temperature, $T_{\mathrm{e}}$([O~{\sc iii}]), is higher in our study compared to WL07. Meanwhile, the measured [Ar~{\sc iv}] line intensities show large variations. The discrepancy in Ar$^{3+}$ abundances could be attributed to the differences in electron temperatures, and it is also possible that the higher Ar$^{3+}$/H$^{+}$ values are due to the use of different atomic data, as discussed in \citet{dere2019chianti}}. 
\\\\
\textit{Chlorine} - the abundance of chlorine is a challenging to determine due to the low line fluxes of its dominant ion in the optical region. In column~8 of Table~\ref{tab:lit_diff}, we present the median differences in Cl/H between our data and literature data, which are usually the largest among all elements 
examined. 
\label{sec:chlorine}
To derive the chlorine abundance, we used the ICF scheme for chlorine given as 
$$\frac{\mathrm{Cl}}{\mathrm{H}}=\frac{\mathrm{S}}{\mathrm{S}^{2+}} \times \frac{\mathrm{Cl}^{2+}}{\mathrm{H}^{+}}$$ 
in \citet{liu2000ngc}, which was also used in WL07. Upon comparing our results with WL07, as illustrated 
in Fig.~\ref{sub_fig:abun_cl}, we found that the Cl$^{2+}$ ionic abundances generally align well except for 
PNG~357.1+03.6. The differences in the Cl/H are mainly due to differences in line intensities of [Cl~{\sc iii}]~$\lambda$5517,37 and are slightly related to variations in ICFs arising from small differences in S$^{2+}$/S ratios. 
\\\\
In conclusion, we believe that our newly estimated bulge PN abundances for elements He, N, O, Ne, S, Ar and Cl 
should be more accurate than current literature values for several reasons: 
i) our spectra exhibit  higher s/n ratios, 
enabling more emission lines to be used for determining ionic abundances compared to prior studies and our enhanced 
detection of weak emission lines further contributes to the accuracy of our results; ii) we ensured tight instrumental consistency throughout our entire sample; iii) {we precisely estimated and incorporated the recombination contribution of specific auroral lines. This is supported by the fact that the chemical abundances derived in this work align most closely with WL07, where recombination abundances were estimated using high-resolution spectra. Furthermore, we used the usually stronger ORLs, specifically, the N~{\sc ii} lines of multiplet V3 measured from our red spectra, which were not employed in WL07;} iv) we employed the updated atomic data in {\sc chianti 9.0} which is proven to have a better agreement with observations. 
Also, most of our measurements align with high-quality literature values in WL07 and GCS09 within a 
2$\sigma$ range, indicating that our independently estimated measurement uncertainties are reasonable. However, we 
did encounter larger uncertainties with some older published data, where spectral fitting and s/r may have been problematic. This was especially true when strong line blending occurred in spectra with lower resolutions, resulting in uncertainties in flux measurement that could exceed 40\%. 

Interestingly, our S/H results show moderate deviation when comparing with literature results using optical 
spectra while significantly systematically lower than the results from mid-infrared spectra in PBS15, this adds 
other evidence to an underestimation of higher ionization stages of sulphur in standard photoionization models as 
pointed out in \citet{kwitter2022planetary}.

\section{General PNe physical conditions and abundance results}
\label{sec:res}
\subsection{Electron densities and temperatures}
The electron densities and temperatures determined from various plasma diagnostic line ratios of 124 PNe in this 
sample are listed in Table.\ref{tab:phy_con}. The letters `L' and `H' denote when the measured 
line ratio exceeds its low- and high-temperature/density limits, respectively. The median electron density for the 
sample ranges from approximately 2000-$3000$~cm$^{-3}$, as determined by different diagnostic lines. These 
relatively high values are unsurprising given the compact and less evolved nature of the bulge PNe sample. 

\setlength{\tabcolsep}{3.4pt}
\renewcommand\arraystretch{1.1}
\begin{table*}
    \centering
    \caption{Estimates of the electron density (columns~2-5) and temperature (columns~6-10) derived 
    from all diagnostics lines used in the abundance determination for our PNe sample. Column~1 shows 
    the PNG number for each object. Electron density values are reported in units of cm$^{-3}$, and 
    electron temperatures are given in units of K. The upper and lower limits were calculated using 
    the Monte Carlo scheme implemented in \textsc{neat}. We use the letters `L' and `H' to denote 
    low- and high-temperature/density limits, respectively, when the measured line ratio exceeds these values.}
    \label{tab:phy_con}
\begin{tabular}{l@{\hskip 0.2in}lllll@{\hskip 0.6in}lllll}
\hline
\multicolumn{1}{c}{\multirow{2}{*}{PN G}} & \multicolumn{4}{c}{{Density diagnostics}  [cm$^{-3}$]} &  & \multicolumn{4}{c}{{Temperature diagnostics}  [K]}  \\ 
\cline{2-5} \cline{7-10} \multicolumn{1}{c}{} 
& [O~{\sc ii}] & \multicolumn{1}{l}{[S~{\sc ii}]} & [Cl~{\sc iii}] & [Ar~{\sc iv}] & & [N~{\sc ii}] & [O~{\sc ii}] & [S~{\sc ii}] & [O~{\sc iii}] \\ \hline
 000.1+02.6 &                              - &      915$_{-427}^{+315}$ &                                - &   1730$_{-1440}^{+1270}$ &    &                               - &                                - &                                - &              9620$\pm$300 \\
 000.1+04.3 &                              - &  10700$_{-2100}^{+1600}$ &   12400$_{-2100}^{+1800}$ &  14800$_{-1500}^{+1300}$ &    &               13400$\pm$500 &                                - &  28200$_{-6000}^{+13900}$ &             11600$\pm$200 \\
 000.1-02.3 &                              - &     1560$_{-190}^{+170}$ &       443$_{-366}^{+325}$ &     1800$_{-580}^{+520}$ &    &                               - &                                - &                                - &             11500$\pm$200 \\
 000.2-01.9 &                              - &     1330$_{-160}^{+140}$ &                                - &                               - &    &                 6750$\pm$90 &                                - &                                - &              6690$\pm$120 \\
 000.2-04.6 &                              - &                  605$\pm$30 &                                L &                               L &    &                7690$\pm$120 &                                - &      6740$_{-340}^{+330}$ &              8160$\pm$130 \\
 000.3+06.9 &                              - &                  219$\pm$56 &                                - &                               - &    &                               - &                                - &                                - &                             - \\
 000.3-04.6 &                              - &     1350$_{-160}^{+140}$ &      1600$_{-350}^{+310}$ &     1560$_{-570}^{+530}$ &    &                7720$\pm$110 &                                - &      7350$_{-560}^{+520}$ &              9100$\pm$170 \\
 000.4-01.9 &                              - &     4350$_{-580}^{+510}$ &      6170$_{-790}^{+700}$ &                               - &    &                8830$\pm$170 &                                - &   11000$_{-1900}^{+1400}$ &              7210$\pm$120 \\
 000.4-02.9 &                              - &    2440$_{-1820}^{+980}$ &                                - &                               L &    &                8470$\pm$270 &                                - &                                - &              8130$\pm$120 \\
 000.7+03.2 &                              - &     2230$_{-790}^{+550}$ &      3230$_{-370}^{+330}$ &     2430$_{-860}^{+820}$ &    &                7340$\pm$270 &                                - &                                - &              8590$\pm$160 \\
 000.7-02.7 &                              - &        407$_{-97}^{+86}$ &    7380$_{-2640}^{+1950}$ &     6540$_{-960}^{+840}$ &    &               13200$\pm$600 &                                - &                                H &             12600$\pm$300 \\
 000.7-07.4 &                              - &      392$_{-340}^{+249}$ &       448$_{-376}^{+320}$ &                               L &    &                9000$\pm$230 &                                - &      7520$_{-980}^{+870}$ &              8970$\pm$170 \\
 000.9-02.0 &                              - &     2730$_{-470}^{+400}$ &      1860$_{-490}^{+460}$ &     3290$_{-680}^{+620}$ &    &   7610$_{-3720}^{+1890}$ &                                - &                                - &              9280$\pm$190 \\
 000.9-04.8 &                              - &     1440$_{-160}^{+140}$ &       519$_{-340}^{+305}$ &     1670$_{-550}^{+520}$ &    &               10700$\pm$300 &                                - &      9180$_{-880}^{+800}$ &             11100$\pm$200 \\
 001.1-01.6 &                              - &     1050$_{-310}^{+240}$ &      3140$_{-410}^{+360}$ &                               L &    &                8190$\pm$140 &                                - &      7200$_{-600}^{+560}$ &               8210$\pm$90 \\
 001.2+02.1 &                              - &     3160$_{-380}^{+340}$ &     3510$_{-1060}^{+850}$ &      585$_{-569}^{+533}$ &    &               10700$\pm$400 &                                - &  22100$_{-12900}^{+6100}$ &              8500$\pm$120 \\
 001.2-03.0 &                              - &     3840$_{-660}^{+570}$ &                                - &                               - &    &                5920$\pm$100 &                                - &     8550$_{-1090}^{+910}$ &                             - \\
 001.3-01.2 &                              - &    2830$_{-1770}^{+940}$ &    9560$_{-9560}^{+4210}$ &                               - &    &     6640$_{-140}^{+150}$ &                                - &                                - &                             - \\
 001.4+05.3 &                              - &     2920$_{-390}^{+340}$ &      1140$_{-820}^{+700}$ &                               - &    &                7800$\pm$160 &                                - &                                - &              7940$\pm$180 \\
 001.6-01.3 &                              - &   8160$_{-8160}^{+3240}$ &      2050$_{-930}^{+760}$ &   7500$_{-1370}^{+1160}$ &    &    10200$_{-600}^{+800}$ &                                - &                                - &              9220$\pm$220 \\
 001.7+05.7 &                              - &     1030$_{-130}^{+110}$ &                                L &     1600$_{-510}^{+480}$ &    &               12500$\pm$600 &                                - &                                - &             16000$\pm$400 \\
 001.7-04.4 &                              - &     2720$_{-410}^{+350}$ &    4140$_{-1680}^{+1190}$ &                               - &    &                 5870$\pm$90 &                                - &      6270$_{-430}^{+410}$ &                             - \\
 002.0-06.2 &    2200$_{-350}^{+300}$ &     2700$_{-340}^{+300}$ &      1140$_{-710}^{+570}$ &                               - &    &                9460$\pm$190 &                                H &      6820$_{-310}^{+300}$ &              8110$\pm$110 \\
 002.1-02.2 &                              - &     5090$_{-770}^{+670}$ &      3470$_{-370}^{+340}$ &     7250$_{-650}^{+600}$ &    &               12100$\pm$300 &                                - &      6280$_{-550}^{+500}$ &             10600$\pm$200 \\
 002.1-04.2 &                              - &     8350$_{-590}^{+550}$ &  24100$_{-22300}^{+8600}$ &                               - &    &               13600$\pm$300 &                                - &     H &  10300$_{-200}^{+300}$ \\
 002.2-09.4 &                              - &     3300$_{-630}^{+530}$ &     3930$_{-1200}^{+920}$ &                               - &    &                9040$\pm$210 &                                - &   10700$_{-1400}^{+1300}$ &              8880$\pm$110 \\
 002.3+02.2 &                              - &      672$_{-208}^{+173}$ &                                - &                               - &    &                8950$\pm$220 &                                - &                                - &                             - \\
 002.5-01.7 &     976$_{-273}^{+213}$ &      479$_{-152}^{+137}$ &                                - &                               L &    &                9200$\pm$260 &   12500$_{-1100}^{+1000}$ &                                - &              8490$\pm$120 \\
 002.6+02.1 &                              - &     1760$_{-320}^{+270}$ &   14300$_{-5400}^{+4100}$ &              24100$\pm$1100 &    &                 8980$\pm$60 &                                - &                                - &   7590$_{-150}^{+210}$ \\
 002.7-04.8 &                              - &     1020$_{-120}^{+110}$ &      1150$_{-360}^{+330}$ &                               L &    &                9970$\pm$170 &                                - &                                - &              9000$\pm$160 \\
 002.8+01.7 &  6360$_{-1760}^{+1250}$ &     4350$_{-720}^{+620}$ &      3350$_{-460}^{+400}$ &                               - &    &                7100$\pm$100 &     9330$_{-1100}^{+910}$ &                  5000$\pm$10 &                             - \\
 002.8+01.8 &     682$_{-315}^{+234}$ &      575$_{-147}^{+127}$ &                                - &                               - &    &                9150$\pm$200 &      8950$_{-740}^{+680}$ &                  5000$\pm$10 &                             - \\
 002.9-03.9 &    1830$_{-460}^{+370}$ &  12400$_{-8500}^{+3900}$ &                                L &     1110$_{-640}^{+590}$ &    &                               - &   H &       5000$_{-380}^{+10}$ &             13200$\pm$200 \\
 003.2-06.2 &   4990$_{-1210}^{+890}$ &     2870$_{-400}^{+350}$ &      4270$_{-610}^{+540}$ &     4140$_{-790}^{+660}$ &    &     9610$_{-250}^{+240}$ &   12400$_{-1900}^{+1500}$ &   11000$_{-2000}^{+1500}$ &              8800$\pm$160 \\
 003.6-02.3 &     864$_{-198}^{+161}$ &      931$_{-180}^{+151}$ &                                L &                               L &    &                8510$\pm$140 &   H &      6680$_{-350}^{+330}$ &              8840$\pm$210 \\
 003.7+07.9 &                              - &       563$_{-107}^{+90}$ &                                - &                               - &    &                 5000$\pm$10 &                                - &                                - &             13700$\pm$200 \\
 003.7-04.6 &                              - &     3420$_{-590}^{+500}$ &      1150$_{-580}^{+520}$ &                               L &    &  19400$_{-3000}^{+2600}$ &                                - &   14400$_{-6200}^{+3200}$ &             10900$\pm$200 \\
 003.8-04.3 &                              - &     2030$_{-240}^{+220}$ &      1400$_{-240}^{+210}$ &     1680$_{-480}^{+470}$ &    &                9330$\pm$160 &                                - &      8240$_{-530}^{+500}$ &             10800$\pm$200 \\
 003.9-02.3 &                              - &     3860$_{-500}^{+450}$ &      5750$_{-900}^{+780}$ &     4850$_{-770}^{+670}$ &    &                8400$\pm$100 &                                - &      8770$_{-830}^{+760}$ &              8460$\pm$100 \\
 003.9-03.1 &                              - &     2680$_{-910}^{+610}$ &    3030$_{-1590}^{+1170}$ &     1310$_{-490}^{+470}$ &    &                               - &                                - &                                - &             13200$\pm$300 \\
 004.0-03.0 &                              - &     3030$_{-400}^{+350}$ &      2370$_{-390}^{+330}$ &                               - &    &               14000$\pm$400 &                                - &      9720$_{-930}^{+850}$ &                             H \\
 004.1-03.8 &                              - &     1630$_{-180}^{+160}$ &      1310$_{-480}^{+410}$ &      829$_{-498}^{+467}$ &    &               10600$\pm$200 &                                - &      7870$_{-680}^{+620}$ &             11100$\pm$200 \\
 004.2-03.2 &                              - &     1440$_{-700}^{+490}$ &                                L &      601$_{-513}^{+472}$ &    &  13300$_{-1600}^{+1400}$ &                                - &                                - &             11400$\pm$300 \\
 004.2-04.3 &                              - &     4110$_{-500}^{+450}$ &    3560$_{-1290}^{+1110}$ &     1610$_{-480}^{+460}$ &    &  21400$_{-2400}^{+2000}$ &                                - &                                - &              9740$\pm$190 \\
 004.6+06.0 &                              - &     3610$_{-450}^{+400}$ &      2300$_{-690}^{+530}$ &                               - &    &                8270$\pm$140 &                                - &                                - &              6690$\pm$160 \\
 004.8+02.0 &                              - &   5610$_{-2130}^{+1340}$ &                                L &                               - &    &     9570$_{-290}^{+300}$ &                                - &                                - &              8200$\pm$270 \\
 004.8-05.0 &                              - &     1650$_{-460}^{+360}$ &       565$_{-295}^{+279}$ &                               L &    &               14900$\pm$700 &                                - &                                - &              8570$\pm$210 \\
 005.0-03.9 &    1460$_{-580}^{+410}$ &        387$_{-97}^{+86}$ &                                - &   1460$_{-1330}^{+1210}$ &    &               17800$\pm$900 &   33700$_{-1300}^{+8800}$ &                                - &            23400$\pm$1100 \\
 005.2+05.6 &                              - &     4040$_{-600}^{+530}$ &       844$_{-239}^{+224}$ &     1260$_{-490}^{+460}$ &    &               10700$\pm$200 &                                - &      7970$_{-840}^{+760}$ &              8490$\pm$140 \\
 005.5+06.1 &                              - &      565$_{-239}^{+193}$ &    4750$_{-1630}^{+1250}$ &                               - &    &                6670$\pm$140 &                                - &                                - &                             - \\
 005.5-04.0 &                              - &     1390$_{-190}^{+160}$ &    1540$_{-1540}^{+1150}$ &      763$_{-437}^{+401}$ &    &                               - &                                - &                                - &             10200$\pm$300 \\
 005.8-06.1 &    2660$_{-600}^{+490}$ &     2890$_{-370}^{+330}$ &      2860$_{-500}^{+430}$ &     2560$_{-460}^{+440}$ &    &                8720$\pm$140 &  19800$_{-10900}^{+4100}$ &      7620$_{-520}^{+490}$ &              9060$\pm$150 \\
 006.1+08.3 &                              - &     6050$_{-980}^{+850}$ &   10800$_{-1400}^{+1200}$ &  16600$_{-1600}^{+1400}$ &    &               10900$\pm$300 &                                - &   H &             10400$\pm$200 \\
 006.4+02.0 &                              - &   7840$_{-1430}^{+1210}$ &   11700$_{-1600}^{+1400}$ &   9330$_{-1250}^{+1100}$ &    &                9490$\pm$200 &                                - &  24300$_{-10700}^{+9700}$ &              7690$\pm$100 \\
 006.4-04.6 &                              - &     1240$_{-150}^{+130}$ &       123$_{-123}^{+122}$ &     1020$_{-540}^{+530}$ &    &               11100$\pm$500 &                                - &                                - &             12800$\pm$300 \phantom{-}  \vspace{0.07cm} \\
\hline
\end{tabular}
\end{table*}
\setlength{\tabcolsep}{3.4pt}
\renewcommand\arraystretch{1.1}
\begin{table*}
    \centering
    \contcaption{}
\begin{tabular}{l@{\hskip 0.2in}lllll@{\hskip 0.6in}lllll}
\hline
\multicolumn{1}{c}{\multirow{2}{*}{PN G}} & \multicolumn{4}{c}{{Density diagnostics}  [cm$^{-3}$]} &  & \multicolumn{4}{c}{{Temperature diagnostics}  [K]}  \\ 
\cline{2-5} \cline{7-10} \multicolumn{1}{c}{} 
& [O~{\sc ii}] & \multicolumn{1}{l}{[S~{\sc ii}]} & [Cl~{\sc iii}] & [Ar~{\sc iv}] & & [N~{\sc ii}] & [O~{\sc ii}] & [S~{\sc ii}] & [O~{\sc iii}] \\ \hline
006.8+02.3 &                              - &      3040$_{-360}^{+320}$ &     1340$_{-450}^{+420}$ &    4860$_{-1050}^{+860}$ &    &               12200$\pm$300 &                                - &   11000$_{-1700}^{+1300}$ &             14200$\pm$400 \\
 006.8-03.4 &  3950$_{-1680}^{+1050}$ &      7540$_{-470}^{+450}$ &                               - &     4150$_{-310}^{+290}$ &    &    10400$_{-300}^{+200}$ &    34600$_{-400}^{+8100}$ &      9600$_{-890}^{+880}$ &             12200$\pm$100 \\
 007.0+06.3 &                              - &    4660$_{-1540}^{+1050}$ &     2390$_{-680}^{+530}$ &                               L &    &     9070$_{-290}^{+280}$ &                                - &     6510$_{-1040}^{+860}$ &              7510$\pm$120 \\
 007.0-06.8 &   3070$_{-1210}^{+750}$ &      2080$_{-670}^{+510}$ &     2110$_{-880}^{+740}$ &    6570$_{-1060}^{+910}$ &    &                8220$\pm$340 &   20700$_{-5200}^{+4100}$ &   13000$_{-3300}^{+2300}$ &              8180$\pm$150 \\
 007.5+07.4 &                              - &       706$_{-162}^{+132}$ &                               L &                               L &    &                8020$\pm$140 &                                - &      9860$_{-960}^{+870}$ &              9290$\pm$170 \\
 007.6+06.9 &                              - &      1280$_{-170}^{+150}$ &      824$_{-320}^{+300}$ &     1090$_{-510}^{+490}$ &    &                9760$\pm$160 &                                - &      8670$_{-820}^{+750}$ &             10600$\pm$100 \\
 007.8-03.7 &                              - &        685$_{-108}^{+94}$ &   8260$_{-3120}^{+2640}$ &     1120$_{-220}^{+200}$ &    &                7620$\pm$170 &                                - &      7730$_{-530}^{+500}$ &              9110$\pm$120 \\
 007.8-04.4 &   4180$_{-1240}^{+850}$ &    7200$_{-2100}^{+1410}$ &                               - &                               - &    &                5930$\pm$150 &                 6760$\pm$440 &      6520$_{-610}^{+560}$ &                             - \\
 008.2+06.8 &                              - &    9750$_{-2110}^{+1730}$ &     2540$_{-460}^{+390}$ &                               - &    &               13400$\pm$400 &                                - &  33900$_{-1100}^{+12200}$ &  19700$_{-600}^{+500}$ \\
 008.4-03.6 &                              - &       570$_{-193}^{+168}$ &                               - &                               - &    &                5980$\pm$100 &                                - &       5000$_{-180}^{+10}$ &                             - \\
 008.6-02.6 &                              - &    8650$_{-8440}^{+3190}$ &     1160$_{-590}^{+490}$ &   7310$_{-1670}^{+1360}$ &    &    10100$_{-500}^{+700}$ &                                - &                                - &             10900$\pm$200 \\
 009.4-09.8 &                              - &    2870$_{-1670}^{+1060}$ &     1990$_{-380}^{+350}$ &     1510$_{-490}^{+460}$ &    &               10800$\pm$500 &                                - &                                - &              8780$\pm$140 \\
 009.8-04.6 &                              - &      1360$_{-380}^{+300}$ &                               L &     1530$_{-580}^{+540}$ &    &                9370$\pm$260 &                                - &   12800$_{-2200}^{+1700}$ &              9870$\pm$240 \\
 351.1+04.8 &                              - &      4990$_{-960}^{+810}$ &     5480$_{-900}^{+770}$ &                               - &    &                7900$\pm$130 &                                - &   12100$_{-2700}^{+1900}$ &              7540$\pm$100 \\
 351.2+05.2 &                              - &      1390$_{-190}^{+170}$ &     2220$_{-280}^{+250}$ &                               - &    &                 5990$\pm$60 &                                - &                  5000$\pm$10 &              6870$\pm$160 \\
 351.6-06.2 &                              - &      1440$_{-180}^{+160}$ &     1290$_{-290}^{+260}$ &     1180$_{-450}^{+430}$ &    &                8020$\pm$170 &                                - &      7610$_{-430}^{+400}$ &             10100$\pm$100 \\
 351.9+09.0 &                              - &     1160$_{-1130}^{+620}$ &      336$_{-336}^{+335}$ &     1100$_{-620}^{+570}$ &    &                               - &                                - &                                - &             12800$\pm$200 \\
 351.9-01.9 &  8130$_{-6250}^{+2690}$ &      5230$_{-940}^{+800}$ &     3510$_{-220}^{+210}$ &  21600$_{-1700}^{+1600}$ &    &    11900$_{-300}^{+400}$ &   16000$_{-4100}^{+3800}$ &   14900$_{-7600}^{+4800}$ &               9520$\pm$70 \\
 352.0-04.6 &                              - &      3750$_{-480}^{+430}$ &     6010$_{-770}^{+680}$ &     4440$_{-800}^{+680}$ &    &                8640$\pm$160 &                                - &   10600$_{-1600}^{+1200}$ &              8830$\pm$150 \\
 352.1+05.1 &    3620$_{-750}^{+560}$ &      3100$_{-130}^{+120}$ &     4700$_{-350}^{+330}$ &     5010$_{-340}^{+320}$ &    &                8290$\pm$100 &   20700$_{-2900}^{+2200}$ &                10100$\pm$600 &               8980$\pm$50 \\
 352.6+03.0 &  3110$_{-1930}^{+1000}$ &      4070$_{-260}^{+240}$ &   8880$_{-1650}^{+1390}$ &               14100$\pm$900 &    &                8290$\pm$130 &                                H &   10500$_{-1600}^{+1500}$ &               7960$\pm$70 \\
 353.2-05.2 &                 164$\pm$56 &         113$_{-56}^{+51}$ &                               - &                               L &    &                7720$\pm$130 &                                H &      5880$_{-390}^{+360}$ &              9320$\pm$170 \\
 353.3+06.3 &                              - &      5340$_{-910}^{+780}$ &   3620$_{-2240}^{+1520}$ &   7920$_{-1820}^{+1480}$ &    &    10600$_{-800}^{+900}$ &                                - &  19200$_{-14900}^{+4800}$ &              9960$\pm$140 \\
 353.7+06.3 &     851$_{-376}^{+261}$ &       514$_{-310}^{+234}$ &                               L &                               - &    &                6950$\pm$160 &    10300$_{-1000}^{+900}$ &      6290$_{-470}^{+440}$ &               7070$\pm$90 \\
 354.5+03.3 &                              - &   11300$_{-7400}^{+3500}$ &  11900$_{-1400}^{+1300}$ &  10200$_{-1100}^{+1000}$ &    &   15200$_{-900}^{+1300}$ &                                - &   32400$_{-2600}^{+4100}$ &             10600$\pm$100 \\
 354.9+03.5 &                              - &      2730$_{-390}^{+340}$ &     1930$_{-520}^{+450}$ &                               - &    &                7690$\pm$180 &                                - &      6710$_{-500}^{+460}$ &                             - \\
 355.4-02.4 &    3580$_{-840}^{+680}$ &      3610$_{-470}^{+420}$ &     6740$_{-900}^{+790}$ &  10700$_{-1100}^{+1000}$ &    &                8330$\pm$130 &   17100$_{-2200}^{+1900}$ &   10500$_{-1400}^{+1100}$ &              8410$\pm$120 \\
 355.9+03.6 &                              - &  18000$_{-16100}^{+5900}$ &                               - &                               - &    &  14300$_{-1100}^{+2000}$ &                                - &   28300$_{-3500}^{+2200}$ &  10400$_{-200}^{+300}$ \\
 355.9-04.2 &                              - &      5930$_{-260}^{+250}$ &    4340$_{-1230}^{+960}$ &                               - &    &                6380$\pm$260 &                                - &      7100$_{-300}^{+290}$ &               6290$\pm$70 \\
 356.1-03.3 &                              - &         317$_{-68}^{+56}$ &      697$_{-147}^{+137}$ &     3730$_{-760}^{+710}$ &    &                8870$\pm$130 &                                - &      8840$_{-540}^{+510}$ &              9130$\pm$180 \\
 356.3-06.2 &                              - &        235$_{-102}^{+97}$ &                               L &     1390$_{-900}^{+820}$ &    &                8730$\pm$150 &                                - &                                - &              9430$\pm$130 \\
 356.5-03.6 &                              - &     3570$_{-1060}^{+820}$ &     3120$_{-600}^{+510}$ &     3660$_{-700}^{+650}$ &    &                8030$\pm$200 &                                - &                                - &              9040$\pm$190 \\
 356.8+03.3 &                              - &      4440$_{-420}^{+380}$ &                               - &                               - &    &                6850$\pm$100 &                                - &                                - &                             - \\
 356.8-05.4 &                              - &       336$_{-112}^{+107}$ &                               L &                               L &    &                7860$\pm$160 &                                - &                                - &              9100$\pm$140 \\
 356.9+04.4 &    3370$_{-870}^{+630}$ &      6560$_{-820}^{+730}$ &  14900$_{-4200}^{+3300}$ &  33800$_{-5600}^{+4800}$ &    &               14000$\pm$400 &                                H &                                H &             12200$\pm$300 \\
 357.0+02.4 &                              - &      1870$_{-270}^{+240}$ &     1610$_{-390}^{+340}$ &                               L &    &                8650$\pm$130 &                                - &                                - &              8830$\pm$190 \\
 357.1+03.6 &                              - &      4010$_{-730}^{+620}$ &     1470$_{-890}^{+700}$ &                               - &    &                8430$\pm$180 &                                - &  H &              7220$\pm$130 \\
 357.1+04.4 &                              - &      1200$_{-270}^{+220}$ &                               - &                               L &    &                8260$\pm$360 &                                - &                                - &              7880$\pm$320 \\
 357.1-04.7 &    2130$_{-240}^{+220}$ &    6680$_{-1510}^{+1100}$ &                               - &                               - &    &                5590$\pm$190 &      9720$_{-810}^{+750}$ &   15700$_{-3900}^{+2800}$ &                             - \\
 357.2+02.0 &                              - &      3110$_{-670}^{+550}$ &     1090$_{-480}^{+420}$ &     2800$_{-710}^{+670}$ &    &     9730$_{-270}^{+260}$ &                                - &    9710$_{-1430}^{+1140}$ &             10100$\pm$200 \\
 357.3+04.0 &                              - &    6900$_{-2000}^{+1380}$ &  18400$_{-6700}^{+4700}$ &     2120$_{-640}^{+580}$ &    &    11900$_{-700}^{+600}$ &                                - &                                - &              8160$\pm$100 \\
 357.5+03.1 &                              - &      1450$_{-280}^{+240}$ &                               - &                               - &    &   13300$_{-1000}^{+900}$ &                                - &                                - &                             - \\
 357.5+03.2 &                              - &      1610$_{-340}^{+280}$ &                               - &                               L &    &                8860$\pm$220 &                                - &                                - &             11000$\pm$300 \\
 357.6-03.3 &                              - &        352$_{-107}^{+97}$ &                               - &                               - &    &                7370$\pm$220 &                                - &                                - &                             - \\
 357.9-03.8 &                              - &    2770$_{-1800}^{+1030}$ &                               - &                               L &    &  21700$_{-2000}^{+1800}$ &                                - &                                - &             21100$\pm$700 \\
 357.9-05.1 &                              - &       872$_{-113}^{+100}$ &      590$_{-274}^{+269}$ &      926$_{-605}^{+569}$ &    &                8620$\pm$190 &                                - &      8360$_{-580}^{+550}$ &              9160$\pm$130 \\
 358.0+09.3 &                              - &       377$_{-377}^{+269}$ &      631$_{-631}^{+630}$ &                               L &    &                               - &                                - &                                - &             10200$\pm$200 \\
 358.2+03.5 &   1850$_{-1850}^{+850}$ &     6770$_{-1020}^{+880}$ &     5400$_{-820}^{+710}$ &    9410$_{-1080}^{+970}$ &    &               13300$\pm$400 &  23800$_{-11200}^{+6300}$ &                                H &             11100$\pm$300 \\
 358.2+04.2 &    1970$_{-290}^{+250}$ &    8530$_{-2110}^{+1540}$ &     3610$_{-590}^{+500}$ &   5620$_{-3540}^{+3120}$ &    &                9350$\pm$200 &   11300$_{-1200}^{+1100}$ &   15100$_{-5800}^{+3200}$ &               7980$\pm$90 \\
 358.5+02.9 &                              - &      2350$_{-370}^{+320}$ &                               - &     1650$_{-710}^{+590}$ &    &                               - &                                - &                                - &             11900$\pm$200 \\
 358.5-04.2 &  6820$_{-3200}^{+1750}$ &    8120$_{-1630}^{+1250}$ &  12400$_{-2700}^{+2200}$ &                               - &    &               12600$\pm$400 &   22800$_{-8700}^{+5200}$ &    H &              9660$\pm$160 \\
 358.6+07.8 &    3170$_{-820}^{+650}$ &      2390$_{-340}^{+300}$ &     1050$_{-380}^{+340}$ &     1600$_{-460}^{+440}$ &    &               12400$\pm$400 &                                H &   11900$_{-1600}^{+1300}$ &               8520$\pm$90 \\
 358.6-05.5 &     518$_{-144}^{+112}$ &          97$_{-50}^{+45}$ &                               L &   1970$_{-1290}^{+1140}$ &    &                8810$\pm$210 &                                H &      7600$_{-390}^{+370}$ &              9400$\pm$120 \\
 358.7+05.2 &   2490$_{-1530}^{+830}$ &    6550$_{-1400}^{+1040}$ &                               - &                               - &    &     6250$_{-120}^{+130}$ &     7310$_{-7310}^{+710}$ &      7190$_{-950}^{+840}$ &                             - \\
 358.8+03.0 &                              - &      1980$_{-200}^{+190}$ &     2820$_{-990}^{+850}$ &     1290$_{-560}^{+520}$ &    &                8000$\pm$180 &                                - &      8730$_{-690}^{+640}$ &             10400$\pm$200 \\
 358.9+03.4 &                              - &                10700$\pm$600 &  11300$_{-1100}^{+1000}$ &                               - &    &                9850$\pm$180 &                                - &  32700$_{-1200}^{+12200}$ &               7000$\pm$50 \\
 359.0-04.1 &                              - &       965$_{-166}^{+141}$ &                               L &      814$_{-630}^{+533}$ &    &                8450$\pm$140 &                                - &      7820$_{-540}^{+500}$ &              8650$\pm$180 \\
 359.1-02.9 &    1150$_{-190}^{+170}$ &      1070$_{-270}^{+210}$ &               1330$\pm$1330 &                               L &    &                7420$\pm$210 &                                H &                  5000$\pm$10 &              7990$\pm$150  \vspace{0.07cm} \\
\hline
\end{tabular}
\end{table*}
\setlength{\tabcolsep}{3.4pt}
\renewcommand\arraystretch{1.1}
\begin{table*}
    \centering
    \contcaption{}
\begin{tabular}{l@{\hskip 0.2in}lllll@{\hskip 0.6in}lllll}
\hline
\multicolumn{1}{c}{\multirow{2}{*}{PN G}} & \multicolumn{4}{c}{{Density diagnostics}  [cm$^{-3}$]} &  & \multicolumn{4}{c}{{Temperature diagnostics}  [K]}  \\ 
\cline{2-5} \cline{7-10} \multicolumn{1}{c}{} 
& [O~{\sc ii}] & \multicolumn{1}{l}{[S~{\sc ii}]} & [Cl~{\sc iii}] & [Ar~{\sc iv}] & & [N~{\sc ii}] & [O~{\sc ii}] & [S~{\sc ii}] & [O~{\sc iii}] \\ \hline
 359.2+04.7 &  5410$_{-1640}^{+1100}$ &  8790$_{-8010}^{+3070}$ &  10700$_{-1900}^{+1600}$ &                               - &    &     9690$_{-300}^{+400}$ &  H &    7440$_{-1400}^{+1420}$ &                  - \\
 359.3-01.8 &  9070$_{-5880}^{+2710}$ &   5290$_{-1190}^{+890}$ &                               L &                               - &    &     6420$_{-260}^{+280}$ &     5860$_{-440}^{+540}$ &     6950$_{-940}^{+1050}$ &                  - \\
 359.6-04.8 &                              - &     362$_{-178}^{+152}$ &                               - &     1030$_{-650}^{+570}$ &    &  19400$_{-1200}^{+1100}$ &                               - &                                - &  13000$\pm$300 \\
 359.7-01.8 &   1830$_{-1450}^{+740}$ &   1780$_{-1210}^{+720}$ &                               L &     1520$_{-680}^{+660}$ &    &   7710$_{-2630}^{+2110}$ &  H &                                - &  12500$\pm$300 \\
 359.8+02.4 &                              - &  5870$_{-1580}^{+1100}$ &  12700$_{-9200}^{+4400}$ &                               - &    &                6700$\pm$100 &                               - &                  5000$\pm$10 &                  - \\
 359.8+03.7 &                              - &    5290$_{-830}^{+720}$ &    2330$_{-1340}^{+980}$ &                               - &    &     9800$_{-660}^{+610}$ &                               - &                                - &   8830$\pm$160 \\
 359.8+05.2 &                  67$\pm$55 &       219$_{-31}^{+27}$ &                               - &                               - &    &                               - &                7440$\pm$310 &                                - &                  - \\
 359.8+05.6 &    3910$_{-960}^{+770}$ &  7520$_{-2130}^{+1400}$ &   6410$_{-6070}^{+3280}$ &                               - &    &                6270$\pm$260 &     7660$_{-710}^{+600}$ &     7870$_{-1000}^{+890}$ &                  - \\
 359.8+06.9 &     999$_{-361}^{+265}$ &     826$_{-178}^{+147}$ &   4360$_{-1690}^{+1220}$ &                               - &    &                9900$\pm$160 &    10100$_{-900}^{+800}$ &      6120$_{-430}^{+410}$ &   9630$\pm$120 \\
 359.8-07.2 &                              - &  8260$_{-3680}^{+2120}$ &    1100$_{-1000}^{+800}$ &     3770$_{-640}^{+600}$ &    &  16100$_{-3200}^{+2500}$ &                               - &                                - &  11800$\pm$200 \\
 359.9-04.5 &  5560$_{-1470}^{+1030}$ &    4620$_{-770}^{+660}$ &   9710$_{-1460}^{+1270}$ &  16600$_{-2000}^{+1800}$ &    &                9560$\pm$230 &                               H &  20300$_{-14700}^{+5800}$ &   8430$\pm$110   \phantom{-} \vspace{0.07cm} 
\\ \hline
\end{tabular}
\end{table*}

\begin{figure}
    \centering
    \includegraphics[width = 0.43\textwidth]{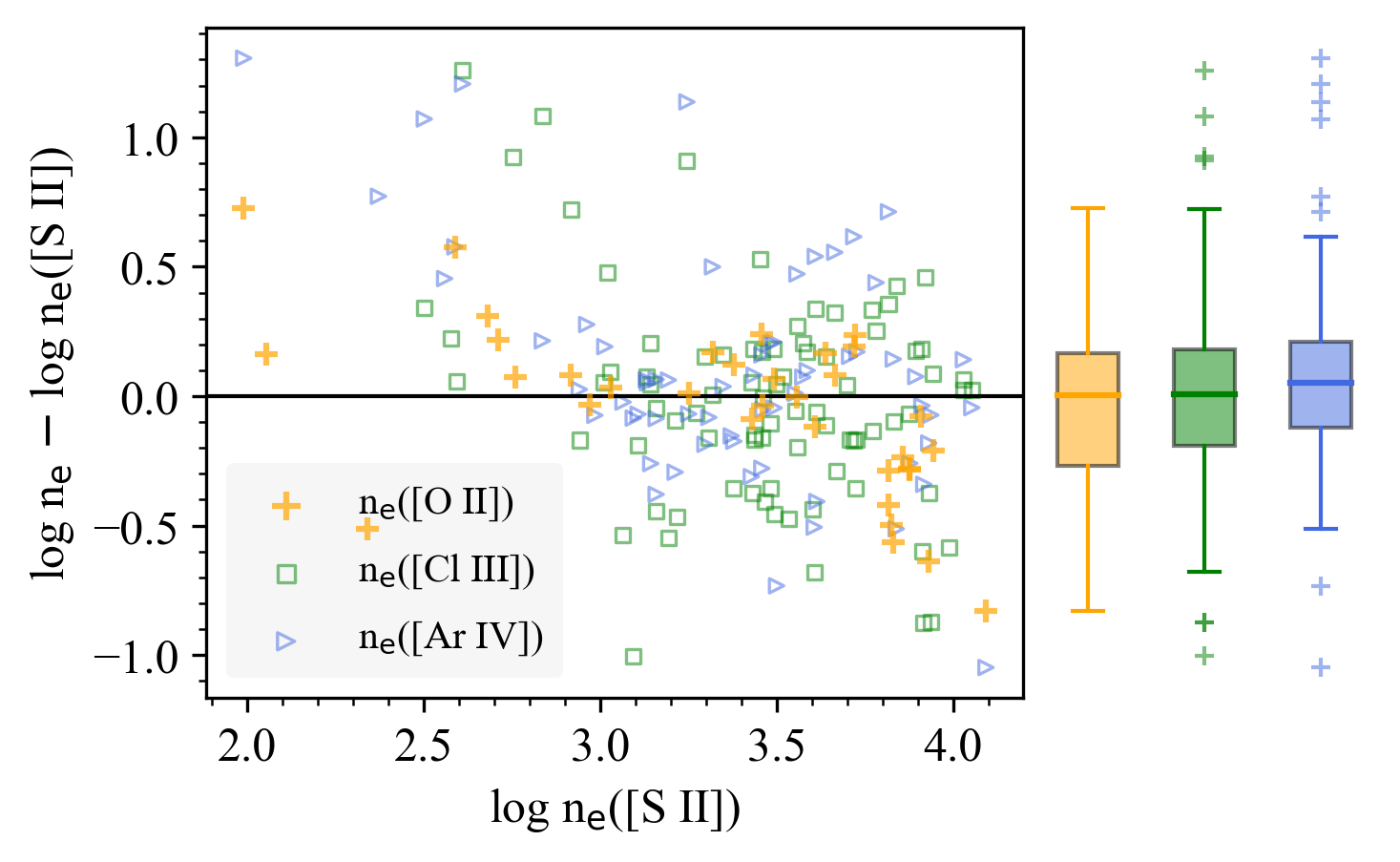}
    \caption{Comparison of electron density estimates using different doublet lines with those derived using [S~{\sc ii}] doublet lines. Electron densities in the units of cm$^{-3}$ are shown on a log scale, with values beyond the low or high limits excluded. The scatter plot 
    displays the difference between electron density estimates using [O~{\sc ii}] (orange crosses), [Cl~{\sc iii}] 
    (green squares) and [Ar~{\sc iv}] (blue triangles) compared to estimates from the [S~{\sc ii}] 
    lines, plotted against electron density 
    estimates using [S~{\sc ii}] lines. The black horizontal line marks the zero point. The box plots on the right show the distribution of the differences in electron density estimates, color-coded accordingly. The boxes depict the 25th to 75th percentiles, with the median indicated by a horizontal line. The whiskers extend to the 10th and 90th percentile, and outliers are represented by individual crosses.}
    \label{fig:ne_comp}
\end{figure}
\begin{figure}
    \centering
    \includegraphics[width = 0.43\textwidth]{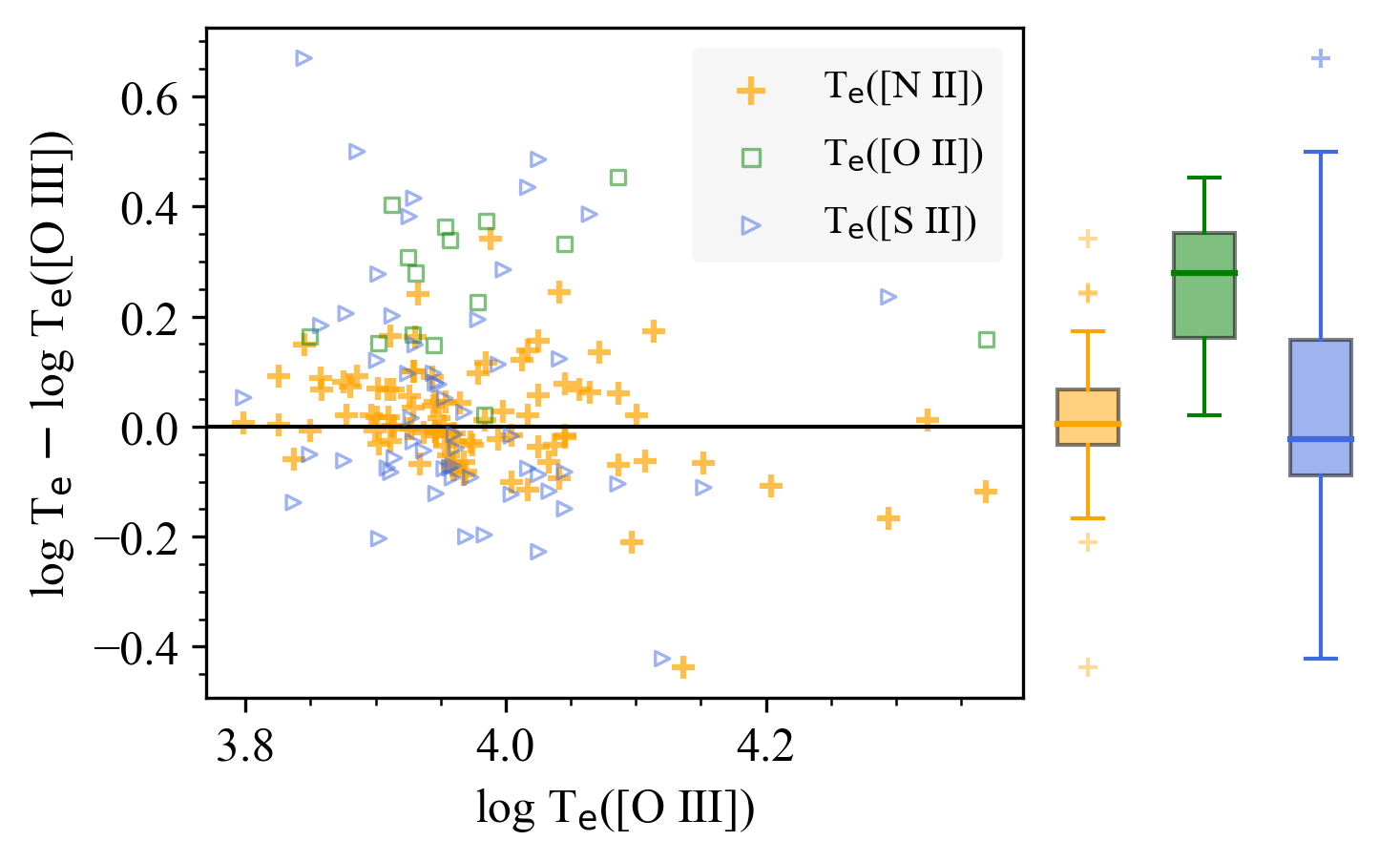}
    \caption{Comparison of electron temperature estimates using different diagnostic lines with those derived using [O~{\sc iii}] 
    lines. Electron temperatures in the units of K are shown on a log scale, with values beyond the low or high limits excluded. The scatter plot 
    displays the difference between electron temperature estimates using [N~{\sc ii}] (orange crosses), [O~{\sc ii}] (green squares) and [S~{\sc ii}] (blue triangles) compared to estimates from the [O~{\sc iii}] lines, plotted against electron 
    temperature estimates using [O~{\sc iii}] lines. The box plots on the right show the distribution of the differences in 
    electron temperature estimates, color-coded accordingly. The boxes depict the 25th to 75th percentiles, with the median indicated by a horizontal line. The whiskers extend to the 10th and 90th percentile, and outliers are represented by 
    individual crosses.}
    \label{fig:Te_comp}
\end{figure}

Figure~\ref{fig:ne_comp} displays a comparison of electron density estimates derived from different 
doublet lines with those obtained using [S~{\sc ii}] doublet lines, which serve as a reference due 
to their typical {observational accessibility}. The figure also showcases the distribution of differences between these estimates. Although there is a good overall agreement between the different {emission-}line diagnostics and the [S~{\sc ii}] lines within 0.05~dex, several outliers are observed above and below the 1:1 agreement line at 0, particularly for estimates using [Cl~{\sc iii}] and [Ar~{\sc iv}] {lines}. {The discrepancies can be attributed to different sensitivity ranges of their line ratios, as illustrated in Fig.~2 in \citet{mendez2023density}. When compared to [S~{\sc ii}], both the [Cl~{\sc iii}] and [Ar~{\sc iv}] line ratios show limited sensitivity to $n_{\mathrm{e}}$ for $n_{\mathrm{e}}<1000$~cm$^{-3}$, resulting in an overestimation of $n_{\mathrm{e}}$ while they exhibit greater sensitivity at $n_{\mathrm{e}}>10,000\ \text{cm}^{-3}$, a range where [S~{\sc ii}] lines tend to underestimate $n_{\mathrm{e}}$.}

The Pearson's correlation coefficient is 0.48 between the low-ionization zone density diagnostics, $n_{\mathrm{e}}$([O~{\sc ii}]) and $n_{\mathrm{e}}$([S~{\sc ii}]), and a value of 0.64 between the 
medium-ionization zone diagnostics, $n_{\mathrm{e}}$([Ar~{\sc iv}]) and $n_{\mathrm{e}}$([Cl~{\sc iii}]). The median difference between $n_{\mathrm{e}}$([S~{\sc ii}]) and $n_{\mathrm{e}}$([O~{\sc ii}]) is 
$-$0.004 dex while at higher $n_{\mathrm{e}}$([S~{\sc ii}]) ($>7000$~cm$^{-3}$), $n_{\mathrm{e}}$([O~{\sc ii}]) is lower by 0.1-0.85~dex, resulting in a weak overall correlation. This discrepancy may be due to sensitivity limits of the line ratios as clearly demonstrated by Fig.~2 in \citet{mendez2023density}, their different dependence on electron temperatures or the atomic data, as discussed in \citet{kisielius2009electron}. The median difference between $n_{\mathrm{e}}$ derived from [Cl~{\sc iii}] and [Ar~{\sc iv}] lines is $-0.18$~dex, indicating a systematic difference. A typical uncertainty in $n_{\mathrm{e}}$([Cl~{\sc iii}]) and $n_{\mathrm{e}}$([Ar~{\sc iv}]) is around 0.12~dex. The discrepancy could be due to the measurement 
uncertainties, or it may indicate that power-law hydrogen density structures are not favoured, according to the photoionization models in \citet{juan2021atomic}.  

Figure~\ref{fig:Te_comp} presents a comparison between electron temperatures obtained from different diagnostic lines, namely the [N~{\sc ii}], [O~{\sc ii}] and [S~{\sc ii}], and those derived from the [O~{\sc iii}] lines, which are usually reliably measured and not largely impacted by the recombination contributions. In our results, $T_{\mathrm{e}}$([N~{\sc ii}]) is lower than $T_{\mathrm{e}}$([O~{\sc iii}]) for 54\% of the objects, with a median difference of 75~K between the two measures. The discrepancies are larger for higher values of  $T_{\mathrm{e}}$ ($>13,000$~K). $T_{\mathrm{e}}$([O~{\sc iii}]) and $T_{\mathrm{e}}$([N~{\sc ii}]) exhibit moderate correlation with a coefficient of $r = 0.59$. The dependence of the ratio $T_{\mathrm{e}}$([O~{\sc iii}])/$T_{\mathrm{e}}$([N~{\sc ii}]) on nebular excitation was examined in \citet{kaler1986electron}, where a correlation between $T_{\mathrm{e}}$([O~{\sc 
iii}])/$T_{\mathrm{e}}$([N~{\sc ii}]) 
and the intensity of He~{\sc ii}~$\lambda$4686 (which is only seen for high excitation PNe) was found. 
This correlation was later confirmed in KB94 and WL07 with Pearson's correlation coefficients of $r = 0.69$ and 0.70 
respectively. Similarly, a correlation between $T_{\mathrm{e}}$([O~{\sc iii}])/$T_{\mathrm{e}}$([N~{\sc ii}]) 
and the nebular excitation class (EC) defined in \citet{dopita1990evolutionary} was found in  \citet{danehkar2021physical} 
with $r = 0.69$ for a sample of PNe surrounding Wolf–Rayet and weak-emission-line nuclei. 

In our analysis, we exclude the object PNG~004.0-03.0, which was found to have a high temperature limit based on the 
[O~{\sc iii}] diagnostic lines, as observed in both our study and previous literature. We also excluded objects with 
weak line fluxes for [O~{\sc iii}]~$\lambda$4363, He~{\sc ii}~$\lambda$4686, and [N~{\sc ii}]~$\lambda$5755, where 
F($\lambda$) $<10^{-16} \mathrm{erg~cm^{-2}s^{-1}}$. Despite these exclusions, we found that the correlations 
between $T_{\mathrm{e}}$([O~{\sc iii}])/$T_{\mathrm{e}}$([N~{\sc ii}]) and the intensity of He~{\sc 
ii}~$\lambda$4686, as well as the excitation class (EC), were weak, with r-values of 0.54 and 0.31, respectively. We 
also evaluated the PN excitation class estimation scheme proposed by \citet{reid2010evaluation}, which is expected 
to provide a better estimate of the central star effective temperatures, and used a linear interpolation for 
continuous estimates of the EC. The resulting r-value was 0.13. We suspect that the weaker correlation may be due to 
the overestimation of the recombination corrections of [N~{\sc ii}]~$\lambda$5755 lines, which may have been 
contaminated by fluorescence, as suggested by previous studies \citep{escalante2005n,escalante2012excitation}. When 
considering $T_{\mathrm{e}}$([N~{\sc ii}]) without recombination [O~{\sc iii}]~$\lambda$4363 and [N~{\sc 
ii}]~$\lambda$5755 emission contribution, the correlation coefficients became slightly stronger, with values of 
0.69, 0.47, and 0.26, respectively. However, we should note that the weaker correction with EC in our data may be 
due to a larger range of $T_{\mathrm{e}}$([N~{\sc ii}]) in our study compared to the literature.

\subsection{Elemental abundances determined for our sample PNe}
The total elemental abundances of He, N, O, Ne, S, Ar, and Cl (relative to H) for each PN, along with their 
estimated uncertainties, are provided in Table.\ref{tab:abun_table}. For all element except Ne, over 96\% of the computed 
chemical abundances display uncertainties below 0.2 dex, and the median uncertainty is less than 0.05 dex. For Ne, 84\% of the calculated abundances have uncertainties below 0.2 dex. PNe with 
an asterisk indicate those for which an abundance has been determined for the first time.

\setlength{\tabcolsep}{6.8pt}
\begin{table*}
    \centering
    \caption{Our measured PNe chemical abundances. Column~1 gives the PNG identification of the PN where an asterisk 
    in front indicates the PN has an abundance determined for the first time; Column~2 is the 
    logarithmic interstellar extinction `c' at H$\beta$ derived from our spectra; Columns~3-9 give the determined 
    abundances from the He/H, N/H, O/H, Ne/H, S/H, Ar/H, Cl/H ratios on a scale such that $\log$(H)$=12$. {The ionic abundances for the sample objects are available online.}}
    \label{tab:abun_table}
\begin{tabular}{lllllllll}
\hline
$\phantom{^{\star}}$PN G &  c(H$\beta$) & He/H &  N/H   &  O/H & Ne/H & S/H &  Ar/H & Cl/H\\
\hline
${^{\star}}$000.1+02.6 &              2.14$\pm$0.08 &              11.14$\pm$0.01 &  8.34$_{-0.07}^{+0.06}$ &  8.98$_{-0.06}^{+0.05}$ &  8.46$_{-0.07}^{+0.06}$ &  7.18$_{-0.11}^{+0.10}$ &              6.76$\pm$0.05 &  4.84$_{-0.13}^{+0.12}$ \\
 $\phantom{^{\star}}$000.1+04.3 &              3.00$\pm$0.07 &              10.96$\pm$0.01 &  8.01$_{-0.04}^{+0.03}$ &              8.55$\pm$0.03 &              7.99$\pm$0.03 &  6.73$_{-0.06}^{+0.05}$ &  6.29$_{-0.04}^{+0.03}$ &              5.01$\pm$0.04 \\
 $\phantom{^{\star}}$000.1-02.3 &              1.54$\pm$0.06 &              11.07$\pm$0.01 &  8.28$_{-0.04}^{+0.03}$ &              8.90$\pm$0.04 &              8.21$\pm$0.04 &              7.34$\pm$0.04 &              6.65$\pm$0.03 &  5.21$_{-0.04}^{+0.03}$ \\
 $\phantom{^{\star}}$000.2-01.9 &              1.58$\pm$0.08 &              10.93$\pm$0.02 &  8.12$_{-0.04}^{+0.03}$ &  9.06$_{-0.06}^{+0.05}$ &              8.61$\pm$0.05 &  7.02$_{-0.07}^{+0.06}$ &  6.47$_{-0.06}^{+0.05}$ &  5.07$_{-0.07}^{+0.06}$ \\
 ${^{\star}}$000.2-04.6 &  1.44$_{-0.06}^{+0.07}$ &              11.19$\pm$0.01 &  8.58$_{-0.03}^{+0.02}$ &  9.01$_{-0.05}^{+0.04}$ &              8.23$\pm$0.04 &  7.05$_{-0.05}^{+0.04}$ &              6.61$\pm$0.04 &              5.14$\pm$0.03 \\
 ${^{\star}}$000.3+06.9 &              1.25$\pm$0.08 &              11.03$\pm$0.02 &                              - &              8.18$\pm$0.01 &              7.85$\pm$0.02 &              6.34$\pm$0.03 &              6.33$\pm$0.03 &                              - \\
 $\phantom{^{\star}}$000.3-04.6 &              1.48$\pm$0.09 &              11.22$\pm$0.02 &              8.58$\pm$0.03 &  8.89$_{-0.06}^{+0.05}$ &              8.13$\pm$0.04 &  6.99$_{-0.07}^{+0.06}$ &  6.55$_{-0.05}^{+0.04}$ &  5.26$_{-0.04}^{+0.03}$ \\
 $\phantom{^{\star}}$000.4-01.9 &              1.86$\pm$0.06 &  11.08$_{-0.02}^{+0.01}$ &              8.40$\pm$0.04 &              8.86$\pm$0.04 &  8.00$_{-0.05}^{+0.04}$ &  7.40$_{-0.07}^{+0.06}$ &              6.83$\pm$0.05 &              5.57$\pm$0.04 \\
 $\phantom{^{\star}}$000.4-02.9 &  1.03$_{-0.08}^{+0.07}$ &              11.23$\pm$0.02 &  7.68$_{-0.08}^{+0.07}$ &              8.74$\pm$0.04 &              8.16$\pm$0.03 &  6.26$_{-0.16}^{+0.10}$ &              6.49$\pm$0.04 &              4.57$\pm$0.05 \\
 $\phantom{^{\star}}$000.7+03.2 &              1.94$\pm$0.08 &              11.17$\pm$0.01 &  8.47$_{-0.05}^{+0.04}$ &  9.03$_{-0.08}^{+0.06}$ &              8.31$\pm$0.04 &  7.10$_{-0.06}^{+0.05}$ &  6.81$_{-0.04}^{+0.03}$ &              5.50$\pm$0.04 \\
 $\phantom{^{\star}}$000.7-02.7 &              1.35$\pm$0.08 &              10.97$\pm$0.01 &  7.36$_{-0.06}^{+0.05}$ &              8.43$\pm$0.04 &              7.74$\pm$0.03 &              6.04$\pm$0.06 &              5.78$\pm$0.03 &              4.47$\pm$0.04 \\
 ${^{\star}}$000.7-07.4 &              0.62$\pm$0.09 &              11.12$\pm$0.02 &  8.50$_{-0.06}^{+0.05}$ &              8.75$\pm$0.05 &              8.26$\pm$0.04 &  7.18$_{-0.09}^{+0.07}$ &              6.47$\pm$0.05 &              5.03$\pm$0.04 \\
 $\phantom{^{\star}}$000.9-02.0 &              2.43$\pm$0.17 &              11.06$\pm$0.03 &  7.17$_{-0.31}^{+0.18}$ &  8.64$_{-0.37}^{+0.07}$ &  8.14$_{-0.28}^{+0.05}$ &  6.66$_{-0.16}^{+0.13}$ &  6.30$_{-0.08}^{+0.06}$ &  5.01$_{-0.09}^{+0.07}$ \\
 $\phantom{^{\star}}$000.9-04.8 &              1.91$\pm$0.07 &              11.01$\pm$0.01 &  8.37$_{-0.07}^{+0.06}$ &  9.03$_{-0.05}^{+0.04}$ &              8.52$\pm$0.05 &  7.42$_{-0.06}^{+0.05}$ &  6.70$_{-0.04}^{+0.03}$ &  5.13$_{-0.04}^{+0.03}$ \\
 ${^{\star}}$001.1-01.6 &              1.98$\pm$0.07 &              11.13$\pm$0.02 &              8.71$\pm$0.03 &  8.99$_{-0.05}^{+0.04}$ &              8.39$\pm$0.03 &  7.11$_{-0.06}^{+0.05}$ &              6.65$\pm$0.04 &  4.89$_{-0.08}^{+0.07}$ \\
 $\phantom{^{\star}}$001.2+02.1 &              2.56$\pm$0.07 &              11.00$\pm$0.02 &  7.99$_{-0.06}^{+0.05}$ &              8.67$\pm$0.04 &              8.16$\pm$0.03 &              7.00$\pm$0.05 &              6.36$\pm$0.04 &  5.05$_{-0.05}^{+0.04}$ \\
 ${^{\star}}$001.2-03.0 &              1.48$\pm$0.08 &  10.04$_{-0.05}^{+0.03}$ &  8.39$_{-0.06}^{+0.05}$ &  8.58$_{-0.10}^{+0.08}$ &                              - &  6.91$_{-0.06}^{+0.05}$ &              5.94$\pm$0.06 &                              - \\
 ${^{\star}}$001.3-01.2 &              2.89$\pm$0.07 &              11.09$\pm$0.01 &  8.42$_{-0.06}^{+0.05}$ &  8.77$_{-0.09}^{+0.07}$ &                              - &  6.99$_{-0.13}^{+0.11}$ &  6.61$_{-0.06}^{+0.05}$ &  5.56$_{-0.07}^{+0.06}$ \\
 $\phantom{^{\star}}$001.4+05.3 &              1.73$\pm$0.07 &              11.03$\pm$0.02 &  7.57$_{-0.04}^{+0.03}$ &              8.70$\pm$0.05 &  7.82$_{-0.06}^{+0.05}$ &  6.71$_{-0.07}^{+0.06}$ &              6.24$\pm$0.05 &              5.02$\pm$0.05 \\
 $\phantom{^{\star}}$001.6-01.3 &              3.38$\pm$0.11 &              11.09$\pm$0.02 &  8.58$_{-0.07}^{+0.06}$ &  8.81$_{-0.06}^{+0.04}$ &  8.18$_{-0.08}^{+0.06}$ &  6.95$_{-0.09}^{+0.07}$ &              6.47$\pm$0.05 &  5.43$_{-0.13}^{+0.11}$ \\
 $\phantom{^{\star}}$001.7+05.7 &              1.76$\pm$0.06 &  10.86$_{-0.02}^{+0.01}$ &              7.43$\pm$0.06 &              8.19$\pm$0.04 &              7.42$\pm$0.03 &              6.11$\pm$0.05 &              5.71$\pm$0.03 &              4.33$\pm$0.04 \\
 $\phantom{^{\star}}$001.7-04.4 &              0.65$\pm$0.07 &  10.62$_{-0.03}^{+0.02}$ &              8.49$\pm$0.05 &  9.07$_{-0.09}^{+0.08}$ &                              - &  6.95$_{-0.05}^{+0.04}$ &  6.60$_{-0.07}^{+0.06}$ &              5.34$\pm$0.05 \\
 $\phantom{^{\star}}$002.0-06.2 &              0.43$\pm$0.03 &              11.00$\pm$0.02 &              7.95$\pm$0.05 &              8.72$\pm$0.04 &              8.06$\pm$0.03 &              6.93$\pm$0.05 &              6.42$\pm$0.07 &              4.94$\pm$0.04 \\
 $\phantom{^{\star}}$002.1-02.2 &              1.56$\pm$0.08 &              10.96$\pm$0.02 &              7.79$\pm$0.04 &              8.56$\pm$0.03 &              7.92$\pm$0.02 &              6.68$\pm$0.04 &              6.10$\pm$0.04 &              4.93$\pm$0.04 \\
 $\phantom{^{\star}}$002.1-04.2 &              1.48$\pm$0.07 &              10.91$\pm$0.02 &  7.42$_{-0.05}^{+0.03}$ &  8.15$_{-0.05}^{+0.03}$ &  7.55$_{-0.04}^{+0.02}$ &  6.50$_{-0.07}^{+0.05}$ &  5.96$_{-0.05}^{+0.04}$ &  4.89$_{-0.15}^{+0.06}$ \\
 $\phantom{^{\star}}$002.2-09.4 &  0.36$_{-0.04}^{+0.05}$ &              11.10$\pm$0.01 &  8.71$_{-0.07}^{+0.06}$ &  8.75$_{-0.04}^{+0.03}$ &              8.35$\pm$0.03 &  7.09$_{-0.11}^{+0.10}$ &  6.66$_{-0.04}^{+0.03}$ &              5.36$\pm$0.04 \\
 ${^{\star}}$002.3+02.2 &              2.69$\pm$0.07 &              11.20$\pm$0.02 &  8.34$_{-0.06}^{+0.05}$ &  8.61$_{-0.06}^{+0.05}$ &  8.26$_{-0.12}^{+0.10}$ &  6.82$_{-0.06}^{+0.05}$ &              6.34$\pm$0.04 &  4.51$_{-0.06}^{+0.05}$ \\
 $\phantom{^{\star}}$002.5-01.7 &              2.16$\pm$0.05 &              11.14$\pm$0.01 &              8.83$\pm$0.04 &              8.80$\pm$0.03 &  8.48$_{-0.08}^{+0.07}$ &              7.30$\pm$0.05 &  6.73$_{-0.04}^{+0.03}$ &  5.16$_{-0.13}^{+0.12}$ \\
 ${^{\star}}$002.6+02.1 &              2.53$\pm$0.04 &              11.11$\pm$0.01 &              9.15$\pm$0.03 &  9.14$_{-0.02}^{+0.01}$ &              8.68$\pm$0.02 &  7.63$_{-0.06}^{+0.05}$ &  7.06$_{-0.02}^{+0.01}$ &              5.27$\pm$0.04 \\
 $\phantom{^{\star}}$002.7-04.8 &  1.05$_{-0.07}^{+0.06}$ &              11.24$\pm$0.01 &  8.32$_{-0.04}^{+0.03}$ &  8.53$_{-0.04}^{+0.03}$ &              8.09$\pm$0.04 &              6.87$\pm$0.04 &  6.39$_{-0.04}^{+0.03}$ &  5.06$_{-0.04}^{+0.03}$ \\
 $\phantom{^{\star}}$002.8+01.7 &              2.46$\pm$0.05 &              10.92$\pm$0.01 &  8.24$_{-0.04}^{+0.03}$ &  8.64$_{-0.05}^{+0.04}$ &                              - &  6.86$_{-0.05}^{+0.04}$ &  6.47$_{-0.05}^{+0.04}$ &  5.29$_{-0.04}^{+0.03}$ \\
 ${^{\star}}$002.8+01.8 &              2.47$\pm$0.07 &              11.05$\pm$0.02 &              8.52$\pm$0.04 &              8.81$\pm$0.04 &  8.27$_{-0.08}^{+0.06}$ &              7.25$\pm$0.04 &              6.52$\pm$0.04 &              4.00$\pm$0.06 \\
 $\phantom{^{\star}}$002.9-03.9 &              1.13$\pm$0.05 &              11.03$\pm$0.01 &  7.94$_{-0.16}^{+0.09}$ &              8.43$\pm$0.03 &              7.76$\pm$0.03 &  6.89$_{-0.43}^{+0.15}$ &              6.13$\pm$0.03 &              4.96$\pm$0.04 \\
 $\phantom{^{\star}}$003.2-06.2 &  0.62$_{-0.08}^{+0.09}$ &              11.13$\pm$0.02 &  8.45$_{-0.06}^{+0.05}$ &              8.70$\pm$0.04 &              8.20$\pm$0.04 &  7.00$_{-0.14}^{+0.12}$ &  6.55$_{-0.06}^{+0.05}$ &  5.30$_{-0.05}^{+0.04}$ \\
 $\phantom{^{\star}}$003.6-02.3 &  1.41$_{-0.06}^{+0.05}$ &              11.18$\pm$0.01 &              8.47$\pm$0.04 &  8.68$_{-0.04}^{+0.03}$ &              8.11$\pm$0.05 &  6.76$_{-0.08}^{+0.07}$ &  6.49$_{-0.05}^{+0.04}$ &              4.97$\pm$0.05 \\
 ${^{\star}}$003.7+07.9 &              2.56$\pm$0.03 &              10.91$\pm$0.01 &              9.43$\pm$0.03 &              9.00$\pm$0.01 &              8.86$\pm$0.04 &              7.49$\pm$0.03 &              6.60$\pm$0.02 &                              - \\
 $\phantom{^{\star}}$003.7-04.6 &              0.81$\pm$0.10 &              10.94$\pm$0.02 &  8.42$_{-0.11}^{+0.10}$ &  8.66$_{-0.04}^{+0.03}$ &              8.05$\pm$0.02 &  7.07$_{-0.08}^{+0.07}$ &              6.41$\pm$0.04 &              5.67$\pm$0.07 \\
 $\phantom{^{\star}}$003.8-04.3 &              0.74$\pm$0.04 &              11.12$\pm$0.01 &  8.37$_{-0.06}^{+0.05}$ &  8.76$_{-0.05}^{+0.04}$ &  8.06$_{-0.04}^{+0.03}$ &  6.96$_{-0.06}^{+0.05}$ &              6.66$\pm$0.03 &  5.31$_{-0.04}^{+0.03}$ \\
 $\phantom{^{\star}}$003.9-02.3 &              1.86$\pm$0.05 &              11.16$\pm$0.01 &  8.42$_{-0.03}^{+0.02}$ &              8.71$\pm$0.03 &              8.24$\pm$0.03 &  6.85$_{-0.06}^{+0.05}$ &              6.59$\pm$0.04 &              5.24$\pm$0.03 \\
 $\phantom{^{\star}}$003.9-03.1 &              1.32$\pm$0.05 &              10.99$\pm$0.01 &  7.99$_{-0.06}^{+0.05}$ &              8.72$\pm$0.03 &              7.91$\pm$0.05 &              6.66$\pm$0.04 &              6.22$\pm$0.03 &              4.75$\pm$0.04 \\
 $\phantom{^{\star}}$004.0-03.0 &              1.12$\pm$0.05 &              10.66$\pm$0.01 &              6.58$\pm$0.02 &              6.95$\pm$0.02 &              6.57$\pm$0.02 &              4.93$\pm$0.02 &  5.00$_{-0.04}^{+0.03}$ &  3.52$_{-0.07}^{+0.06}$ \\
 $\phantom{^{\star}}$004.1-03.8 &  1.54$_{-0.09}^{+0.08}$ &              11.02$\pm$0.02 &              7.88$\pm$0.04 &  8.41$_{-0.04}^{+0.03}$ &              8.01$\pm$0.03 &  6.63$_{-0.06}^{+0.05}$ &  6.05$_{-0.05}^{+0.04}$ &  4.83$_{-0.04}^{+0.03}$ \\
 $\phantom{^{\star}}$004.2-03.2 &              1.55$\pm$0.13 &  11.02$_{-0.03}^{+0.02}$ &  7.27$_{-0.16}^{+0.11}$ &  8.43$_{-0.05}^{+0.04}$ &              7.78$\pm$0.03 &  6.43$_{-0.13}^{+0.09}$ &              5.94$\pm$0.05 &  4.75$_{-0.08}^{+0.07}$ \\
 $\phantom{^{\star}}$004.2-04.3 &  1.10$_{-0.13}^{+0.14}$ &              11.01$\pm$0.03 &  8.07$_{-0.06}^{+0.05}$ &              8.60$\pm$0.04 &              8.02$\pm$0.03 &  7.31$_{-0.15}^{+0.11}$ &  6.11$_{-0.07}^{+0.06}$ &  4.86$_{-0.07}^{+0.06}$ \\
 ${^{\star}}$004.6+06.0 &              0.66$\pm$0.05 &              11.08$\pm$0.01 &  8.13$_{-0.06}^{+0.05}$ &  8.78$_{-0.06}^{+0.05}$ &  8.25$_{-0.06}^{+0.05}$ &  6.95$_{-0.10}^{+0.08}$ &              6.84$\pm$0.05 &              5.26$\pm$0.06 \\
 $\phantom{^{\star}}$004.8+02.0 &              2.37$\pm$0.06 &              10.88$\pm$0.01 &  7.45$_{-0.05}^{+0.04}$ &  8.33$_{-0.07}^{+0.06}$ &                              - &  6.50$_{-0.09}^{+0.08}$ &  6.00$_{-0.06}^{+0.05}$ &  5.76$_{-0.13}^{+0.10}$ \\
 $\phantom{^{\star}}$004.8-05.0 &              0.78$\pm$0.05 &              11.15$\pm$0.01 &  8.03$_{-0.08}^{+0.07}$ &  8.65$_{-0.06}^{+0.05}$ &  8.29$_{-0.06}^{+0.05}$ &  6.87$_{-0.08}^{+0.07}$ &              6.50$\pm$0.04 &  5.05$_{-0.04}^{+0.03}$ \\
 ${^{\star}}$005.0-03.9 &              1.22$\pm$0.07 &  11.16$_{-0.09}^{+0.03}$ &  7.20$_{-0.06}^{+0.05}$ &  7.41$_{-0.05}^{+0.04}$ &  6.98$_{-0.05}^{+0.04}$ &              5.84$\pm$0.06 &              5.54$\pm$0.05 &  3.43$_{-0.13}^{+0.12}$ \\
 ${^{\star}}$005.2+05.6 &              1.01$\pm$0.04 &              11.14$\pm$0.01 &              8.18$\pm$0.05 &  8.61$_{-0.05}^{+0.04}$ &  8.14$_{-0.04}^{+0.03}$ &              6.83$\pm$0.05 &  6.52$_{-0.04}^{+0.03}$ &  5.19$_{-0.05}^{+0.08}$ \\
 ${^{\star}}$005.5+06.1 &              1.35$\pm$0.07 &              11.03$\pm$0.02 &  8.21$_{-0.06}^{+0.05}$ &  8.70$_{-0.11}^{+0.09}$ &  8.00$_{-0.10}^{+0.08}$ &  6.63$_{-0.05}^{+0.04}$ &  6.19$_{-0.07}^{+0.06}$ &              5.29$\pm$0.06 \\
 ${^{\star}}$005.5-04.0 &              1.13$\pm$0.05 &              11.15$\pm$0.01 &              7.65$\pm$0.03 &  8.92$_{-0.05}^{+0.04}$ &                              - &              7.21$\pm$0.05 &              6.77$\pm$0.03 &  5.65$_{-0.06}^{+0.05}$ \\
 $\phantom{^{\star}}$005.8-06.1 &              0.53$\pm$0.05 &              11.12$\pm$0.01 &              8.71$\pm$0.05 &              9.02$\pm$0.04 &              8.38$\pm$0.04 &              7.30$\pm$0.06 &  6.84$_{-0.05}^{+0.04}$ &  5.48$_{-0.04}^{+0.03}$ \\
 $\phantom{^{\star}}$006.1+08.3 &              1.26$\pm$0.05 &              11.01$\pm$0.02 &              7.60$\pm$0.03 &              8.53$\pm$0.03 &              7.92$\pm$0.03 &  6.39$_{-0.07}^{+0.06}$ &              5.98$\pm$0.04 &  4.76$_{-0.04}^{+0.03}$ \\
 $\phantom{^{\star}}$006.4+02.0 &  1.92$_{-0.10}^{+0.09}$ &              11.12$\pm$0.02 &              8.54$\pm$0.03 &              8.95$\pm$0.03 &  8.40$_{-0.03}^{+0.02}$ &  7.27$_{-0.05}^{+0.04}$ &  6.80$_{-0.05}^{+0.04}$ &              5.37$\pm$0.03 \\
 $\phantom{^{\star}}$006.4-04.6 &              1.01$\pm$0.12 &              11.12$\pm$0.01 &  7.89$_{-0.09}^{+0.07}$ &              8.62$\pm$0.04 &                              - &  6.86$_{-0.07}^{+0.06}$ &              6.56$\pm$0.05 &  5.38$_{-0.05}^{+0.06}$ \\
 $\phantom{^{\star}}$006.8+02.3 &              2.14$\pm$0.08 &              11.03$\pm$0.01 &  7.87$_{-0.05}^{+0.04}$ &  8.54$_{-0.05}^{+0.04}$ &              7.79$\pm$0.03 &              6.46$\pm$0.05 &              5.92$\pm$0.04 &              4.85$\pm$0.04 \\
 $\phantom{^{\star}}$006.8-03.4 &              1.41$\pm$0.02 &              10.90$\pm$0.01 &              7.35$\pm$0.02 &              8.29$\pm$0.02 &              7.65$\pm$0.01 &  6.13$_{-0.06}^{+0.05}$ &              5.78$\pm$0.01 &              4.30$\pm$0.01 \vspace{0.07cm} \\
 \hline
\end{tabular}
\end{table*}
\begin{table*}
\centering
\contcaption{}
\begin{tabular}{lllllllll}
\hline
$\phantom{^{\star}}$PN G &  c(H$\beta$) & He/H &  N/H   &  O/H & Ne/H & S/H &  Ar/H & Cl/H\\
\hline
 $\phantom{^{\star}}$007.0+06.3 &              1.61$\pm$0.10 &              11.03$\pm$0.02 &  8.16$_{-0.07}^{+0.06}$ &  8.73$_{-0.05}^{+0.04}$ &              8.18$\pm$0.03 &  7.09$_{-0.07}^{+0.06}$ &  6.61$_{-0.06}^{+0.05}$ &  5.37$_{-0.05}^{+0.06}$ \\
 $\phantom{^{\star}}$007.0-06.8 &              0.72$\pm$0.08 &              11.13$\pm$0.02 &  8.28$_{-0.07}^{+0.06}$ &  8.78$_{-0.05}^{+0.04}$ &  8.23$_{-0.05}^{+0.04}$ &  7.02$_{-0.10}^{+0.08}$ &  6.58$_{-0.06}^{+0.05}$ &  5.36$_{-0.06}^{+0.05}$ \\
 ${^{\star}}$007.5+07.4 &  0.71$_{-0.07}^{+0.06}$ &              11.16$\pm$0.01 &              8.28$\pm$0.03 &  8.81$_{-0.05}^{+0.04}$ &              8.07$\pm$0.04 &              6.82$\pm$0.04 &  6.43$_{-0.04}^{+0.03}$ &  5.10$_{-0.04}^{+0.03}$ \\
 ${^{\star}}$007.6+06.9 &              1.03$\pm$0.04 &              11.14$\pm$0.01 &              7.92$\pm$0.03 &              8.49$\pm$0.03 &              7.94$\pm$0.02 &  6.53$_{-0.06}^{+0.05}$ &              6.40$\pm$0.02 &              5.11$\pm$0.03 \\
 $\phantom{^{\star}}$007.8-03.7 &  1.17$_{-0.07}^{+0.06}$ &              11.14$\pm$0.01 &              8.35$\pm$0.03 &              8.77$\pm$0.04 &              8.05$\pm$0.03 &              7.08$\pm$0.10 &              6.43$\pm$0.03 &  5.52$_{-0.06}^{+0.05}$ \\
 $\phantom{^{\star}}$007.8-04.4 &              1.02$\pm$0.04 &  10.18$_{-0.06}^{+0.02}$ &  8.34$_{-0.06}^{+0.05}$ &  8.48$_{-0.12}^{+0.09}$ &                              - &  6.66$_{-0.08}^{+0.06}$ &              5.41$\pm$0.12 &                              - \\
 $\phantom{^{\star}}$008.2+06.8 &              0.82$\pm$0.05 &               9.91$\pm$0.02 &  7.07$_{-0.05}^{+0.04}$ &  7.67$_{-0.06}^{+0.05}$ &                              - &  5.53$_{-0.06}^{+0.04}$ &              5.08$\pm$0.04 &              3.86$\pm$0.04 \\
 $\phantom{^{\star}}$008.4-03.6 &              1.49$\pm$0.06 &  10.99$_{-0.06}^{+0.02}$ &  8.33$_{-0.05}^{+0.04}$ &  9.01$_{-0.11}^{+0.09}$ &  8.65$_{-0.17}^{+0.13}$ &  6.90$_{-0.05}^{+0.04}$ &  6.50$_{-0.07}^{+0.06}$ &  4.90$_{-0.09}^{+0.08}$ \\
 $\phantom{^{\star}}$008.6-02.6 &              2.31$\pm$0.06 &              10.96$\pm$0.01 &                              - &              8.33$\pm$0.04 &              7.62$\pm$0.03 &  6.19$_{-0.06}^{+0.05}$ &              5.77$\pm$0.03 &  4.58$_{-0.04}^{+0.03}$ \\
 $\phantom{^{\star}}$009.4-09.8 &              0.85$\pm$0.08 &              11.09$\pm$0.02 &  8.06$_{-0.08}^{+0.06}$ &              8.64$\pm$0.04 &              8.14$\pm$0.03 &  6.83$_{-0.06}^{+0.05}$ &              6.41$\pm$0.04 &              5.15$\pm$0.08 \\
 $\phantom{^{\star}}$009.8-04.6 &              0.83$\pm$0.05 &              11.06$\pm$0.01 &  8.39$_{-0.07}^{+0.06}$ &              8.87$\pm$0.04 &  8.32$_{-0.05}^{+0.04}$ &  7.25$_{-0.07}^{+0.06}$ &              6.68$\pm$0.04 &              5.45$\pm$0.05 \\
 $\phantom{^{\star}}$351.1+04.8 &              1.15$\pm$0.06 &              11.05$\pm$0.02 &  8.14$_{-0.04}^{+0.03}$ &              8.85$\pm$0.04 &              8.22$\pm$0.03 &  7.05$_{-0.07}^{+0.06}$ &  6.62$_{-0.05}^{+0.04}$ &  5.32$_{-0.04}^{+0.03}$ \\
 $\phantom{^{\star}}$351.2+05.2 &              1.00$\pm$0.05 &              10.97$\pm$0.01 &              8.51$\pm$0.03 &  8.78$_{-0.06}^{+0.05}$ &  7.84$_{-0.06}^{+0.05}$ &              6.58$\pm$0.04 &              6.19$\pm$0.05 &              5.03$\pm$0.04 \\
 $\phantom{^{\star}}$351.6-06.2 &              0.70$\pm$0.05 &              11.15$\pm$0.01 &              8.68$\pm$0.03 &  8.88$_{-0.04}^{+0.03}$ &              8.17$\pm$0.03 &  7.24$_{-0.04}^{+0.03}$ &  6.83$_{-0.03}^{+0.02}$ &              5.55$\pm$0.03 \\
 ${^{\star}}$351.9+09.0 &              0.51$\pm$0.05 &              11.02$\pm$0.01 &  7.29$_{-0.14}^{+0.10}$ &              8.56$\pm$0.03 &              8.03$\pm$0.03 &  6.49$_{-0.06}^{+0.05}$ &              6.11$\pm$0.02 &              4.90$\pm$0.04 \\
 $\phantom{^{\star}}$351.9-01.9 &              2.93$\pm$0.05 &              11.04$\pm$0.01 &              8.24$\pm$0.02 &              8.66$\pm$0.02 &              8.20$\pm$0.01 &              6.92$\pm$0.04 &              6.35$\pm$0.03 &  5.36$_{-0.03}^{+0.16}$ \\
 $\phantom{^{\star}}$352.0-04.6 &              1.53$\pm$0.07 &              11.14$\pm$0.02 &  8.88$_{-0.04}^{+0.03}$ &              9.02$\pm$0.04 &              8.42$\pm$0.04 &  7.30$_{-0.06}^{+0.05}$ &  6.92$_{-0.04}^{+0.03}$ &  5.43$_{-0.04}^{+0.03}$ \\
 $\phantom{^{\star}}$352.1+05.1 &  1.49$_{-0.01}^{+0.02}$ &              11.10$\pm$0.01 &              8.56$\pm$0.02 &              8.76$\pm$0.01 &              8.32$\pm$0.01 &  7.01$_{-0.05}^{+0.04}$ &              6.62$\pm$0.02 &              5.37$\pm$0.01 \\
 $\phantom{^{\star}}$352.6+03.0 &              2.58$\pm$0.06 &              11.18$\pm$0.01 &              8.66$\pm$0.02 &              8.84$\pm$0.02 &              8.49$\pm$0.01 &              7.08$\pm$0.08 &  6.74$_{-0.04}^{+0.03}$ &              5.50$\pm$0.02 \\
 ${^{\star}}$353.2-05.2 &              0.80$\pm$0.06 &              11.24$\pm$0.01 &  8.50$_{-0.04}^{+0.03}$ &              8.69$\pm$0.03 &              8.09$\pm$0.05 &  6.64$_{-0.06}^{+0.05}$ &  6.41$_{-0.04}^{+0.03}$ &              4.47$\pm$0.05 \\
 $\phantom{^{\star}}$353.3+06.3 &              0.91$\pm$0.05 &              10.96$\pm$0.02 &  7.72$_{-0.09}^{+0.08}$ &  8.46$_{-0.05}^{+0.04}$ &  7.94$_{-0.08}^{+0.05}$ &  6.56$_{-0.08}^{+0.07}$ &  6.00$_{-0.05}^{+0.04}$ &  4.97$_{-0.06}^{+0.05}$ \\
 $\phantom{^{\star}}$353.7+06.3 &              0.81$\pm$0.07 &              11.05$\pm$0.02 &              8.10$\pm$0.05 &              8.81$\pm$0.04 &              8.02$\pm$0.10 &  6.88$_{-0.06}^{+0.05}$ &              6.38$\pm$0.04 &              5.24$\pm$0.03 \\
 $\phantom{^{\star}}$354.5+03.3 &              3.48$\pm$0.07 &              10.96$\pm$0.02 &              8.56$\pm$0.03 &              8.70$\pm$0.02 &              8.22$\pm$0.02 &              6.87$\pm$0.04 &              6.48$\pm$0.04 &  5.08$_{-0.04}^{+0.03}$ \\
 $\phantom{^{\star}}$354.9+03.5 &              2.33$\pm$0.07 &              11.08$\pm$0.01 &              8.23$\pm$0.05 &  8.62$_{-0.07}^{+0.06}$ &  8.35$_{-0.08}^{+0.07}$ &  6.90$_{-0.09}^{+0.08}$ &  6.41$_{-0.06}^{+0.05}$ &  5.15$_{-0.05}^{+0.04}$ \\
 $\phantom{^{\star}}$355.4-02.4 &              1.98$\pm$0.06 &              11.13$\pm$0.01 &              8.79$\pm$0.03 &              8.81$\pm$0.03 &              8.30$\pm$0.03 &              7.12$\pm$0.06 &  6.74$_{-0.06}^{+0.05}$ &              5.45$\pm$0.03 \\
 $\phantom{^{\star}}$355.9+03.6 &              1.47$\pm$0.07 &              10.71$\pm$0.02 &  7.13$_{-0.14}^{+0.05}$ &  8.04$_{-0.20}^{+0.07}$ &  7.60$_{-0.20}^{+0.06}$ &  6.53$_{-0.06}^{+0.05}$ &  5.68$_{-0.07}^{+0.05}$ &  4.46$_{-0.12}^{+0.07}$ \\
 $\phantom{^{\star}}$355.9-04.2 &              1.19$\pm$0.03 &              11.17$\pm$0.02 &              8.58$\pm$0.02 &  9.11$_{-0.15}^{+0.09}$ &  7.94$_{-0.04}^{+0.03}$ &              7.39$\pm$0.11 &              6.85$\pm$0.05 &  5.76$_{-0.04}^{+0.03}$ \\
 ${^{\star}}$356.1-03.3 &              1.20$\pm$0.05 &              11.14$\pm$0.01 &              8.55$\pm$0.02 &              8.90$\pm$0.04 &              8.20$\pm$0.04 &              7.00$\pm$0.04 &              6.57$\pm$0.04 &  5.50$_{-0.06}^{+0.05}$ \\
 $\phantom{^{\star}}$356.3-06.2 &              0.43$\pm$0.05 &              11.11$\pm$0.01 &  8.49$_{-0.08}^{+0.05}$ &  8.70$_{-0.07}^{+0.06}$ &  7.83$_{-0.04}^{+0.03}$ &  6.84$_{-0.06}^{+0.05}$ &  6.40$_{-0.04}^{+0.03}$ &              5.07$\pm$0.04 \\
 ${^{\star}}$356.5-03.6 &              2.46$\pm$0.10 &              11.13$\pm$0.02 &  8.47$_{-0.07}^{+0.06}$ &  8.78$_{-0.05}^{+0.04}$ &  7.78$_{-0.07}^{+0.06}$ &  7.05$_{-0.07}^{+0.06}$ &  6.63$_{-0.05}^{+0.04}$ &              5.44$\pm$0.04 \\
 ${^{\star}}$356.8+03.3 &              2.24$\pm$0.06 &              10.82$\pm$0.01 &              8.36$\pm$0.04 &  8.54$_{-0.07}^{+0.06}$ &                              - &  6.65$_{-0.11}^{+0.10}$ &  6.39$_{-0.06}^{+0.05}$ &              5.45$\pm$0.08 \\
 $\phantom{^{\star}}$356.8-05.4 &              0.84$\pm$0.06 &              11.12$\pm$0.01 &              8.14$\pm$0.03 &              8.86$\pm$0.04 &              7.87$\pm$0.04 &              6.84$\pm$0.04 &  6.49$_{-0.04}^{+0.03}$ &              5.03$\pm$0.03 \\
 $\phantom{^{\star}}$356.9+04.4 &              1.94$\pm$0.05 &              10.94$\pm$0.01 &  8.44$_{-0.06}^{+0.05}$ &  8.67$_{-0.06}^{+0.05}$ &  8.01$_{-0.04}^{+0.03}$ &  6.99$_{-0.07}^{+0.06}$ &              6.55$\pm$0.04 &              5.07$\pm$0.04 \\
 $\phantom{^{\star}}$357.0+02.4 &              2.32$\pm$0.06 &              11.12$\pm$0.01 &  8.55$_{-0.04}^{+0.03}$ &              8.88$\pm$0.03 &              8.23$\pm$0.04 &              7.23$\pm$0.06 &  6.78$_{-0.04}^{+0.03}$ &  5.55$_{-0.05}^{+0.04}$ \\
 $\phantom{^{\star}}$357.1+03.6 &              0.99$\pm$0.05 &              11.00$\pm$0.01 &  8.06$_{-0.05}^{+0.04}$ &              8.73$\pm$0.04 &              8.40$\pm$0.04 &  7.17$_{-0.07}^{+0.06}$ &              6.58$\pm$0.04 &  5.03$_{-0.06}^{+0.05}$ \\
 ${^{\star}}$357.1+04.4 &  1.81$_{-0.05}^{+0.06}$ &              11.16$\pm$0.01 &  7.36$_{-0.09}^{+0.07}$ &  9.03$_{-0.08}^{+0.07}$ &  8.51$_{-0.11}^{+0.08}$ &  6.69$_{-0.15}^{+0.10}$ &  6.54$_{-0.07}^{+0.06}$ &  4.90$_{-0.07}^{+0.06}$ \\
 $\phantom{^{\star}}$357.1-04.7 &  1.02$_{-0.05}^{+0.04}$ &              10.21$\pm$0.02 &  8.50$_{-0.08}^{+0.06}$ &  8.50$_{-0.17}^{+0.11}$ &                              - &  6.86$_{-0.08}^{+0.07}$ &                              - &  5.64$_{-0.10}^{+0.07}$ \\
 ${^{\star}}$357.2+02.0 &              2.34$\pm$0.07 &              10.98$\pm$0.01 &              7.87$\pm$0.04 &              8.83$\pm$0.04 &              8.16$\pm$0.03 &              6.65$\pm$0.05 &              6.30$\pm$0.03 &              4.88$\pm$0.04 \\
 ${^{\star}}$357.3+04.0 &              1.65$\pm$0.05 &              11.05$\pm$0.01 &  8.18$_{-0.07}^{+0.06}$ &              8.81$\pm$0.03 &              8.26$\pm$0.03 &  7.05$_{-0.07}^{+0.06}$ &              6.55$\pm$0.03 &  4.98$_{-0.04}^{+0.03}$ \\
 ${^{\star}}$357.5+03.1 &              1.98$\pm$0.07 &                               - &  6.98$_{-0.08}^{+0.07}$ &  7.09$_{-0.20}^{+0.13}$ &                              - &  5.94$_{-0.07}^{+0.06}$ &                              - &  4.89$_{-0.14}^{+0.11}$ \\
 $\phantom{^{\star}}$357.5+03.2 &              2.30$\pm$0.08 &              11.11$\pm$0.01 &              8.86$\pm$0.05 &              8.93$\pm$0.04 &  8.47$_{-0.06}^{+0.05}$ &  7.37$_{-0.06}^{+0.05}$ &  6.78$_{-0.05}^{+0.04}$ &  5.43$_{-0.08}^{+0.07}$ \\
 $\phantom{^{\star}}$357.6-03.3 &              1.59$\pm$0.06 &              11.16$\pm$0.01 &  7.95$_{-0.05}^{+0.04}$ &  8.53$_{-0.11}^{+0.09}$ &  8.43$_{-0.17}^{+0.12}$ &              6.51$\pm$0.04 &  6.09$_{-0.07}^{+0.06}$ &                              - \\
 ${^{\star}}$357.9-03.8 &              1.56$\pm$0.05 &              11.25$\pm$0.01 &  6.78$_{-0.10}^{+0.08}$ &              7.63$\pm$0.04 &              7.27$\pm$0.04 &  5.31$_{-0.07}^{+0.06}$ &              5.82$\pm$0.04 &  3.96$_{-0.05}^{+0.04}$ \\
 $\phantom{^{\star}}$357.9-05.1 &              1.32$\pm$0.07 &              11.07$\pm$0.01 &  8.61$_{-0.03}^{+0.02}$ &              8.78$\pm$0.04 &              8.23$\pm$0.03 &  6.97$_{-0.04}^{+0.03}$ &              6.47$\pm$0.03 &              5.16$\pm$0.03 \\
 ${^{\star}}$358.0+09.3 &  0.63$_{-0.11}^{+0.12}$ &              11.11$\pm$0.02 &  7.40$_{-0.10}^{+0.08}$ &  8.51$_{-0.05}^{+0.04}$ &              7.94$\pm$0.04 &  6.11$_{-0.07}^{+0.06}$ &  6.12$_{-0.05}^{+0.04}$ &  4.63$_{-0.08}^{+0.07}$ \\
 $\phantom{^{\star}}$358.2+03.5 &  2.69$_{-0.11}^{+0.10}$ &              10.96$\pm$0.02 &  7.91$_{-0.07}^{+0.06}$ &  8.47$_{-0.06}^{+0.05}$ &              7.61$\pm$0.04 &  6.56$_{-0.09}^{+0.07}$ &  5.97$_{-0.05}^{+0.04}$ &              4.85$\pm$0.05 \\
 $\phantom{^{\star}}$358.2+04.2 &              2.15$\pm$0.06 &              11.09$\pm$0.01 &  8.26$_{-0.04}^{+0.03}$ &              8.75$\pm$0.03 &              8.32$\pm$0.03 &  6.95$_{-0.10}^{+0.09}$ &  6.65$_{-0.04}^{+0.03}$ &              5.40$\pm$0.04 \\
 ${^{\star}}$358.5+02.9 &              2.09$\pm$0.06 &              10.95$\pm$0.02 &  7.34$_{-0.04}^{+0.03}$ &              8.37$\pm$0.04 &              7.78$\pm$0.03 &  6.31$_{-0.06}^{+0.05}$ &              5.85$\pm$0.03 &              4.57$\pm$0.04 \\
 $\phantom{^{\star}}$358.5-04.2 &              1.52$\pm$0.05 &              10.96$\pm$0.03 &              7.66$\pm$0.04 &  8.34$_{-0.04}^{+0.03}$ &  7.75$_{-0.04}^{+0.03}$ &  6.77$_{-0.06}^{+0.05}$ &  6.15$_{-0.04}^{+0.03}$ &  4.81$_{-0.04}^{+0.03}$ \\
 ${^{\star}}$358.6+07.8 &              0.98$\pm$0.05 &              11.04$\pm$0.01 &              8.07$\pm$0.04 &              8.72$\pm$0.03 &  8.13$_{-0.03}^{+0.02}$ &              7.09$\pm$0.05 &              6.36$\pm$0.03 &  5.42$_{-0.04}^{+0.03}$ \\
 $\phantom{^{\star}}$358.6-05.5 &              0.80$\pm$0.06 &              11.18$\pm$0.01 &  8.40$_{-0.04}^{+0.03}$ &  8.51$_{-0.03}^{+0.02}$ &              8.01$\pm$0.03 &  6.71$_{-0.07}^{+0.06}$ &              6.39$\pm$0.04 &              4.73$\pm$0.06 \\
 $\phantom{^{\star}}$358.7+05.2 &              2.57$\pm$0.10 &              10.26$\pm$0.02 &  8.32$_{-0.06}^{+0.05}$ &  8.51$_{-0.11}^{+0.09}$ &                              - &  6.90$_{-0.08}^{+0.06}$ &  6.28$_{-0.08}^{+0.07}$ &  5.96$_{-0.18}^{+0.15}$ \\
 $\phantom{^{\star}}$358.8+03.0 &              1.89$\pm$0.06 &              11.11$\pm$0.01 &              8.45$\pm$0.03 &              8.88$\pm$0.03 &              8.19$\pm$0.03 &              7.14$\pm$0.04 &              6.71$\pm$0.03 &  5.25$_{-0.04}^{+0.03}$ \\
 $\phantom{^{\star}}$358.9+03.4 &              2.54$\pm$0.07 &              11.11$\pm$0.01 &              8.83$\pm$0.02 &              8.91$\pm$0.02 &              8.48$\pm$0.02 &  7.57$_{-0.06}^{+0.05}$ &  6.97$_{-0.04}^{+0.03}$ &  5.77$_{-0.02}^{+0.03}$ \\
 $\phantom{^{\star}}$359.0-04.1 &              1.16$\pm$0.07 &              11.14$\pm$0.01 &              8.71$\pm$0.03 &              8.88$\pm$0.04 &  8.31$_{-0.05}^{+0.04}$ &              7.14$\pm$0.05 &              6.69$\pm$0.04 &  5.25$_{-0.04}^{+0.03}$ \\
 $\phantom{^{\star}}$359.1-02.9 &              1.31$\pm$0.06 &              11.13$\pm$0.01 &  8.73$_{-0.04}^{+0.03}$ &              8.85$\pm$0.04 &  8.24$_{-0.05}^{+0.04}$ &  6.85$_{-0.08}^{+0.06}$ &              6.54$\pm$0.04 &  5.26$_{-0.10}^{+0.09}$ \\
 ${^{\star}}$359.2+04.7 &              2.10$\pm$0.07 &              11.24$\pm$0.02 &  7.43$_{-0.09}^{+0.05}$ &  8.28$_{-0.12}^{+0.07}$ &                              - &  5.87$_{-0.17}^{+0.08}$ &  6.43$_{-0.08}^{+0.05}$ &  5.54$_{-0.07}^{+0.05}$ \\
 $\phantom{^{\star}}$359.3-01.8 &              3.28$\pm$0.05 &              10.20$\pm$0.02 &  8.30$_{-0.12}^{+0.09}$ &  8.54$_{-0.21}^{+0.13}$ &                              - &  6.95$_{-0.15}^{+0.09}$ &  5.99$_{-0.09}^{+0.08}$ &  5.64$_{-0.18}^{+0.11}$ \\
 $\phantom{^{\star}}$359.6-04.8 &              0.95$\pm$0.04 &              11.28$\pm$0.01 &              7.33$\pm$0.06 &              8.13$\pm$0.03 &              7.65$\pm$0.03 &  6.35$_{-0.07}^{+0.06}$ &              6.13$\pm$0.03 &  4.64$_{-0.05}^{+0.04}$  \vspace{0.07cm} \\
\hline
\end{tabular}
\end{table*}
\begin{table*}
\centering
\contcaption{}
\begin{tabular}{lllllllll}
\hline
$\phantom{^{\star}}$PN G &  c(H$\beta$) & He/H &  N/H   &  O/H & Ne/H & S/H &  Ar/H & Cl/H\\ \hline
 $\phantom{^{\star}}$359.7-01.8 &              2.47$\pm$0.13 &  10.97$\pm$0.02 &  7.47$_{-0.43}^{+0.17}$ &  8.59$_{-0.43}^{+0.09}$ &  7.48$_{-0.05}^{+0.04}$ &  6.40$_{-0.43}^{+0.13}$ &  6.06$_{-0.05}^{+0.04}$ &  4.84$_{-0.07}^{+0.06}$ \\
 ${^{\star}}$359.8+02.4 &              2.85$\pm$0.08 &  10.79$\pm$0.01 &  8.44$_{-0.05}^{+0.04}$ &  8.78$_{-0.07}^{+0.06}$ &                              - &  6.96$_{-0.08}^{+0.07}$ &  6.43$_{-0.07}^{+0.06}$ &  5.51$_{-0.06}^{+0.05}$ \\
 $\phantom{^{\star}}$359.8+03.7 &              2.32$\pm$0.09 &  10.98$\pm$0.02 &  7.63$_{-0.07}^{+0.06}$ &  8.53$_{-0.06}^{+0.05}$ &                              - &  6.63$_{-0.08}^{+0.07}$ &  6.16$_{-0.06}^{+0.05}$ &  5.08$_{-0.06}^{+0.05}$ \\
 ${^{\star}}$359.8+05.2 &              1.33$\pm$0.04 &  11.03$\pm$0.01 &  7.99$_{-0.05}^{+0.04}$ &  8.37$_{-0.12}^{+0.09}$ &                              - &              6.72$\pm$0.04 &  6.07$_{-0.06}^{+0.05}$ &  6.20$_{-0.13}^{+0.11}$ \\
 $\phantom{^{\star}}$359.8+05.6 &  0.72$_{-0.08}^{+0.07}$ &  10.11$\pm$0.03 &  8.26$_{-0.09}^{+0.07}$ &  8.66$_{-0.17}^{+0.11}$ &                              - &  7.03$_{-0.15}^{+0.11}$ &  6.45$_{-0.09}^{+0.07}$ &  5.86$_{-0.14}^{+0.11}$ \\
 $\phantom{^{\star}}$359.8+06.9 &  1.42$_{-0.07}^{+0.06}$ &  11.02$\pm$0.02 &              8.57$\pm$0.04 &  8.78$_{-0.03}^{+0.02}$ &              8.23$\pm$0.03 &  7.02$_{-0.05}^{+0.04}$ &              6.48$\pm$0.04 &              5.15$\pm$0.03 \\
 $\phantom{^{\star}}$359.8-07.2 &  0.67$_{-0.09}^{+0.10}$ &  10.95$\pm$0.02 &  7.48$_{-0.09}^{+0.07}$ &              8.44$\pm$0.03 &              7.83$\pm$0.02 &              6.27$\pm$0.05 &              5.88$\pm$0.03 &  5.05$_{-0.07}^{+0.06}$ \\
 $\phantom{^{\star}}$359.9-04.5 &  1.97$_{-0.10}^{+0.09}$ &  11.16$\pm$0.02 &              8.76$\pm$0.03 &              8.78$\pm$0.03 &              8.36$\pm$0.02 &  7.21$_{-0.09}^{+0.08}$ &  6.78$_{-0.05}^{+0.04}$ &  5.55$_{-0.04}^{+0.15}$     \vspace{0.07cm} \\
\hline
\end{tabular}
\end{table*}

We compared our data with results of PNe in the Galactic bulge given in \citet{chiappini2009abundances} 
CGS09, which combined the data sets from \citet{gorny2004populations} which is itself a merger of PNe abundance 
measures from \citet{cuisinier1999observations}, \citet{escudero2001abundances}, and \citet{escudero2004new}, 
\citet{gorny2009planetary} and \citet{wang2007elemental}. We limited our comparison to high-quality measurements 
with reported uncertainties of less than 0.3 dex (removing of 25\% of their data set), as done in their analysis. 
The median, 25th and 75th percentile values of He/H, N/H, O/H, Ar/H, Ne/H, S/H, Cl/H derived in this 
work and in CGS09 are presented in Table.\ref{tab:median_cgs09} together with the number of PNe with each 
measurement available. 

\begin{table*}
\caption{Statistics of chemical abundances for 
all the Bulge PNe measured in this work and in CGS09. Columns~2-5 provide information about the median, 
25th and 75th percentile values (in parentheses in columns~2-3), and the number of PNe measurements used 
for the calculation (in brackets in column~5). Columns~6-9 present results obtained from the combined 
literature data sets in CGS09, expressed in the same manner. Column~10 displays solar abundances taken from 
\citet{asplund2009chemical}. The abundances are given on a log scale where $\log$(H)$=12$}
\label{tab:median_cgs09}
    \begin{tabular}{ll@{\hskip 0.1in}l@{\hskip 0.15in}lcl@{\hskip 0.1in}l@{\hskip 0.15in}cc}
\hline & \multicolumn{3}{l}{This work}& & \multicolumn{3}{l}{CGS09} & Sun \\
\hline He/H & 11.06 & [10.97, 11.14] & (123) & & 11.11 & [11.05, 11.19] & (144) & 10.93$\pm$0.01 \\
N/H & 8.27 & [7.88, 8.49] & (122)& & 8.11 & [7.76, 8.50] & (123) & 7.83$\pm$0.05 \\
O/H & 8.70 & [8.53, 8.83] & (124)& & 8.57 & [8.40, 8.72] & (117) & 8.69$\pm$0.05 \\
Ne/H & 8.14 & [7.92, 8.29] & (105)& & 7.93 & [7.71, 8.17] & (77) & 7.93$\pm$0.10 \\
S/H & 6.90 & [6.63, 7.08] & (124)& & 6.79 & [6.54, 7.04] & (94) & 7.12$\pm$0.03 \\
Ar/H & 6.44 & [6.14, 6.61] & (122)& & 6.34 & [6.05, 6.56] & (120) & 6.40$\pm$0.13 \\
Cl/H & 5.14 & [4.89, 5.41] & (119)& & 6.22 & [6.00, 6.50] & (47) & 5.50$\pm$0.30 \\
\hline
\end{tabular}
\end{table*}


In comparison to the CGS09 findings, our results show higher median abundances for N and the alpha elements (O, Ne, 
S, and Ar), with differences ranging from 0.10-0.30~dex. Notably, the median O/H, Ne/H, Ar/H values in our 
data are slightly super-solar. However, the median abundance for S, other $\alpha$ elements, remains sub-solar due to the well-known 'sulphur anomaly' \citep{henry2004sulfur, henry2012curious}. Our results also show a slightly lower median abundance for He, with a tail towards lower values, by approximately 0.05 dex. In terms of chlorine, our 
median abundance value is lower than the literature value by 1.09~dex. This discrepancy is primarily due to the 
CGS09 study, where most of the Cl/H values are derived from GCS09. We discuss the reasons for this discrepancy in 
Section~\ref{sec:in_abun_comp}. The Cl/H results in CGS09 are mostly extremely super-solar, which is also contrary to 
the behaviour of other alpha elements in their results. In contrast, 90\% of the Cl/H values in our 
results are sub-solar, and the difference between the PN Cl abundances and solar abundance becomes smaller. Thus, we 
believe that our Cl/H estimation is more reliable.


In summary, our data offers a highly accurate and precise representation of the general abundance pattern of PNe in 
the Galactic bulge, thanks to the consistently reduced and analysed high s/n VLT spectra, the availability of 
emission lines for a broader range of ionic species, and the incorporation of updated atomic data and the ICF scheme 
introduced in DMS14. Our results reveal a systematic increase in median values for elemental abundances across the 
majority of elements. The median abundances of alpha elements, O, Ne, and Ar, are found to be super-solar in our 
assessment.

\section{Discussion}
\label{sec:discussion}
It is clear from this work that accurate PN physical conditions and abundance measures, even determined from 
excellent, high quality s/n spectra is an inherently difficult process as is the comparison with previous work. 
There are many reasons for this. Prime among these is the assumption that chemical abundances determined for any 
individual PN are representative of the entire nebula whereas we know that internal abundance variations do exist, 
albeit with small variations (e.g., \citet{1988A&A...191..128M,  2013MNRAS.434.1513D}). 
Nevertheless, for the comparisons 
made here we use a single, overall abundance estimate as measured from an integrated 1-D spectrum {\bf from our compact PNe that are all $\leq$10~arcseconds across. Hence decent central, representative fractions of all our PNe are always sampled by the spectrograph slit.} We also know 
that PNe are complex, resolved 3-D objects with internal condensations, density, temperature and ionisation 
variations etc, e.g. \citet{monteiro2004three, akras2016deciphering}. An important secondary factor, especially when 
comparing with other work, are issues such as the chosen telescope and spectrograph, the slit widths and 
positions in terms of orientation on the projected PN image and offsets from any CSPN and of course the observing 
conditions (seeing, transparency, airmass etc.) all of which vary between data sets. 

The future of PNe abundance determinations is with deep IFU work such as in \citet{2013MNRAS.434.1513D, 
garcia2022muse}. 
IFUs have already shown how physical conditions and chemical abundances can vary across a given PN. However, until 
direct point-to-point mapping of physical parameters and chemical abundances based on IFU observations are available 
for large samples of PNe from different studies, this type of research remains essential. Nevertheless, the 
currently best plasma diagnostics used to determine the same property for some individual PNe 
do not show the levels of agreement hoped for and for all the reasons given above. Nevertheless, we believe we have 
produced the most internally self-consistent and robust data set to date for PNe abundance studies.

\subsection{Correlations between the abundances \texorpdfstring{$\alpha$}--element}
\label{sec:alpha}
\noindent
The $\alpha$-elements, including O, Ne, S, and Ar, are produced via the alpha processes during the evolution of massive 
stars, where nuclear fusion transforms helium into elements heavier than carbon. As these elements share common synthesis 
sites in massive stars \citep[e.g.][]{henry2008multiwavelength,esteban2020carbon}, their elemental abundances are expected to vary in lockstep. Therefore, in low- to- intermediate-mass stars, including the PN progenitors, their abundance ratios should remain constant. Similarly, Cl is produced by single particle (proton or neutron) capture by an S or Ar isotope, 
and therefore, indicating that it should exhibit similar behaviour \citep{esteban2015radial}. Thus, generally, the 
abundances of alpha elements observed in PNe reflect the amounts that were initially present in the star-forming gas of the progenitor star, which enables PNe to serve as a tool for establishing the metallicity of the ISM during the formation of PN progenitor stars \citep{kwitter2022planetary}. 

\begin{figure}
    \centering
    \caption{Abundances for $\log$(Ne/H), $\log$(Ar/H) and $\log$(Cl/H) for all PNe in our sample that 
        provide an estimate plotted against $\log$(O/H), with error bars are from our measurement uncertainties. The black solid line in each panel represents the best-fit to the data. The best-fit parameters and associated uncertainties are presented in Table.\ref{tab:alpha_line_fit}. For Ne and Ar, the dashed orange lines show the fit to abundances for H~{\sc ii} regions and BCGs from the literature. The orange dashed line for Cl presents an analogous fit based on abundances of Galactic disk H~{\sc ii} regions from \citet{arellano2020galactic}. The red asterisks denote the solar abundances.}
    \label{fig:alpha_line}
    \includegraphics[width=0.235\textwidth]{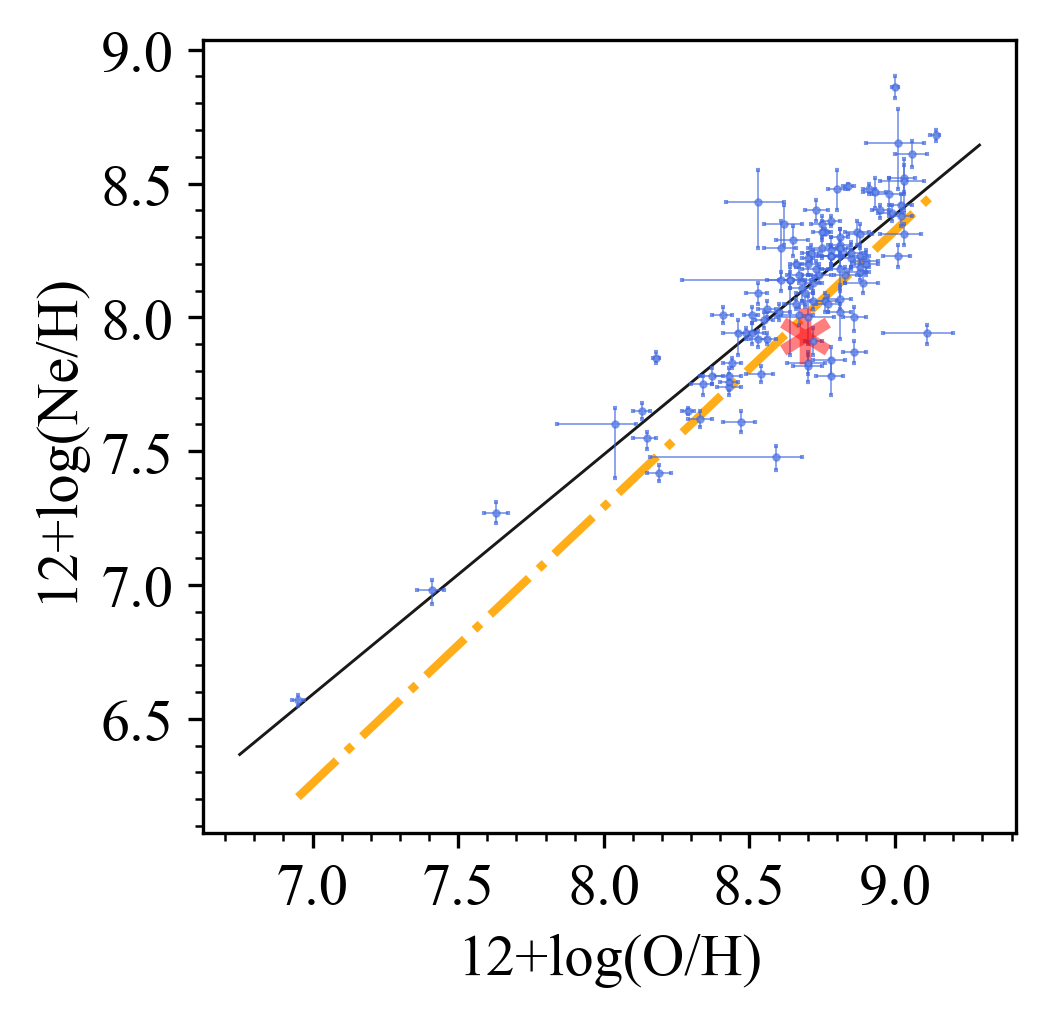}
    \includegraphics[width=0.235\textwidth]{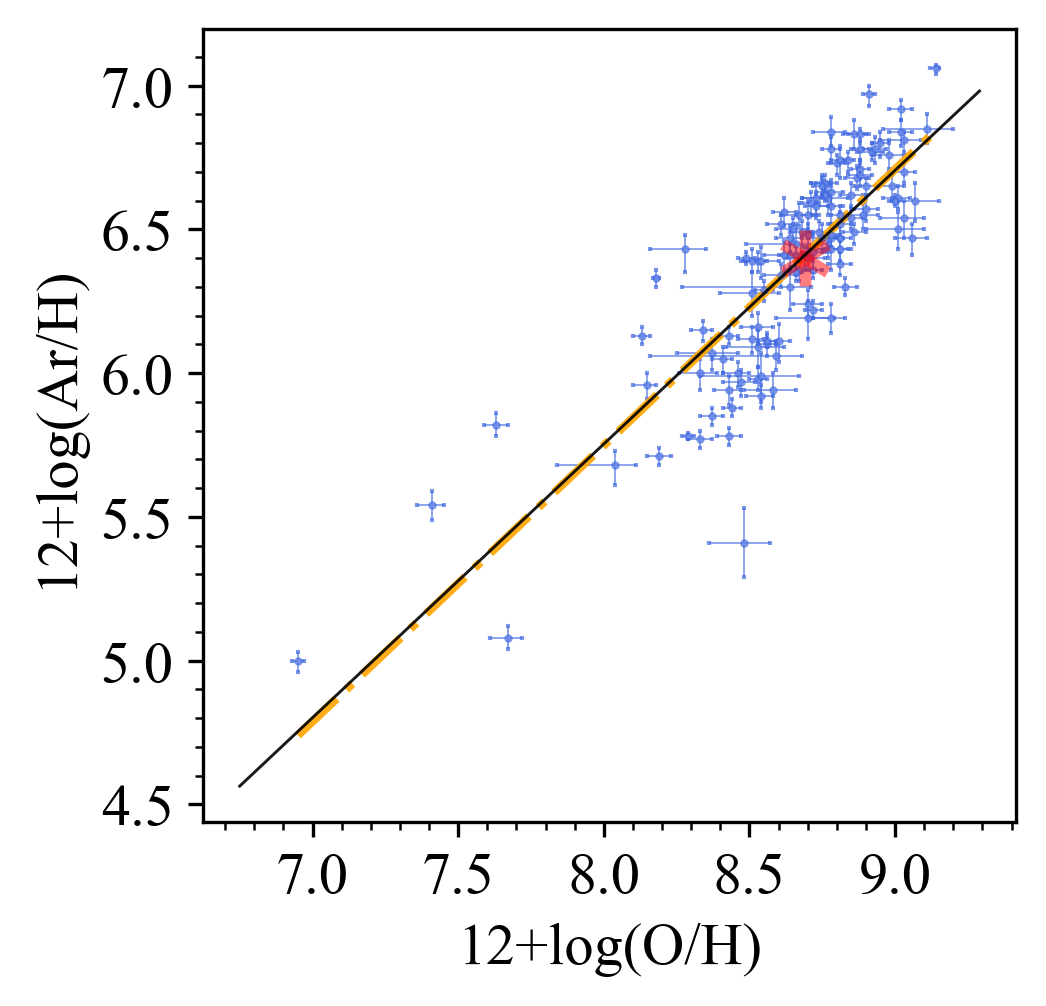}
    \hfill
    \includegraphics[width=0.235\textwidth]{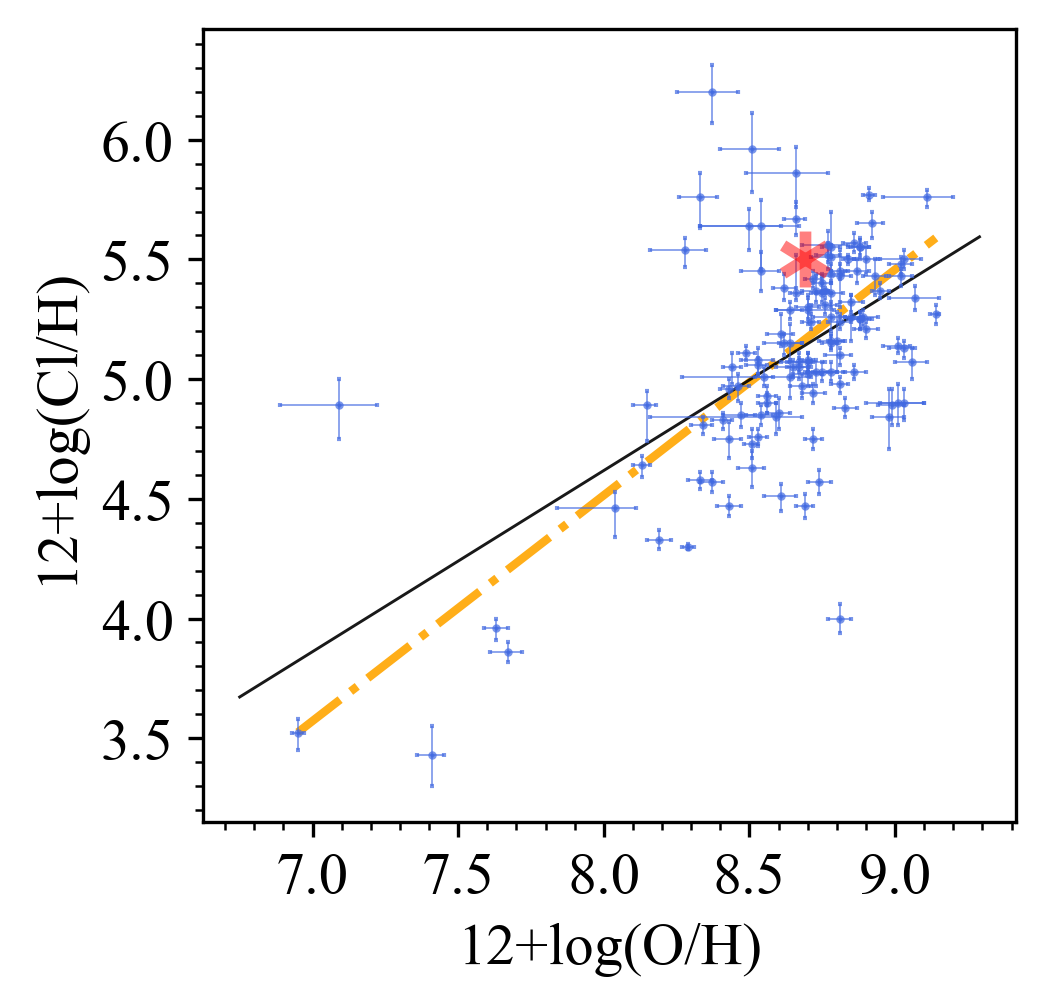}
\end{figure}

Among these elements, O is commonly used as a proxy for metallicity in ionised gases around stars, and a lockstep behaviour between $\alpha$ elements has been demonstrated in H~{\sc ii} regions and blue compact galaxies (BCG), with a strong correlation between elemental abundances of Ne, Ar and S, and O/H and a slope close to unity on a log scale. However, previous studies on PN have observed a slightly weaker correlation. To investigate the lockstep behaviour with PN abundances compiled in this work, we compared the abundances of $\alpha$-elements Ne and Ar, along with Cl, with that of oxygen in Fig.~\ref{fig:alpha_line}. {The plot showing the relationship between S/H and O/H will be presented in the next paper in the series (Tan~\&~Parker, submitted), where we specifically address the well-known `sulfur anomaly' observed in PNe. Thus, it is not included in the current paper.}

Each element-versus-oxygen plot was fitted with a straight line using the least squares fitting method. The slopes, intercepts of best-fit lines, and correlation coefficients are summarised in Table~\ref{tab:alpha_line_fit}. For Ne and Ar, the dashed line represents the best-fit to a collection of published H~{\sc ii} region and BCG abundances (H2BCG) that clearly demonstrating the lockstep behaviour of $\alpha$ elements, with abundances of H~{\sc ii} regions from \citet{kennicutt2003composition} and BCGs compiled by \citet{izotov1999heavy}. The analogous fit for Cl was inferred from a study of Galactic disk H~{\sc ii} regions by \citet{arellano2020galactic}. The red asterisks in the figure denote solar abundances. 

Previous studies on PN abundances \citep[e.g.]{henry2004sulfur, milingo2010alpha} have observed the lockstep behaviour with correlation coefficients of around 0.7 of Ne and Ar with a slope around 0.9 \citep{kwitter2022planetary}. However, this study obtained a much stronger correlation than those in the literature of around 0.85. The best-fit line for Ar/h versus O/H is highly consistent with the H2BCG results, suggesting that the previous less apparent lockstep behaviour observed in PNe could largely be attributed to observational and measurement errors. However, the Ne/H results from PNe do not agree with the best-fit of H2BCG results at extremely sub-solar abundances, which could be due to uncertainties in ICFs used to account for unobserved ionization stages more dominant than Ne$^{2+}$. As a result, the slope derived for Ne/H versus O/H is smaller than unity.

As previously mentioned in Sec.~\ref{sec:chlorine}, the determination of Cl/H is typically subject to uncertainty. The Cl/H versus O/H plot appears messier, exhibiting large scatters and increased measurement uncertainty at extremely sub-solar and super-solar abundances. Among all elements, the correlation between Cl/H and O/H is the weakest, with a correlation coefficient of $r=0.59$.

\setlength{\tabcolsep}{3.5pt}
\begin{table}
\centering
\begin{tabular}{cccccc}
\hline X/H vs. O/H & Slope & y-Intercept  & r & \# of PNe \\
\hline Ne & $0.90 \pm 0.05$ & $\ \ 0.32 \pm 0.46$ & 0.86 & 105 \\
Ar & $0.95 \pm 0.06$ & $-1.86 \pm 0.48$ & 0.84 & 122 \\
Cl & $0.76 \pm 0.10$ & $-1.44 \pm 0.83$ & 0.59 & 119 \\
\hline
\end{tabular}
\caption{Least squares fit parameters and Pearson's r values for $\alpha$-element correlations. The straight line was fit to $12+\log$(O/H) and $12+\log$(X/H). The number of PNe used is shown in the last column.}
\label{tab:alpha_line_fit}
\end{table}

To conclude, the expected lockstep behaviour for alpha elements is more prominent in our data compared to previous 
PN abundance studies. We observe a strong correlation between O/H and both Ar/H and Ne/H on a log scale, as reported in the literature. The lockstep association between Cl and O is not clearly evident in our data, which we attribute to large measurement uncertainties.

\subsection{He and N abundances}
{He and N are elements that could be produced in more massive PN progenitors as suggested by
\citet{kaler1978enrichment} and \citet{kaler1990relation}.
We find 104 out of 124 PNe exhibit He/H measurements greater than solar (i.e. $>10.93$) while with several objects with measured $12+\log$(He/H) lower than 10.3. The derived N enrichment, $\log$(N/O), range from $-1.67$ to 0.43, with a median value of $-0.48_{-0.37}^{+0.29}$~dex. This is 0.09 and 0.15 dex lower than the results compiled in KB94 for southern PNe and in CGS09 for the bulge PNe, respectively. We remark there that the KB94 ICF scheme we adopted for N abundance in this study uses N/O$=$N$^{+}$/O$^{+}$ and we observe large discrepancies in N/H when comparing with the literature. This discrepancy could be due to the subtraction of the recombination contribution of [N~{\sc ii}] and [O~{\sc ii}] auroral lines. In this case, ignoring the recombination correction could lead to significant systematic uncertainties.}

\begin{figure}
    \centering
    \includegraphics[width = 0.43\textwidth]{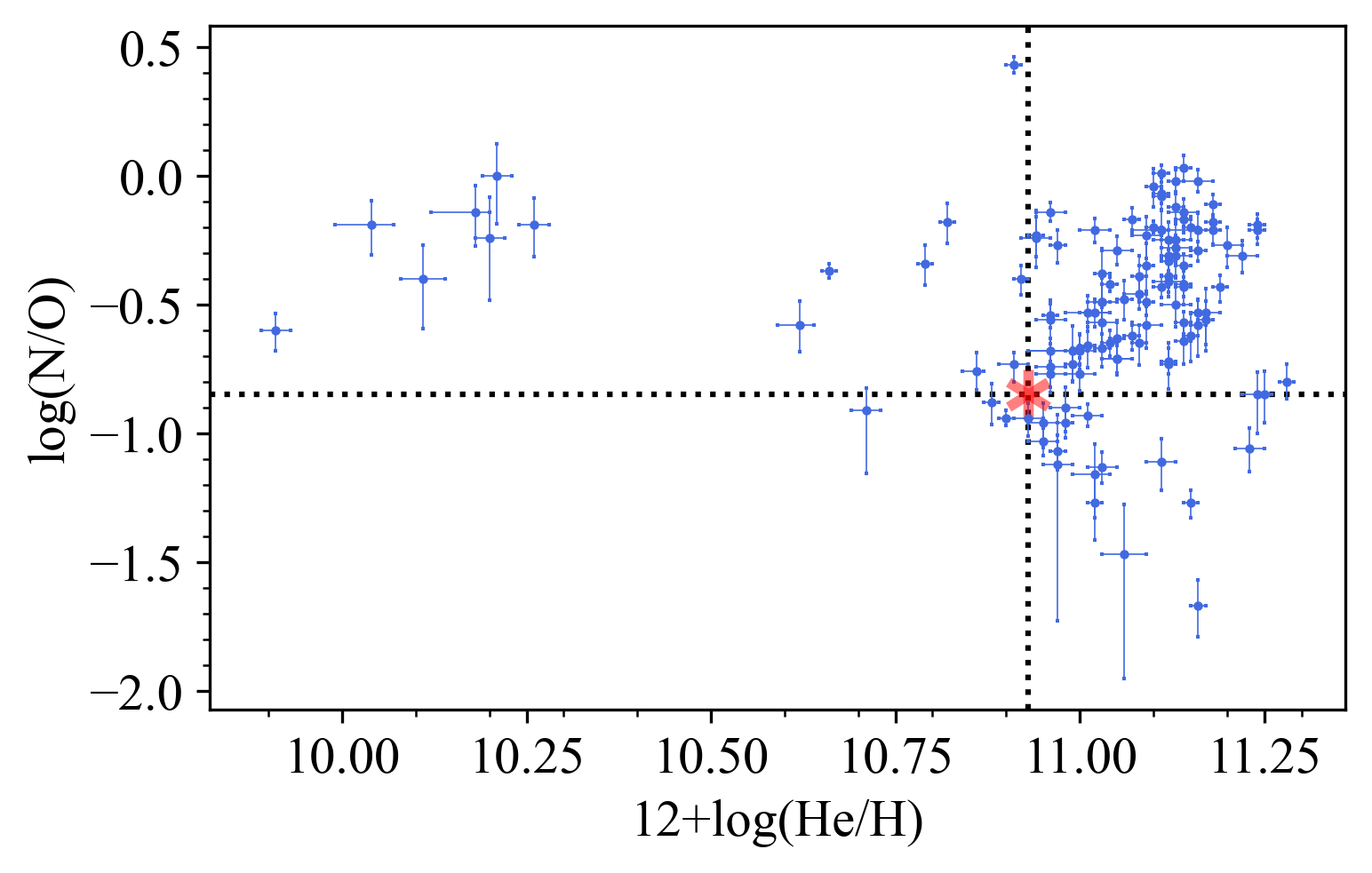}
    \caption{The N enrichment relative to O, $\log$(N/O), versus the he abundances $12+\log$(He/H), of 121 PNe in this sample. The horizontal black dotted line indicates the solar $12+\log$(He/H) values of 10.93 while the vertical line represents the solar nitrogen enrichment of $-0.85$. The red asterisks also denote the solar values.}
    \label{fig:no_he_plot}
\end{figure}

{Fig.~\ref{fig:no_he_plot} presents the relationship between N/O and He/H for all objects with measurable abundances. A significant fractions of objects in our sample exhibit N enrichment above the solar value. In addition to a few objects show extremely sub-solar He/H values, which are presumably underestimated, as well as a few objects with low N/O ratios in which the large uncertainties might present as discussed in \citet{wesson2018confirmation}, the overall pattern suggests that high N/O values are associated with high He/H. We will further use the measurements of N enrichment and He/H to investigate the central star properties of PNe in Paper~V in this series of papers.}

\subsection{Oxygen-poor PNe in this bulge sample}
Three PNe, PNG~004.0-03.0, PNG~005.0-03.9 and PNG~357.5+03.1 are measured with extremely 
sub-solar oxygen abundances, with 
$12+\log$(O/H)~$<7.5$~dex (~15.5 times lower than the solar abundance). Their $12+\log$(O/H) estimates are $6.95\pm0.02$, 
$7.41^{+0.04}_{-0.05}$ and $7.09^{+0.13}_{-0.20}$~dex, respectively. Despite large uncertainties in O/H for 
PNG~357.5+03.1, where the lower error is 46\%, the O/H values measured for PNG~004.0-03.0 and 
PNG~005.0-03.9 have good precision, with uncertainties below 0.1~dex, confirming their oxygen poor status based on our spectroscopic observations. 
The extremely sub-solar oxygen abundance of PNG~004.0-03.0 has been noticed in previous studies, such as \citet
{pena1991halo} and \citet{exter2004abundance}. \citet{torres1997hst} and \citet{miszalski2009binary} found that this anomalously low abundance is primarily due to low spatial resolution ground-based observations averaging over the strong spatial variation of [O~{\sc iii}~$\lambda$4363/$\lambda$5007. \citet{torres1997hst} measured an $12+\log$(O/H) value of 8.5~dex from \emph{HST} spectroscopy of the inner nebula, while \citet{miszalski2011influence} calculated an $12+\log$(O/H) of 8.3~dex in the outer nebula using VLT/FLAMES IFU observations.
PNG~005.0-03.9 has a complex structure classified as Eamrs according to \citet{tan2023morphologies}, all the 
temperature diagnostic lines, including [O~{\sc ii}], [N~{\sc ii}] and [O~{\sc iii}], give an estimate of electron 
temperatures greater than 15,000~K, which is similar to PNG~004.0-03.0. The low oxygen abundance measured with the VLT spectroscopy could be due to the low spatial resolution as for PNG~004.0-03.0.

PNG~357.5+03.1 is a very compact and low-excitation nebula with no [O~{\sc iii}]~$\lambda$4363 line detected in its spectra. Only the weak [O~{\sc iii}]~$\lambda$5007 lines of O$^{2+}$ were well-detected with low s/n.

\section{Conclusions}
\label{sec:conclusion}
We have presented logarithmic extinction coefficients, plasma diagnostics of $n_{\mathrm{e}}$ and $T_{\mathrm{e}}$ and elemental abundances of He, N, O, Ne, S, Ar and Cl for a well-defined sample of 124 Galactic bulge PNe with high-quality, wide-wavelength coverage optical spectra obtained using the ESO VLT/FORS2 with consistent instrumental configurations and observing strategies. We believe this represents the most self-consistent and well characterised sample of PNe available with high s/n spectra, from which well-determined abundances are compiled. For 34 PNe, the abundance determinations are presented for the first time, significantly adding ($\sim14$\%) to the Bulge PNe population abundances. Moreover, for an additional 6 Bulge PNe, we provide the first reliable abundances. Comparing our work with the best previous Bulge PNe abundances in the literature for PNe overlapping with our sample demonstrates the robustness and overall reliability and consistency of this sample.

The physical parameters in PNe exhibit a wider range in this study. The low-to-medium resolution spectroscopic observation from the VLT exhibit excellent consistency with previous high-resolution spectra with 2-m class telescope, demonstrating the reliability of weak line detection, the line deblending methods and our line flux measurement, particularly for weak recombination lines. Our abundance compilation, which adopted updated atomic data and ICF schemes introduced in DMS14, results in overall higher abundances of alpha elements than solar compared to the general abundance pattern in the literature. The lockstep behaviour of alpha elements, which was less evident in previous studies, is clearly observed in our results. 

The further analysis and investigation of these abundances form the basis of the following paper in the series.

\section*{Acknowledgements}
\noindent
{We are very grateful to the anonymous referee for his/her insightful comments that have significantly contributed to the improvement of the final version of this paper.} ST thanks HKU and QAP for provision of an MPhil scholarship and a research assistant position. QAP thanks the Hong Kong Research Grants Council for GRF research support under grants 17326116 and 17300417. AAZ acknowledges support from STFC under grant ST/T000414/1. 

This paper is based on observations made with the ESO VLT under programme IDs 095.D-0270(A), 097.D-0024(A), 099.D-0163(A), and 0101.D-0192(A) (PI: Rees). This research made use of NASA’s Astrophysics Data System; the SIMBAD database, operated at CDS, Strasbourg, France; \textsc{alfa}: an automated line fitting algorithm \citep{wesson2016alfa}; \textsc{neat}: nebular empirical analysis tool \citep{wesson2012understanding}; APLpy, an open-source plotting package for Python hosted at \url{http://aplpy.github.io}; Astropy, a community-developed core Python package for Astronomy \citep{robitaille2013astropy}; MATPLOTLIB, a Python library for publication quality graphics \citep{barrett2005matplotlib}; {\sc chianti}, a collaborative project involving George Mason University, the University of Michigan (USA), and the University of Cambridge (UK).

\section*{Data Availability}
\noindent
The raw VLT/FORS2 spectroscopic data can be accessed through the ESO archive facility at \url{http://archive.eso.org/}. The codes for the analysis, as well as intermediate data products are available from ST under reasonable request.



\bibliographystyle{mnras}
\bibliography{Bulge} 







\bsp	
\label{lastpage}
\end{document}